\documentclass[rmp,aps,preprint,nofootinbib,endfloats]{revtex4}

\usepackage{graphicx}
\usepackage{epstopdf}

\makeatletter
\def\underbracket{%
\@ifnextchar[{\@underbracket}{\@underbracket [\@bracketheight]}%
}
\def\@underbracket[#1]{%
\@ifnextchar[{\@under@bracket[#1]}{\@under@bracket[#1][0.4em]}%
}
\def\@under@bracket[#1][#2]#3{%\message {Underbracket: #1,#2,#3}
\mathop{\vtop{\m@th \ialign {##\crcr $\hfil \displaystyle {#3}\hfil $%
\crcr \noalign {\kern 3\p@ \nointerlineskip }\upbracketfill {#1}{#2}
   \crcr \noalign {\kern 3\p@ }}}}\limits}
\def\upbracketfill#1#2{$\m@th \setbox \z@ \hbox {$\braceld$}
    \edef\@bracketheight{\the\ht\z@}\bracketend{#1}{#2}
    \leaders \vrule \@height #1 \@depth \z@ \hfill
    \leaders \vrule \@height #1 \@depth \z@ \hfill \bracketend {#1}{#2}$}
\def\bracketend#1#2{\vrule height #2 width #1\relax}
\makeatother

\def\inbar{\,\vrule height1.5ex width.4pt depth0pt}
\def\IR{\relax{\rm I\kern-.18em R}}
\def\IC{\relax\hbox{$\inbar\kern-.3em{\rm C}$}}

\def\nuc#1#2{${}^{#1}$#2}
\def\mee{$\langle m_{\beta\beta} \rangle$}

\def\ml{$m_{lightest}$}

\def\mb{$\langle m_{\beta} \rangle$}
\def\BBz{$\beta\beta(0\nu)$}
\def\BBm{$\beta\beta(0\nu,\chi)$}
\def\BBt{$\beta\beta(2\nu)$}
\def\BB{$\beta\beta$}
\def\Mz{$M_{0\nu}$}
\def\Mt{$M_{2\nu}$}
\def\MtG{$M^{GT}_{2\nu}$}           %Gamov-Teller
\def\MtF{$M^{F}_{2\nu}$}                %Fermi
\def\Tz{$T^{0\nu}_{1/2}$}
\def\Tt{$T^{2\nu}_{1/2}$}

\def\Rz{$\Gamma_{0\nu}$}            %0 nu decay rate
\def\Rt{$\Gamma_{2\nu}$}            %2 nu decay rate
\def\ms{$\delta m_{\rm sol}^{2}$}

\def\ts{$\theta_{\rm sol}$}
\def\ta{$\theta_{\rm atm}$}
\def\tot{$\theta_{13}$}
\def\gpp{$g_{pp}$}                  % the g_pp of QRPA fame
\def\qval{$Q_{\beta\beta}$}                 % The Q-value
\def\MJ{{\sc Majorana}}             %Majorana project name
\def\be{\begin{equation}}
\def\ee{\end{equation}}
\def\today{\space\number\day\space\ifcase\month\or January\or February\or
    March\or April\or May\or June\or July\or August\or September\or October\or
    November\or December\fi\space\number\year}
\newcommand{\simgt}{\ \raisebox{-.25ex}{$\stackrel{>}{\scriptstyle \sim}$}\ }

\begin{document}
\title{Double Beta Decay, Majorana Neutrinos, and Neutrino Mass}

\author{Frank T. Avignone III}
\email{avignone@sc.edu}
\affiliation{Department of Physics and Astronomy, University of South Carolina,
Columbia, SC, USA 29208}
\author{Steven R. Elliott}
\email{elliotts@lanl.gov}
\affiliation{Los Alamos National Laboratory, Los Alamos, NM, USA 87545}
\author{Jonathan Engel}
\email{engelj@physics.unc.edu}
\affiliation{Department of Physics and Astronomy, University of North Carolina,
Chapel Hill, NC, USA 27599-3255}

\begin{abstract}
The theoretical and experimental issues relevant to neutrinoless
double-beta decay are reviewed. The impact that a direct observation
of this exotic process would have on elementary particle physics,
nuclear physics, astrophysics and cosmology is profound.  Now that
neutrinos are known to have mass and experiments are becoming more
sensitive, even the non-observation of neutrinoless double-beta
decay will be useful.  If the process is actually observed, we will
immediately learn much about the neutrino. The status and discovery
potential of proposed experiments are reviewed in this context, with
significant emphasis on proposals favored by recent panel
reviews. The importance of and challenges in the calculation of
nuclear matrix elements that govern the decay are considered in
detail. The increasing sensitivity of experiments and improvements
in nuclear theory make the future exciting for this field at the
interface of nuclear and particle physics.
\end{abstract}

\pacs{11.30.Fs, 14.60.Pq, 23.40.-s}

%\date{May 2001}
\maketitle
\tableofcontents

%\section{To Do List}
%\begin{enumerate}
% %\item In the abstract, it claims we discuss the impact on astrophysics and
%cosmology and we don't. In general the abstract is rather weak. It doesn't draw the
%reader in.
 %\item in introduction overview, need figure of decay scheme and \BB\ diagrams
 %\item in introduction overview or in decay section, add spectrum
 %\item eqn 31 needs help with formatting.
 %\item get permission for figure/photo usage
 %\item have rest of experimental sections vetted
%\item should we include a KKDC spectrum figure in the claim of evidence section?
%\end{enumerate}

\section{INTRODUCTION}
\label{sec:intro}

Neutrinoless double beta decay (\BBz) is a very slow
lepton-number-violating nuclear transition that occurs if
neutrinos have mass (which they do) and are their own antiparticles.  An initial
nucleus $(Z,A)$, with proton number $Z$ and total nucleon number
$A$ decays to $(Z + 2,A)$, emitting two electrons in the process. A
related transition, called two-neutrino double-beta decay (\BBt)
results in the emission of two electron antineutrinos in addition to
the electrons, and occurs whether or not neutrinos are their own
antiparticles.  \BBt\ has in fact been observed in a number of
experiments.  With the exception of one unconfirmed observation, on
the other hand, \BBz\ has never been seen, and searches  for it are
ongoing in a number of laboratories around the world. Other even
slower and more exotic process, including double-positron decay,
double-electron capture, and neutrinoless decay with the emission of
a hypothetical particle called the Majoron (\BBm), have likewise
never been observed.

The development of effective-field theory and grand-unification schemes in the
late 1970's and early 1980's led to the expectation that  neutrinos  are
identical with their antiparticles and have nonzero mass, and engendered
renewed interest in \BBz\ experiments. More recently, neutrino-oscillation
experiments  have yielded compelling evidence that the three observed flavors
of neutrinos are not mass eigenstates but rather linear combinations of those
eigenstates (at least two of which have nonzero mass eigenvalues). These
experiments also allow the electron neutrino to mix significantly with
the heaviest of the mass eigenstates.  If it does, the effective neutrino mass
will be large enough that  \BBz\ may well be observed in
experiments currently under construction or development. An observation would
establish that neutrinos are ``Majorana'' particles ($\nu \equiv
\overline{\nu}$, roughly speaking), and a measurement of the decay rate, when
combined with neutrino-oscillation data, would yield insight into all three
neutrino-mass eigenstates. This article is motivated by the recent developments
in neutrino physics and by the interest among physicists throughout the world
in a coherent experimental \BBz\ program.

\subsection{The Early History}

Double-beta decay was first considered in a 1935 paper by Maria Goeppert-Mayer
\cite{goe35}.  The author, who acknowledged Eugene Wigner for suggesting the
problem, derived an expression for the decay rate and estimated a half-life of
$\sim 10^{17}$ y for a decay with the emission of two electrons and two
anti-neutrinos ($\overline{\nu}$), carrying about 10 MeV of energy. Two years
later Ettore Majorana formulated a theory of neutrinos in which there was no
distinction between $\nu$ and $\overline{\nu}$ \cite{maj37}, and suggested an
experimental test of his hypothesis with a reaction similar to 
 $\overline{\nu}_{e} + ^{37}Cl \rightarrow ^{37}Ar + e^{-}$, which was
later searched for (and not found) by Raymond Davis \cite{dav55}.  It was
Giulio Racah, however, who first suggested testing Majorana's theory with \BBz\
\cite{rac37}. In 1939 Furry calculated approximate rates for \BBz\
\cite{fur39}, and in 1952 Primakoff~\cite{pri52} calculated the
electron-electron angular correlations and electron energy spectra for both
\BBt\ and \BBz, producing a useful tool for distinguishing between the two
processes. These early contributions set the stage for many years of
experimental and theoretical activity.

The review by \textcite{hax84} contains a chronology of experiments from 1948
through 1983.  There were some early claims of observation. \textcite{fir49}
reported observing the \BB\ of $^{124}$Sn in a laboratory experiment, but
retracted the claim later~\cite{fir52}.  The first geochemical observation of
\BB, with an estimated half-life of $T^{\beta\beta}_{1/2}\;(^{130}Te) = 1.4
\times 10^{21}$ y, was reported in 1950~\cite{ing50}. The first actual
laboratory observation of \BBt\ was not made until 1987~\cite{ell87}.  Here, we
concentrate on experiments developed since the late 1980s, referencing earlier
work where appropriate. The early developments have been covered well in other
reviews, for example
\textcite{pri81,hax84,doi85,avi88,tom91,moe94,fae98,suh98,ell02,tre02,zde02b,ell04,avi05,eji05}.

\subsection{Overview of Theory and Recent Experimental Developments}

\begin{figure}
\vspace{9pt}
\begin{center}
\includegraphics[width=6cm]{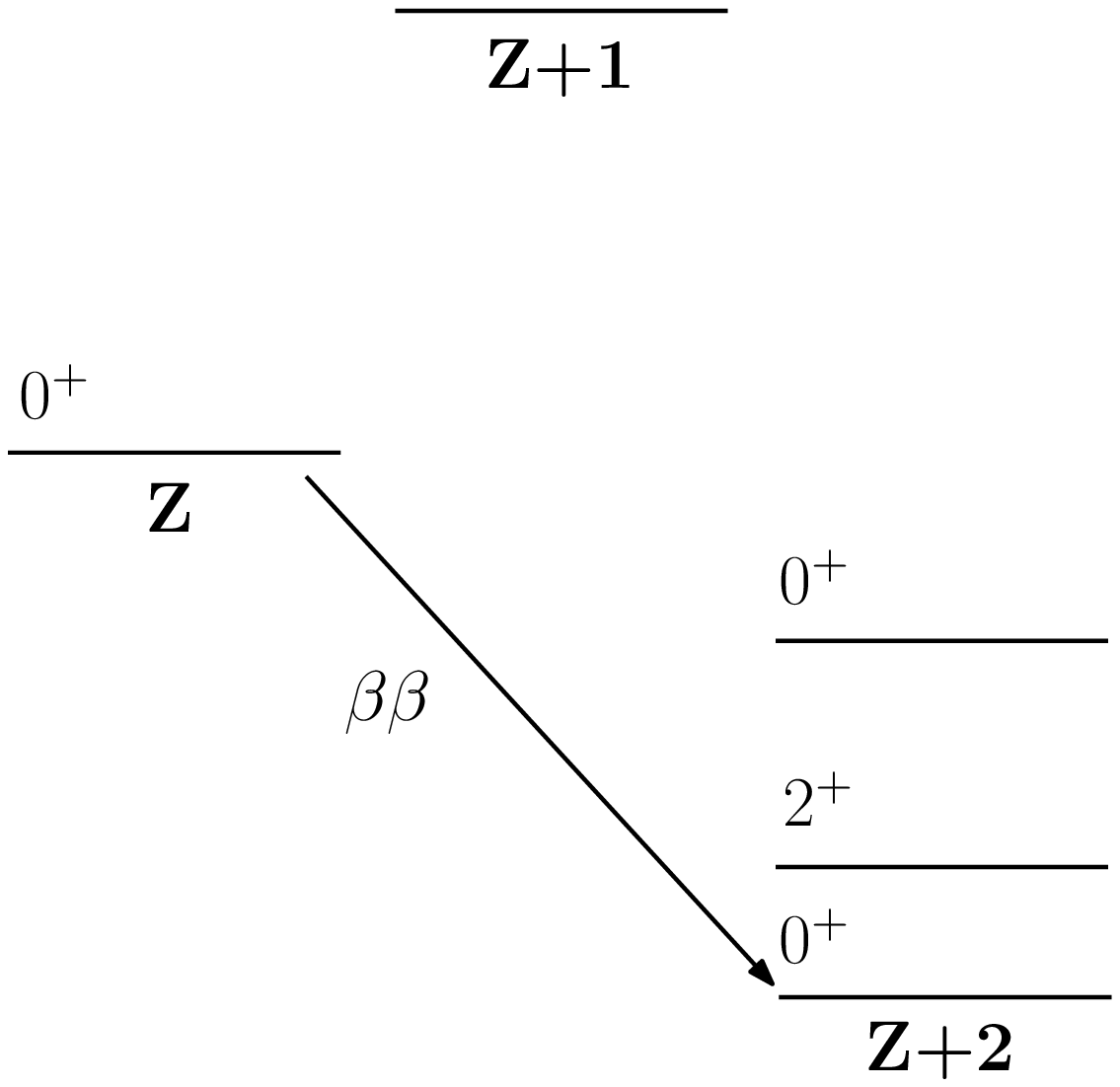}
\end{center}
\caption{A generic level diagram for \BB.}
\label{fig:LevelDiagram}
\end{figure}

\begin{figure}
\vspace{9pt}
\begin{center}
\includegraphics[width=5cm]{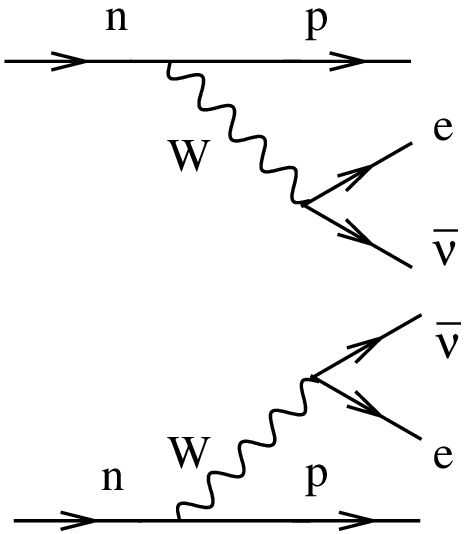}\\[.5cm]
\includegraphics[width=5cm]{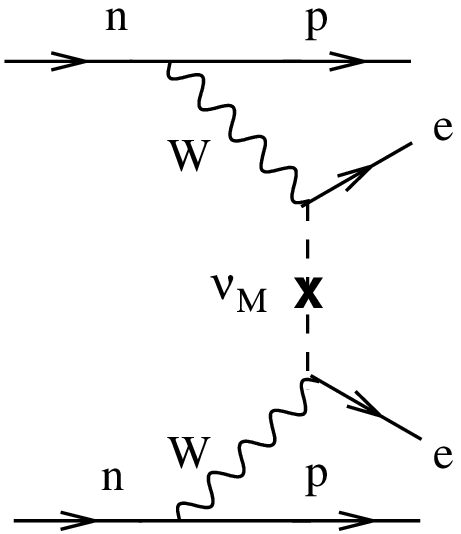}
\end{center}
\caption{Feynman Diagrams for \BBt\ (top) and \BBz\ (bottom).}
\label{fig:FenymanDiagrams}
\end{figure}

A typical \BB\ candidate is an even-even nucleus $(Z,A)$ which
pairing forces make more bound than its (Z+1,A) neighbor, but less
so than the (Z+2,A) nuclide, as shown in Fig.
\ref{fig:LevelDiagram}.  In Fig. \ref{fig:FenymanDiagrams} we depict
\BBt\ and neutrino-exchange-driven \BBz.  The rate of \BBt, which
has been measured in 10 isotopes (see Table~\ref{tab:TwoNeutrino}),
can be written as

\begin{equation}
\label{eq:TwoRate}
(\mbox{\Tt})^{-1} = G_{2\nu}(\mbox{\qval},Z)|\mbox{\Mt}|^{2}~,
\end{equation}

\noindent where $G_{2\nu}(\mbox{\qval},Z)$ is the four-particle phase space
factor, and \Mt\ is a ``nuclear matrix element'' for
this second-order process.  This decay conserves lepton number,
does not discriminate between Majorana and Dirac neutrinos,
and does not depend significantly on the masses of the neutrinos.  The rate of
\BBz, if driven by the
exchange of light Majorana neutrinos, is approximately

\begin{equation}
\label{eq:ZeroRate}
(\mbox{\Tz})^{-1} = G_{0\nu}(\mbox{\qval},Z)|\mbox{\Mz}|^2 \mbox{\mee}^2 ~,
\end{equation}

\noindent where $G_{0\nu}(\mbox{\qval},Z)$ is the phase space factor for the
emission of the two electrons, \Mz\ is another nuclear matrix element, and
\mee\ is the ``effective'' Majorana
mass of the electron neutrino:

\begin{equation}
\label{eq:firstmbb}
\mbox{\mee} \equiv |\sum_k m_k U_{ek}^2|~.
\end{equation}

\noindent Here the $m_k$'s are the masses of the three light neutrinos and $U$
is the matrix that transforms states with well-defined mass into states with well-defined
flavor (\emph{e.g.}, electron, mu, tau). 
Equation~\ref{eq:ZeroRate} gives the \BBz\ rate if the exchange of light 
Majorana neutrinos with left handed interactions
is responsible. Other mechanisms are possible (see Sections~\ref{sec:rate} and \ref{sec:AltMech}), but they
require the existence of new particles and/or interactions in addition to requiring that
neutrinos be Majorana particles. Light-neutrino exchange is therefore, in some sense,
the ``minima'' mechanism and the most commonly considered.

That neutrinos mix and have mass is now accepted wisdom.
Oscillation experiments constrain $U$ fairly well ---
Table~\ref{Tab:MixingParameters} summarizes our current knowledge
--- but they determine only the differences between the squares of
the masses $m_k$ (\emph{e.g.}, $m_2^2-m_1^2$) rather than the masses
themselves. It will turn out that \BBz\ is among the best ways of
getting at the masses (along with cosmology and $\beta$-decay
measurements), and the only practical way to establish that
neutrinos are Majorana particles.

To extract the effective mass from a measurement, it is customary to
define a nuclear structure factor $F_{N}\equiv
G_{0\nu}(\mbox{\qval},Z)|\mbox{\Mz}|^{2}m_e^2$, where  $m_{e}$ is
the electron mass. (The quantity $F_N$ is sometimes written as
$C_{mm}$.) The effective mass \mee\ can be written in terms of the
calculated $F_N$ and the measured half life as

\begin{equation}
\mbox{\mee} = m_{e}[F_{N}\mbox{\Tz}]^{-1/2}~.
\end{equation}

\noindent The range of mixing matrix values given below in
Table~\ref{Tab:MixingParameters}, combined with calculated values for $F_{N}$,
allow us to estimate the half-life a given experiment must be able to
measure in order to be sensitive to a particular value of \mee.  Published
values of $F_N$ are typically between $10^{-13}$ and
$10^{-14}$ y$^{-1}$. To reach a sensitivity of \mee $\approx$ 0.1 eV,
therefore, an experiment must be able to observe a half life of
$10^{26} - 10^{27}$ y.  As we discuss later, at this level of sensitivity an
experiment can draw important conclusions whether or not the decay is observed.

The most sensitive limits thus far are from the Heidelberg-Moscow
experiment: $T^{0\nu}_{1/2}(^{76}Ge)\geq 1.9 \times 10^{25}$ y \cite{bau99}, the
IGEX experiment: $T^{0\nu}_{1/2}(^{76}Ge)\geq 1.6 \times 10^{25}$ y
\cite{aal02a, aal04}, and the CUORICINO experiment
$T^{0\nu}_{1/2}(^{130}Te)\geq 3.0 \times 10^{24}$ y \cite{arn05,arn07}.  These
experiments contained 5 to 10 kg of the parent isotope and ran for several
years. Hence, increasing the half-life sensitivity by a factor of about 100,
the goal of the next generation of experiments, will require hundreds of kg of
parent isotope and a significant decrease in background beyond the present
state of the art (roughly 0.1 counts/(keV kg y).

It is straightforward to derive an approximate analytical expression
for the half-life to which an experiment with a given level of background is
sensitive~\cite{avi05}:

\begin{equation}
\label{eq:Sensitivity}
\mbox{\Tz}(n_{\sigma}) = \frac{4.16 \times 10^{26}y}{n_{\sigma}}\left( \frac{\varepsilon a}{W}\right)\sqrt{\frac{Mt}{b\Delta(E)}}.
\end{equation}

\noindent Here $n_{\sigma}$ is the number of standard deviations corresponding
to a given confidence level (C.L.) --- a CL of 99.73\% corresponds to
$n_{\sigma} = 3$ --- the quantity $\varepsilon$ is the event-detection and
identification efficiency, $a$ is the isotopic abundance, $W$ is the molecular
weight of the source material, and $M$ is the total mass of the source.  The
instrumental spectral-width $\Delta(E)$, defining the signal region, is related to the energy resolution at the
energy of the expected \BBz\ peak, and $b$ is the specific background rate in
counts/(keV kg y), where the mass is that of the source, as opposed to the
isotope.  Equation~(\ref{eq:Sensitivity}) is valid only if the background level
is large enough so that the uncertainty is proportional to
$\sqrt{b\Delta(E)}$.  For a 200-kg \nuc{76}{Ge} experiment with a background
rate of 0.01 counts/(keV kg y) and an energy resolution of 3.5 keV, running for
5 years, the values for these parameters are $Mt=10^{3} kg \cdot y,
\varepsilon=0.95$, $a$ = 0.86, $W$ = 76, and $\Delta(E)$= 3.5 keV. This results
in a 4$\sigma$ half-life sensitivity of $\mbox{\Tz}(4\sigma,\; ^{76}Ge) = 1.9
\times 10^{26}$ y.  The background rate quoted above is conservative for a Ge
experiment,  only a factor of 6 below that of the Heidelberg-Moscow and IGEX
experiments. A background rate of 0.001 counts/(keV kg y) would allow a
4$\sigma$ discovery with \Tz\ = $6 \times 10^{26}$ y.  But an experiment with even
modestly lower efficiency or poorer resolution must attain much lower
background rates to have comparable sensitivity.

These numbers characterize the level future experiments will have to reach to
make a significant contribution to the field. Later, we will discuss a number
of proposed experiments and attempt to estimate their sensitivity.

\subsection{The Claimed Observation}

In 2001, a subset of the Heidelberg-Moscow
collaboration~\cite{kla01a,kla01b,kla03a,kla03b,kla06} claimed to observe evidence
for a \BBz\ peak in the spectrum of their \nuc{76}{Ge}
experiment at 2039 keV.  This claim and later papers by the same group elicited
a number of critical replies, for example Refs.
\textcite{har01,fer02,aal02b,zde02a}.  But whether or not the result is valid,
the experiment was the most sensitive so far.  The parameter values were $Mt$ =
71.7 kg y,  $b$ = 0.11 counts/(keV kg y), $\varepsilon$ = 0.95, $a$ = 0.86, $W$
= 76, and $\Delta(E) = $ 3.27 keV. The number of counts under the  peak at 2039
keV was  $28.75 \pm 6.86$~\cite{kla04b}.  Substitution into
Eq.~(\ref{eq:Sensitivity})  yields $T^{0\nu}_{1/2}(4\sigma, ^{76}Ge) = 1.6
\times 10^{25}$ y, a lifetime comparable to the claimed most probable value, $2.23
\times 10^{25}$ y.  At least nominally, the experiment had a 4$\sigma$ discovery
potential, and cannot be dismissed out of hand. Since this analysis does not
account for statistical fluctuations, background structure, or systematic
uncertainties, the actual confidence level could be significantly
different.  But the only certain way to confirm or refute the claim is with
additional experimentation, preferably in \nuc{76}{Ge}.

To this end, the GERDA experiment is under construction in the LNGS
\cite{abt04} and the \MJ\ project \cite{gai03} is being developed in the
U.S. The CUORICINO experiment in the LNGS\cite{arn05} uses \nuc{130}{Te}, and
is the most sensitive experiment currently operating, with a lower bound of
$T^{0\nu}_{1/2}(^{130}Te)\geq 3 \times 10^{24}$ y.  This limit is at or near
the sensitivity needed to test the 2001 claim, but uncertainty in the
calculated value of the nuclear matrix element \Mz\ (or, equivalently, $F_N$)
will preclude a definitive statement.

Foiled by the nuclear matrix elements!  One can see even in this brief overview
how nice it would be to have accurate matrix elements.  We will address the
issue of how well they can be calculated later.

\section{MAJORANA NEUTRINOS}
\label{sec:NuMass}
%Majorana Neutrinos

As we will see, \BBz\ cannot occur unless
neutrinos are Majorana particles, {\emph i.e.} their own antiparticles.  We
therefore briefly review the nature of neutral fermions.  Much of what we say
here is covered in a similar fashion but much greater depth in \textcite{bil87}.

We can project an arbitrary 4-spinor $\Psi$ onto states of definite
chirality, which are singlets under one of the two $SU(2)$ algebras
that make up the Lorentz algebra $SO(3,1)$. We define the left- and
right-handed projections as $\Psi_{L,R}=[(1\mp\gamma_5)/2 ]\Psi$.
Because of the minus sign in the Minkowski metric and the resulting
need to work with $\overline{\Psi} \equiv \Psi^{\dag}\gamma_0$
rather than $\Psi^{\dag}$ alone, a Lorentz scalar cannot be
constructed by contracting two-left handed spinors in the usual way.
Instead one must contract a left-handed spinor with a right-handed
one.  The (scalar) term in the Lagranginan obtained in this way is
called the ``Dirac mass'' term:

\begin{equation}
\label{eq:ld}
\mathcal{L}_D = -m_D \overline{\Psi}\Psi = -m_D \left(
\overline{\Psi_L}\Psi_R + \overline{\Psi_R}\Psi_L \right ) \quad .
\end{equation}

\noindent The terms above contract $\Psi^*_L$ with $\Psi_R$ (and vice versa),
with $\gamma_0$ flipping the chirality so that the contraction can be made.

If charge conservation isn't required, one can form a scalar by
combining $\Psi$ with itself rather than with $\Psi^*$.  Such a term
cannot exist in the Lagrangian for electrons or quarks because it
destroys or creates two particles of the same charge (or destroys a
particle while creating an antiparticle of opposite charge), but
nothing forbids it in the Lagrangian of neutral fermions such as
neutrinos. It violates lepton number, a global $U(1)$ symmetry that
tracks the difference between particle number and antiparticle
number, but there is nothing sacred about global symmetries. Indeed,
\BBz\ can't occur unless lepton-number symmetry is violated.

To construct a Lorentz scalar from two $\Psi$'s, we note that the
``charge conjugate" of $\Psi$, defined up to an arbitrary phase as
$\Psi^c \equiv \gamma^2 \Psi$, transforms in the correct way; its
chirality is opposite that of $\Psi$ itself because $\gamma^2$ and
$\gamma_5$ anticommute.  Thus $(\Psi_L)^c$ is right-handed, and we
can construct left- and right-handed ``Majorana" mass terms of the
form

\begin{equation}
\label{eq:lm}
\mathcal{L}_M = -\frac{m_L}{2} \left( \overline{(\Psi_L)^c} \Psi_L +
h.c.\right)- \frac{m_R}{2} \left( \overline{(\Psi_R)^c} \Psi_R + h.c.\right)~.
\end{equation}

\noindent Although we have used all four Dirac components in this equation, it is possible that only
 the two in $\Psi_L$ actually exist if there is no Dirac mass term.

Equations (\ref{eq:ld}) and (\ref{eq:lm}) can be generalized to include $N$
flavors.  If, following \textcite{bil87} and letting $\Psi$ be the neutrino
field $\nu$, we define

\begin{equation}
\label{eq:ndef}
n_L \equiv \left[ \begin{array}{c}
              \nu_L \\
              (\nu_R)^c
            \end{array} \right]~,
\end{equation}

\noindent where

\begin{equation}
\label{eq:nudef}
\nu_L \equiv \left[ \begin{array}{c}
              \nu_{eL} \\
              \nu_{\mu L}\\
              \nu_{\tau L}\\
              \vdots
            \end{array} \right]~,\ \ \
(\nu_R)^c \equiv \left[ \begin{array}{c}
             (\nu_{eR})^c \\
             (\nu_{\mu R})^c\\
             (\nu_{\tau R})^c\\
              \vdots
            \end{array} \right]~,
\end{equation}

\noindent then we find for the mass part of the Lagrangian,

\begin{equation}
\label{eq:md} \mathcal{L}_{D+M} =  -\frac{1}{2} \overline{(n_L)^c}
\mathcal{M} n_L + h.c.~, ~~~\mathcal{M}=\left(
\begin{array}{cc} \mathcal{M}_L~ \mathcal{M}_D^T\\ \mathcal{M}_D~
\mathcal{M}_R
\end{array}
\right) ~,
\end{equation}

The (generically complex symmetric) $N \times N$ matrices
$\mathcal{M}_L$ and $\mathcal{M}_R$ are the Majorana mass terms and
the $N \times N$ matrix $\mathcal{M}_D$ is the Dirac term. Except in
special cases, the eigenstates of $\mathcal{L}_{D+M}$ will
correspond to Majorana particles. The Lagrangian is invariant under
CPT, a transformation that takes a particle into an antiparticle
with spin direction reversed. Since the eigenstates will in general
consist of $2N$ nondegenerate spinors, the components in each spinor
must transform into each under under CPT, rather than into an
orthogonal spinor, so that the neutrinos will be their own
antiparticles.  To see this more precisely, we note that
$\mathcal{M}$, if nondegenerate, can be rewritten in the form
$\mathcal{M} = (\mathcal{U}^{\dag})^T \hat{m} \mathcal{U}^{\dag}$,
where $\hat{m}$ is a diagonal matrix with $2N$ positive entries
$m_k$ and $\mathcal{U}$ is unitary. The mass Lagrangian then takes
the form:

\begin{equation}
\label{eq:mddiag} \mathcal{L}_{D+M}=
-\frac{1}{2}\sum_{k=1}^{2N} m_k \overline{(n_{kL}^\prime)^c} n_{kL}^\prime + h.c.~
= -\frac{1}{2} \sum_{k=1}^{2N} m_k \overline{\phi_k} \phi_k~,
\end{equation}

\noindent where $n_L=\mathcal{U} n_L^\prime$ and

\begin{equation}
\label{eq:phidef}
\phi_k = n_{kL}^\prime + (n_{kL}^\prime)^c =
\phi_k^c~,
\end{equation}

\noindent with only the cross terms surviving when the $\phi_k$ are written out
as in Eq.~(\ref{eq:phidef}).  Clearly, then, the physical eigenstates $\phi_k$
are Majorana particles.  The interacting left-handed flavor eigenstates are
linear combinations of the left-handed parts of these Majorana eigenstates
(which don't have well-defined chirality).

Dirac neutrinos are a special case, produced if, {\em e.g.},
$\mathcal{M}_L$ and $\mathcal{M}_R$ are zero.  We can see this most
easily in the case of the one-flavor mass matrix where
$\mathcal{M}_D$ is simply $m$;

\begin{equation}
\label{eq:degen}
\mathcal{M} = \left(\begin{array}{cc} 0& m \\ m& 0 \end{array}\right)~.
\end{equation}

\noindent The eigenvalues $m$ and -$m$ are obtained by diagonalizing this real
matrix with a real orthogonal matrix.  To get the two states with positive mass
$m$ we cannot use this real matrix as $\mathcal{U}$ but instead must
incorporate a phase that changes the sign of the second eigenvalue:

\begin{equation}
\label{eq:Udegen}
\mathcal{U} = \left(\begin{array}{rr} 1& i \\ 1& -i \end{array}\right)~.
\end{equation}

\noindent The phase in this matrix that mixes the degenerate Majorana states
$\phi_1$ and $\phi_2$ will cause the two to cancel in the neutrino-exchange
diagram via which Majorana neutrinos directly mediate \BBz\ decay.  And since
they are degenerate, orthogonal linear combinations $\chi_1\equiv
1/\sqrt{2}(\phi_1+i\phi_2)$ and $\chi_2\equiv 1/\sqrt{2}(\phi_1-i\phi_2)$, that
go into one another under charge conjugation, are also eigensates yielding
$\mathcal{L}_D=(-m/2)(\overline{\phi}_1 \phi_1 + \overline{\phi}_2 \phi_2) =
-m(\overline{\chi}_1 \chi_1 + \overline{\chi}_2 \chi_2)$.  The $\chi$'s make up
the lepton-number conserving Dirac neutrino and antineutrino.

We can also use the one-flavor case to illustrate the ``see-saw"
mechanism, a natural explanation for the existence of light Majorana
neutrinos \cite{gel79,yan79,moh80}.  Majorana mass terms for
left-handed neutrinos cannot be accommodated in the standard model
because those terms have different quantum numbers under
$SU(2)_L \times SU(2)_R$ than do the Dirac mass terms. But by
introducing new physics (e.g.\ new Higgs bosons with different
quantum numbers) at a very large mass scale $m_R$, extended models
can avoid this problem.  The result is often a mass matrix
$\mathcal{M}$ (in the one-flavor example, for which $\mathcal{M}_R$,
$\mathcal{M}_L$, and $\mathcal{M}_D$ become numbers $m_R$, $m_L$,
and $m_D$) with $m_R \gg m_D \gg m_L$, where $m_D$, which comes from
our familiar Higgs vacuum expectation value, is on the order of a
typical quark or lepton mass. Diagonalization yields eigenvalues
$m_1 \approx -m_D^2/m_R$ and $m_2 \approx m_R$. The matrix that
converts from eigenstates with positive mass to flavor eigenstates
is

\begin{equation}
\label{eq:seesaw} \mathcal{U} \approx  \left(\begin{array}{cc} i& m_D/m_R \\
-im_D/m_R& 1
\end{array}\right)~.
\end{equation}

\noindent In this scheme, neutrinos that we know about are much lighter than
other leptons because of the existence of other very heavy neutrinos. Even without
heavy neutrinos, the fact that the only  dimension-5 neutrino-mass operator in
standard-model fields is of Majorana form~\cite{wei79} leads one to expect
that massive neutrinos will be Majorana particles .

In the general $N$-flavor case

\begin{equation}
\label{eq:flavors}
\nu_{lL}=  \sum_{k=1}^N \mathcal{U}_{lk} P_L \phi_{k}~, \ \ \ \ \
\nu_{lR} = \sum_{k=1}^N \mathcal{U}_{l^\prime k}^* P_R \phi_{k}~, \ \ \
(l^\prime = l+N)~,
\end{equation}

\noindent where $P_L$ and $P_R$ are projection operators onto states of well-defined
chirality.  We assume here that something like the see-saw with very
large $m_R$ is operating so that half the eigenstates are very heavy.  In that
case the light eigenstates mix nearly entirely among themselves, and the $N
\times N$ matrix $U$ responsible, defined to be the ``upper-left quarter" of
$\mathcal{U}$, is nearly unitary:

\begin{equation}
\nu_{lL} \simeq  \sum_{k=1}^N U_{lk} P_L \phi_{k}~.
\label{eq:nuMix}
\end{equation}

\noindent Although we have used the see-saw to arrive at Eq.~\ref{eq:nuMix}, a mixing matrix U can be defined even if the right-handed sector is light or absent.

The matrix $U$, which we introduced in Eq.\ (\ref{eq:firstmbb}), nominally has $N^2$ parameters, $N(N-1)/2$ angles and $N(N+1)/2$
phases.   $N$ of the phases are unphysical, so that there are $N(N-1)/2$ independent
physical phases. For three active neutrino flavors, the three-phase mixing
matrix can be written in the form

\begin{widetext}
\begin{equation}
\label{eq:MixingMatrix}
U = \left(\begin{array}{ccc} c_{12} c_{13} & s_{12} c_{13} & s_{13} e^{-i
\delta}
  \\
-s_{12} c_{23} - c_{12} s_{23} s_{13} e^{i \delta} & c_{12} c_{23} - s_{12}
s_{23} s_{13} e^{ i \delta} & s_{23} c_{13}
\\
s_{12} s_{23} - c_{12} c_{23} s_{13} e^{i \delta} & -c_{12} s_{23} - s_{12}
c_{23} s_{13} e^{i \delta} & c_{23} c_{13}
\end{array}\right)
\mathrm{diag}\{e^{\frac{i\alpha_1}{2}},e^{\frac{i\alpha_2}{2}},1\}~~,
\end{equation}
\end{widetext}

\noindent where $s_{ij}$ and $c_{ij}$ stand for the sine and cosine of the
angles $\theta_{ij}$, $\delta$ is a ``Dirac" phase analogous to the
unremovable phase in
the CKM matrix, and the other two phases $\alpha_1$ and $\alpha_2$ affect only
Majorana particles. If neutrinos were Dirac particles these two phases
could be absorbed into redefinitions of the fields.  The Majorana mass terms
in the Lagrangian, however,
are not invariant under such redefinitions. \textcite{kob80} gives a detailed discussion of the number
of free parameters in the mixing matrix.

\section{RATE OF DOUBLE-BETA DECAY}
\label{sec:rate}
%Rate of Double-Beta Decay

The neutrino masses and mixing matrix figure prominently in neutrino-mediated
\BBz\ decay. The rate for that process is

\begin{equation}
\label{eq:rate} [T^{0\nu}_{1/2}]^{-1}=\sum_{\textrm{spins}} \int |Z_{0\nu}|^2
\delta(E_{e1}+E_{e2}-\mbox{\qval}) \frac{d^3p_1}{2\pi^3}\frac{d^3p_2}
{2\pi^3}~,
\end{equation}

\noindent where $Z_{0\nu}$ is the amplitude and \qval\ is the Q-value of the
decay.  The amplitude is second order in the weak interaction and depends on
the currents in the effective low-energy semileptonic Hamiltonian ($\mathcal{H}_{\beta}$), which we
assume for the time being is purely left-handed: $\mathcal{H}_{\beta}(x)=
G_F/\sqrt{2}\{ \overline{e}(x) \gamma_{\mu} (1-\gamma_5) \nu_e(x) J_L^{\mu}(x)\} +
h.c.$, with $J_L^{\mu}$ the charge-changing hadronic current. We assume as well
that only the particles we know about exist, or that any others are too heavy
to measurably affect \BBz\ decay.  Then the decay is mediated solely by the exchange
of three light neutrinos and the amplitude contains a lepton part (a function
of $x$ and $y$ to be contracted with a similar hadron part and integrated over
$x$ and $y$) of the form

\begin{widetext}
\begin{eqnarray}
\label{eq:amplitude}
\sum_k \ \overline{e}(x) \gamma_{\mu}
(1-\gamma_5) U_{ek} \phi_k\!\underbracket{(x) \  \overline{e}(y)
\gamma_{\nu} (1-\gamma_5) U_{ek} \phi_k}(y)  \ \ = \ \  \nonumber \\
 -\sum_k \
\overline{e}(x) \gamma_{\mu} (1-\gamma_5) U_{ek}
\phi_k\!\underbracket{(x) \ \overline{\phi_k^c}}(y) \gamma_{\nu}
(1+\gamma_5) U_{ek} e^c(y)~,
\end{eqnarray}
\end{widetext}

\noindent where the underbrackets indicate contraction.  With our convention
$\phi_k^c = \phi_k$, the contraction of $\phi_k$ with $\overline{\phi_k^c}$
turns out to be the usual fermion propagator, so that the lepton part above
becomes

\begin{widetext}
\begin{equation}
\label{eq:lep} -\frac{i}{4} \int \sum_k \frac{d^4q}{(2\pi)^4}e^{-i
q\cdot(x-y)} \overline{e}(x) \gamma_{\mu}(1-\gamma_5)
\frac{q^{\rho}\gamma_{\rho}+m_k}{q^2-m_k^2} \gamma_{\nu}(1+\gamma_5)
 e^c(y)~U_{ek}^2~,
\end{equation}
\end{widetext}

\noindent where $q$ is the 4-momentum transfer.  The term with $q^{\rho}$
vanishes and the $m_k$ in the denominator can be neglected for light neutrinos,
so that the amplitude is proportional to

\begin{widetext}
  \begin{equation}
   \label{eq:mass}
    \mbox{\mee} \equiv \left| \sum_k m_k U^2_{ek}\right|
= \left| m_1
    |U_{e1}|^2 + m_2
    |U_{e2}|^2 e^{i(\alpha_2-\alpha_1)} + m_3 |U_{e3}|^2 e^{i
    (-\alpha_1-2\delta)}
    \right|~.
  \end{equation}
\end{widetext}

\noindent The absolute value has been inserted for convenience, since the
quantity inside it is squared in equation (\ref{eq:rate}) and is complex if CP
is violated.  Applying the first expression in Eq.\ (\ref{eq:mass}) to our
one-flavor example, one can see explicitly that a Dirac
neutrino, which is equivalent to a degenerate Majorana pair $\phi_1$ and
$\phi_2$,
cannot contribute to \BBz\ decay because the two states would have $U_{e1}=1$ and
$U_{e2}=i$, as in Eq.\ (\ref{eq:Udegen}).

To complete the calculation, one must multiply the lepton part of the amplitude
by the nuclear matrix element of two time-ordered hadronic currents and
integrate over $x$ and $y$. We can write the  matrix element of a product of
currents between initial and final nuclear states $i$ and $f$ as

\begin{widetext}
\begin{equation}
\label{eq:hadron} \langle f | J_L^{\mu}(x) J_L^{\nu}(y) | i \rangle
= \sum_n \langle f | J_L^{\mu}(\vec{x}) | n \rangle \langle n |
J_L^{\nu}(\vec{y}) | i \rangle e^{-i(E_f-E_n) x_0}e^{-i(E_n-E_i)
y_0}~,
\end{equation}
\end{widetext}

\noindent where the $|n\rangle$'s are a complete set of intermediate nuclear
states, the $E_n$'s are the corresponding energies, and $E_i$ and $E_f$ are the
energies of the initial and final nuclei.  When the two time coordinates, $x_0$ and $y_0$,
are ordered and integrated over, and when the exponential in Eq.~(\ref{eq:hadron}),
is combined with a similar factor from the lepton currents and the $q$-dependence
of the neutrino propagator, the following factor in the amplitude results:

\begin{widetext}
\begin{equation}
\label{eq:preclosure} 2\pi \delta(E_f+E_{e1}+E_{e2}-E) \sum_n
\left[\frac{\langle f |J_L^{\mu}(\vec{x})| n \rangle \langle n |
J_L^{\nu}(\vec{y})|i\rangle}  {q^0(E_n+q^0+E_{e2}-E_i)} +
\frac{\langle f |J_L^{\nu}(\vec{x})| n \rangle \langle n |
J_L^{\mu}(\vec{y})|i\rangle}  {q^0(E_n+q^0+E_{e1}-E_i)}\right] ~.
\end{equation}
\end{widetext}

\noindent We have ignored the neutrino masses, which are small compared to their momenta.
The quantity $q_0 = q$  is the energy of the virtual neutrino, to be
integrated over along with the virtual momenta, and $E_{e1}$, $E_{e2}$ are the energies of the outgoing
electrons.  The energy $q_0$ is typically about an average inverse
spacing between nucleons, 100 MeV or so. This value is much larger than the
excitation energy of states contributing to the decay
amplitude, so the intermediate-state energies are usually replaced by an
estimate $\overline{E}$ of their average value.  Studies show that the resulting amplitude is
not very sensitive to the choice of the average \cite{pan90}, and that the error
it causes in the decay rate is perhaps 15\%.  In this ``closure approximation"
one replaces the (now unweighted) sum over intermediate states by 1, so that
Eq.\ (\ref{eq:preclosure}) above becomes

\begin{widetext}
\begin{equation}
\label{eq:closure} 2\pi \delta(E_f+E_{e1}+E_{e2}-E_i)
\left[\frac{\langle f |J_L^{\mu}(\vec{x})J_L^{\nu}(\vec{y})|i\rangle}
{q^0(\overline{E}+q^0+E_{e2}-E_i)} + \frac{\langle f
|J_L^{\nu}(\vec{x}) J_L^{\mu}(\vec{y})|i\rangle}
{q^0(\overline{E}+q^0+E_{e1}-E_i})\right] ~.
\end{equation}
\end{widetext}

To go any further, we need an expression for the hadronic current $J_L(x)$.  In
``the impulse approximation'', the hadronic current is obtained from that
of free nucleons, and the resulting one-body operator $J_L(x) = \sum_a
\hat{O}_a(x) \tau_a^+$ (where the operator $\hat{O}_a(x)$ acts on space and
spin variables of the $a^\mathrm{th}$ nucleon) is used to evaluate the matrix
element between states in the initial and final nuclei. In this approximation
$J_L^{\mu}(\vec{x})J_L^{\nu}(\vec{y})=J_L^{\nu}(\vec{y})J_L^{\mu}(\vec{x})$
because $(\tau_a^+)^2=0$. The charge-changing hadronic current for a nucleon
(\emph{i.e.} the proton-bra neutron-ket matrix element of the current) is

\begin{widetext}
\begin{equation}
\label{eq:relcur} \langle p | J_L^{\mu}(x)| p^\prime \rangle =  e^{iqx}
\overline{u}(p)\left(
g_V(q^2)\gamma^{\mu} - g_A(q^2)
 \gamma_5 \gamma^{\mu} - i g_M(q^2) \frac{\sigma^{\mu\nu}}{2m_p} q_{\nu}
+ g_P(q^2)
\gamma_5 q^{\mu} \right)\overline{u}(p^{\prime})~,
\end{equation}
\end{widetext}

\noindent where  $q=p^\prime-p$, $g_V \equiv g_V(0)=1$, $g_A \equiv
g_A(0)=1.26$,  conservation of the vector current tells us that
$g_M(q^2)= g_M g_V(q^2)$ with $g_M \equiv g_M(0) \approx 3.70 g_V$,  and the
Goldberger-Triemann relation, accurate enough for our purposes, says that
$g_P(q^2)=2m_p g_A(q^2)/(q^2+m_\pi^2)$, with $m_p$ and $m_\pi$ the proton and
pion masses.  The behavior with $q^2$ of the other coefficients can be
parameterized in a simple way from experimental data:

\begin{equation}
\label{eq:ff}
g_V(q^2)=\frac{g_V}{(1+q^2/\Lambda_V^2)^2}~, \ \ \
g_A(q^2)=\frac{g_A}{(1+q^2/\Lambda_A^2)^2}~,
\end{equation}

\noindent with $\Lambda_V^2=0.71$ (GeV)$^2$ and $\Lambda_A^2=1.09$ (GeV)$^2$.
After reducing to nonrelativistic kinematics to obtain $\hat{O}(x)$,
keeping all terms of $\mathcal{O}(1/m_p)$  except ``recoil'' terms that
depend on $p+p^\prime$ rather than $q$, and omitting 2$^{nd}$ order terms in
$E_{1,2}/q_0$, which takes the electron energies out of the denominators of Eq.\
(\ref{eq:closure}), one can integrate the rate over
electron phase space making the long-wavelength approximation.
Only $0^+ \rightarrow 0^+$ decay is considered so that the electrons
are predominantly in $s$ states and the effect on them of the charged
nucleus as they exit may be approximated via a Fermi function. The rate then
takes the form given in Eq.\ (\ref{eq:ZeroRate}):

\begin{equation}
\label{eq:finalrate}
[T^{0\nu}_{1/2}]^{-1}=G_{0\nu}(\mbox{\qval},Z)
  \left|M_{0\nu}\right|^2 \langle m_{\beta\beta} \rangle^2~,
\end{equation}

\noindent where \qval\  $\equiv E_i-E_f$, $G_{0\nu}(\mbox{\qval},Z)$ comes from the
phase-space integral, which includes the $Z$-dependent Fermi function, and
according to \textcite{rod06,sim99},

\begin{widetext}
\begin{equation}
\label{eq:me}
 M_{0\nu}  = 
 \langle f | \frac{2R}{\pi g_A^2}
 \int_0^\infty \!\!\! q \, dq \sum_{a,b}\frac{  j_0(qr_{ab})\left[ h_F(q)+ h_{GT}(q) \vec{\sigma}_a \cdot
 \vec{\sigma}_b \right]
 +j_2(qr_{ab}) h_T(q)\left[3\vec{\sigma}_j \cdot \hat{r}_{ab}
 \vec{\sigma}_k \cdot \hat{r}_{ab}- \vec{\sigma}_a \cdot \vec{\sigma}_b\right]
 }{q+\overline{E}-(E_i+E_f)/2} \tau^+_a \tau^+_b \nonumber
|i\rangle~.
\end{equation}
\end{widetext}

\noindent Here, the nucleon coordinates are all operators that, like spin and
isospin operators, act on nuclear wave functions. The nuclear radius, $R$, is
inserted to make the matrix element dimensionless, with a compensating factor
in $G_{0\nu}$. (As pointed out in \textcite{cow06}, errors have resulted
from using different values of $R$ in \Mz\ and $G_{0\nu}$.) The internucleon
position vectors are defined by $r_{ab}=|\vec{r}_a-\vec{r}_b|$ and $\hat{r}_{ab} = (\vec{r}_a-\vec{r}_b)/r_{ab}$, while $j_0$ and $j_2$ are spherical Bessel functions, and

\begin{widetext}
\begin{eqnarray}
h_F(q) &\equiv& -g_V^2(q^2)\\
h_{GT}(q)&\equiv& g_A^2(q^2) -\frac{g_A(q^2) g_P(q^2) q^2}{3m_p} +
\frac{g_P^2(q^2)q^4}{12m_p^2}+ \frac{g_M^2(q^2)q^2}{6m_p^2}\\
h_T(q)&\equiv&
\frac{g_A(q^2) g_P(q^2)
q^2}{3m_p}-\frac{g_P^2(q^2)q^4}{12m_p^2}+\frac{g_M^2(q^2)q^2}{12m_p^2}~.
\end{eqnarray}
\end{widetext}

The terms above containing $g_M$ are negligible, but those with $g_P$
typically reduce the matrix element by about 30\%.  In most calculations,
however, even these terms are neglected, so that the matrix element
takes the approximate form

\begin{equation}
\label{eq:approxme}
M_{0\nu} \approx M^{GT}_{0\nu}-\frac{g_V^2}{g_A^2}
M^{F}_{0\nu}
\end{equation}

\noindent with

\begin{eqnarray}
   M^{F}_{0\nu}&=& \langle f |\sum_{a,b} H(r_{ab},\overline{E}) \tau^+_a \tau^+_b
   |i\rangle~ ,\mbox{and} \\
   M^{GT}_{0\nu}&=& \langle f |\sum_{a,b} H(r_{ab},\overline{E}) \vec{\sigma}_a
\cdot \vec{\sigma}_b \tau^+_a \tau^+_b |i\rangle ~.
\end{eqnarray}

\noindent Here the ``neutrino potential'', $H$ is defined as

\begin{equation}
   H(r,\overline{E})\approx \frac{2R}{\pi r} \int_0^{\infty} dq \frac{\sin{qr}}
{q + \overline{E}-(E_i+E_f)/2}~.
\end{equation}

For later reference, we also give an approximate expression for the rate of
\BBt\ decay, which doesn't depend on neutrino mass or charge-conjugation
properties, and involves no neutrino propagator:

\begin{equation}
\label{eq:tworate}
[\mbox{\Tt}]^{-1}=G_{2\nu}(\mbox{\qval},Z)
|\mbox{\MtG}-\frac{g_V^2}{g_A^2}\mbox{\MtF}|^2
\end{equation}

\noindent where$=G_{2\nu}(\mbox{\qval},Z)$ is another phase-space factor,
presented earlier in Eq.\ (\ref{eq:TwoRate}), and

\begin{eqnarray}
  \label{eq:2nu}
   \mbox{\MtF}&=& \sum_n \frac{\langle f |\sum_a \tau^+_a |n\rangle \langle n|
 \sum_b
    \tau^+_b |i\rangle}{E_n-(M_i+M_f)/2}\\
   \mbox{\MtG}&=&\sum_n \frac{\langle f |\sum_a \vec{\sigma}_a \tau^+_a
 |n\rangle \langle n|\sum_b
    \vec{\sigma}_b \tau^+_b |i\rangle}{E_n-(M_i+M_f)/2} ~.
\end{eqnarray}

\noindent Nearly all the Fermi strength goes to the isobar analog state in the
daughter, so that \MtF\ can be neglected.

We know that there are three light neutrinos with largely
left-handed interactions, so it makes sense to calculate the \BBz\
rate that those neutrinos and interactions induce.  But most
theorists believe that unobserved particles and interactions exist
as well.  The most popular explanation of small neutrino masses is
the see-saw mechanism, which implies the existence of heavy
neutrinos that couple to left-handed gauge bosons.  One simple
extension of the standard model that gives rise to a see-saw is the
``left-right'' symmetric model, in which a heavy right-handed weak
boson $W_R$ coexists alongside the familiar, and lighter, $W_L$.
Refs.\ \textcite{hir96,pre03}  give a general analysis of
double-beta decay in such models.  We will not repeat that here, but
instead examine the general question of whether we can expect
physics beyond left-handed weak interactions and light Majorana
neutrinos to generate double-beta decay at a level that competes
with Eq.\ (\ref{eq:finalrate}).

Right-handed currents can cause \BBz\ through the exchange of both
light and heavy neutrinos.  The coupling of $W_R$ to neutrino mass
eigenstates contains a factor $\mathcal{U}_{l^\prime i}$ (where
$l^\prime$ labels the right-handed states with definite flavor),
while the coupling of the usual $W_L$ contains $\mathcal{U}_{li}$
(where $l$ labels the left-handed states), so that the exchange of
light neutrinos with a right-handed $W$ involved is proportional to

\begin{equation}
\label{eq:LRU}
\sum_{k= {\rm light}} m_k
\mathcal{U}^{\dag}_{l^\prime k} \mathcal{U}_{lk}~.
\end{equation}

\noindent As we see in our one-flavor example, Eq.\ (\ref{eq:seesaw}), this
quantity is $\sim m_k m_D/m_R$ and the amplitude is very suppressed.
The largest contribution, not including the one proportional to
\mee\ derived above, generally comes from
the exchange of heavy neutrinos, through two $W_R$'s.  Then there is
no suppression from the mixing matrix, the neutrino propagator is
roughly proportional to $1/m_R$ and, crudely speaking,

\begin{equation}
\label{eq:heavy} Z_{0\nu}^\textrm{heavy} \propto
\frac{G_F^2}{2}\left(\frac{M_{W_L}}{M_{W_R}}\right)^4
\left(\frac{1}{m_R}\right)~.
\end{equation}

\noindent For light neutrino exchange, there are no $W_R$'s in the
dominant term, and the propagator is roughly proportional to
$\mbox{\mee}/\langle q \rangle^2$, where $\langle q \rangle \sim
100$ MeV is a typical virtual-neutrino momentum. Then, instead of
Eq.\ (\ref{eq:heavy}), we have

\begin{equation}
\label{eq:light}
Z_{0\nu}^\textrm{light} \propto\frac{G_F^2 \mbox{\mee}}{2 \langle q \rangle^2}
\end{equation}

\noindent so that the two amplitudes will be approximately equal when
(assuming that $M_{W_R} \approx m_R$)~\cite{moh99,cir04},

\begin{equation}
\label{eq:ratio}
 m_R \approx
 \left(\frac{M_{W_L}^4\langle q \rangle^2}{
  {\mbox{\mee}}}\right)^{\frac{1}{5}}~,
\end{equation}

\noindent which is on the order of 1 TeV for \mee\ $\approx \sqrt{\Delta
m_\textrm{atm}^2}$.  Thus, if the heavy mass scale in left-right symmetric
models is about a TeV or less, it will not be so easy to determine the mass
scale of the light neutrinos from double beta decay.  The same statement is
true of many other kinds of hypothetical lepton-number-violating models
(supersymmetry, leptoquarks, \ldots) because they usually generate double-beta
decay in a similar way, through graphs in which heavy particles of some kind
play the role of the $W_R$'s and heavy neutrinos.

Neutrinoless double beta decay in extra-standard models gives rise
to new nuclear matrix elements.  The presence of a single
right-handed lepton current causes the $q^{\rho}\gamma_{\rho}$ term
in the propagator of Eq.\ (\ref{eq:lep}) to contribute to the
amplitude, giving rise to derivatives of the neutrino potential
presented here or forcing one of the electrons into a $p$ state. The
outgoing $p$ wave leads to a different dependence on the angle
between the two emitted electrons that could in principle be
exploited to distinguish between the action of right-handed currents
and the neutrino mass in light neutrino exchange.  But the
short-range exchange of a heavy particle will not always
manifest something like the $q^{\rho}\gamma_{\rho}$ term, and often
the only way to distinguish such a process from
neutrino-mass-induced decay is to exploit the different nuclear
matrix elements that enter. Provided the matrix elements can be
accurately calculated, analysis of measured lifetimes in several
isotopes or to several states in the same istotope can tell you
whether long or short range is responsible (a little more on this
below).  Of course, as already mentioned, the accuracy with which
nuclear matrix elements can be calculated is a big issue, and we
discuss it later.  A more detailed treatment of the matrix elements
governing the various kinds of double-beta decay can be found in
Refs.~\textcite{sim02,doi85,hax84,tom91}.

\begin{figure}[h]
\vspace{9pt}
\begin{center}
\includegraphics[width=7.5cm]{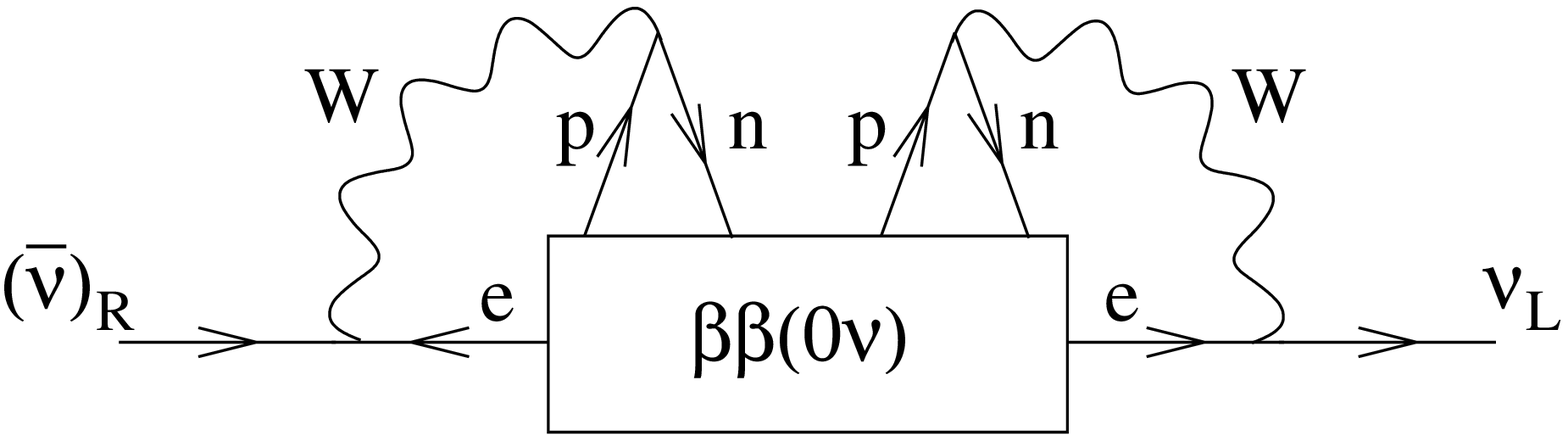}
\end{center}
\caption{Majorana propagator resulting from \BBz\ amplitude~\cite{sch82}.}
\label{fig:BlackBox}
\end{figure}

The implications of some popular extra-standard models for \BBz\
are discussed below.  We close this section with two general points.
First, when lepton number is spontaneously broken, as it is in most models that
result in a see-saw mass matrix, there must exist one or more zero-mass bosons,
called Majorons, that could be emitted along with the two electrons in double
beta decay (\BBm)~\cite{geo81,gel81,chi81}.  Apparently, however, it is difficult for such
a process to
have a very large amplitude.  Second, even if some exotic
lepton-number-violating physics exists and light neutrino exchange is not
responsible for the decay, the occurrence of \BBz\ still implies that neutrinos
are Majorana particles with nonzero mass~\cite{sch82}.  The reason is that
any diagram contributing to the decay can be inserted into a neutrino
propagator, with outgoing electron lines closed appropriately as in Fig.~\ref{fig:BlackBox}.  If \BBz\ decay
is observed, we will know for certain that neutrinos are their own
antiparticles, even if the possibility of exotic physics or uncertainty in the
nuclear matrix elements prevents an accurate extraction of the neutrino mass
scale from the observation.

\section{DOUBLE-BETA DECAY AND NEW PHYSICS}
\label{sec:Phenom}
Over the past few decades, especially the last,
much has been learned about the neutrino mixing angles and mass
eigenvalues. Table~\ref{Tab:MixingParameters} summarizes our
knowledge of these neutrino parameters. These results have increased
the importance of \BBz\ experiments; in the first subsection below,
we explain why.  The other subsections discuss other physics that
might be revealed by \BBz.

\subsection{Neutrino Mass}

If neutrinos are Majorana particles they will mediate \BBz\ at a
rate proportional to the square of \mee, Eq.~(\ref{eq:mass}).  The known
values of the mixing-matrix elements in Eq.~(\ref{eq:MixingMatrix})
allow us to predict the rate of \BBz\ under several scenarios for the
neutrino's mass spectrum.  If we ignore the LSND result (see
Sec.~(\ref{Sec:Steriles})) the oscillation data are consistent with
only 3 such masses, but their spectrum can still take 4 possible
forms:

\begin{enumerate}
\item{Normal Hierarchy-Dirac} The two masses with the smaller splitting
indicated by \ms\ are smaller than the 3$^{rd}$ mass. The neutrinos are Dirac.
\item{Inverted Hierarchy-Dirac} The two masses with the smaller splitting
indicated by \ms\ are larger than the 3$^{rd}$ mass. The neutrinos are Dirac.
\item{Normal Hierarchy-Majorana} The two masses with the smaller splitting
indicated by \ms\ are smaller than the 3$^{rd}$ mass. The neutrinos are Majorana.
\item{Inverted Hierarchy-Majorana} The two masses with the smaller splitting
indicated by \ms\ are larger than the 3$^{rd}$ mass. The neutrinos
are Majorana.
\end{enumerate}

\noindent In addition, since the absolute mass scale is unknown, it
is possible that the differences between the 3 mass eigenvalues are
small compared to the masses themselves. In this arrangement the
neutrinos are referred to as quasi-degenerate or sometimes simply as
degenerate. In the normal hierarchy, the state corresponding to the
largest mass contributes with a small mixing angle. Hence if the
mass of the lightest state, \ml\, is small, \mee\ is also small. By
contrast, in the inverted hierarchy, the heavier neutrinos are large
contributors to \mee.

\begin{figure}
\vspace{9pt}
\begin{center}
\includegraphics{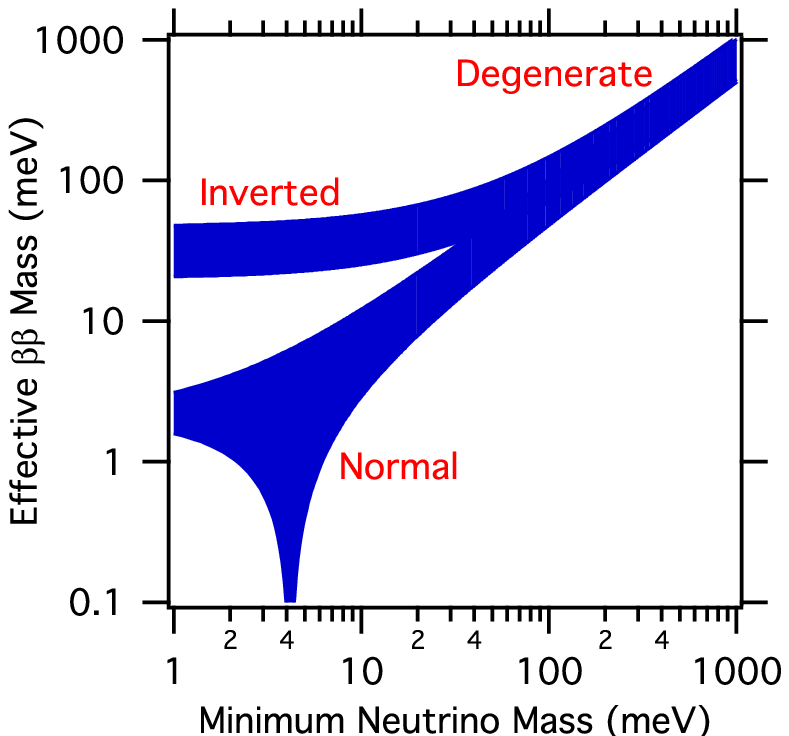}
\end{center}
\caption{The effective Majorana mass \mee\ as a function of the mass 
 of the lightest neutrino, \ml. In making the plot, we have used the best fit values
for the
parameters in Table~\ref{Tab:MixingParameters}. The filled areas represent
the range possible because of the Majorana phases and are irreducible. If one incorporates the uncertainties in
the mixing parameters, the regions widen. See \textcite{bil04a} for an example of how the mixing parameter uncertainty affects the regions.}
\label{fig:EffectiveMass}
\end{figure}

Modern \BB\ research is exciting in part because if nature has
selected possiblilty 4 we should be able to see \BBz\ with the next
generation of experiments.  By the same token, null experiments
would rule it out, and could restrict the parameter space available for
possibility 3.  And if some very sensitive experiment ever saw a very small
\mee, it would demonstrate that possibility 3 is nature's choice.
Actually certifying that possibility 4 is the choice is a
trickier matter, though.  To see this, we show in
Fig.~\ref{fig:EffectiveMass} the dependence of \mee\ on
\ml. To make this plot we have used the best fit values of the
mixing angles (\ts\ = 33.9$^{\circ}$, \ta\ = 45$^{\circ}$ and \tot\
= 0$^{\circ}$) and the $\Delta m^2$'s (Table
\ref{Tab:MixingParameters}). The figure shows, as expected, that \mee\ is
larger in the inverted hierarchy than in the normal one. (The plot shows regions rather than lines due to the unknown Majorana phases.)  But
because there is no way of measuring \ml, \BBz\ will not be able to distinguish
the inverted hierarchy from the quasidegenerate arrangements. (Although in principle it is possible to directly measure \ml\ in a beta decay experiment with ideal energy resolution, in practice it is not feasible.)  A large \mee\
will still not tell us for sure which eigenstate is the lightest.  And we also
won't know for sure that other TeV-scale physics isn't responsible for the
decay.

Other measurements can help, however.  Unlike \BBz, the rate of which
reflects the coherent exchange of virtual neutrinos,
beta decay involves the emission of real neutrinos,
whose mass can alter the beta particle spectrum. The corresponding
effective beta-decay mass \mb\ reflects the incoherent sum of the
mass terms:

 \begin{equation}
   \label{eq:betamass}
    \mbox{\mb}^2 = \sum_j m^2_j \left| U_{ej}\right| ^2
     = m^2_1|U_{e1}|^2 + m^2_2 |U_{e2}|^2 +m^2_3 |U_{e3}|^2~.
 \end{equation}

\noindent In a beta-decay experiment this quantity would
approximate the difference between the endpoint of the electron
spectrum and the $Q$-value. The approximation is valid as long as
the energy resolution is too poor to separate individual endpoints
due to each of the $m_i$. For all presently planned beta decay
experiments, that is the case.

Equations~(\ref{eq:mass}) and ~(\ref{eq:betamass}) depend
differently on the mixing angles, phases and the mass eigenvalues. If
beta-decay experiments find \mb, \BBz\ measures \mee, and \Mz\ ever get accurate enough, they could help constrain
the Majorana phases discussed below.  If \mb\ yielded a result that was
inconsistent with the two known $\Delta m^2$'s and a measured \mee, it could
demonstrate that new physics, either in the form of new particles exchanged
in \BBz\ decay or sterile neutrinos that mix with the 3 active neutrinos, is at
play.

We should note that the neutrino contribution to the mass
density ($\Omega_{\nu}$) of the universe~\cite{han03} constrains a
third combination of the neutrino masses:

  \begin{equation}
   \label{eq:omega}
    \Omega_{\nu}h^2 = \frac{\Sigma}{92.5~eV}~,
  \end{equation}

\noindent where

  \begin{equation}
   \label{eq:sigma}
    \Sigma = \sum_j m_j = m_1 + m_2 + m_3~.
  \end{equation}

\noindent Since no experiment measures the mass eigenstates directly, effective neutrino mass measurements, coupled with the oscillation measurements, are all required to determine a complete set of the parameters describing neutrinos.

\begin{table*}
\caption{Neutrino mixing parameters as summarized by the Particle Data Book
(\protect\textcite{Yao:2006}) based on the individual experimental reference reporting. The
limit on \mb\  and $\Sigma$ are based on the references given. The \mee\ limit comes from
the Ge experiments. The parameter values would be slightly different if determined by a global fit to all oscillation data~\cite{fog06}. }
\label{Tab:MixingParameters}
\begin{tabular}{cccc}
Parameter                          &    Value                       &   Confidence Level    & Reference    \\
\hline
$sin^2(2\theta_{12})$       &    $0.86^{+0.03}_{-0.04}$&    68\%                &  \protect\textcite{Aharmin:2005a}     \\
$sin^2(2\theta_{23})$       &    $>0.92$                 &    90\%                          &  \protect\textcite{Ashie:2005}                \\
$sin^2(2\theta_{13})$       &    $<0.19$                 &    90\%                          &  \protect\textcite{Apollonio:1999}     \\
$\Delta m^{2}_{21}$          &    $8.0^{+0.4}_{-0.3} \times 10^{-5}~eV^2$   &    68\%                         &  \protect\textcite{Aharmin:2005a}     \\
$|\Delta m^{2}_{32}|$        &    $2.4^{+0.6}_{-0.5} \times 10^{-3}~eV^2$   &    90\%                        &  \protect\textcite{Ashie:2004}            \\
\mb\                                      &    $<2~eV$                &  95\%                             &   \protect\textcite{kra05,lob99}            \\
\mee\                                   &    $<0.7~eV$\footnote{Using the matrix element of \protect\textcite{rod06}}  &    90\%  & \protect\textcite{kla01a,aal02a}              \\
$\Sigma$                            &   $<2~eV$                 &  95\%                            &  \protect\textcite{elg03} \\
\end{tabular}
\end{table*}

%\begin{figure}
%\vspace{9pt}
%\begin{center}
%\includegraphics{BBvsBGraph}
%\epsfbox{BBvsBGraph.eps}
%\end{center}
%\caption{The effective double-beta decay
%mass \mee\ as a function of \mb. The shaded
%areas represent
%the range possible because of the Majorana phases.}
%\label{fig:BetaMass}
%\end{figure}

\subsection{The Majorana Phases}
The elements of the mixing matrix may contain many phases, but only
a small number have any physical consequence. In a $3\times3$ mixing
matrix describing Dirac neutrinos, all but one phase (the so-called
Dirac phase $\delta$) can be absorbed into a redefinition of the
neutrino and anti-neutrino fields. If neutrinos are Majorana, the
phases of each neutrino and antineutrino pair are correlated and
fewer phases can be eliminated in this way. For the $3\times3$ case,
two additional phases (the Majorana phases $\alpha_1$, $\alpha_2$ of
Eq.~(\ref{eq:MixingMatrix})) are allowed.  Any of these phases can
result in CP violation if its value differs from a multiple of
$\pi$. In Fig. \ref{fig:EffectiveMass}, the
borders of the shaded regions correspond to phases that are
multiples of $\pi$, resulting in CP conservation. The shaded regions
between the borders correspond to CP violation. By comparing
measurements of $\Sigma$ or \mb\ to \mee, one might be able to
demonstrate that CP is violated in the lepton sector.

The Dirac phase can lead to observable effects in oscillation
experiments, whereas the Majorana phases that appear in
Eq.~(\ref{eq:mass}) have no effect in those experiments. To see
this, note that the transition amplitude for oscillations from a
neutrino of flavor $l$ to flavor $j$ is

\begin{equation}
Z(\nu_l \rightarrow \nu_{j}) = \sum_k
U_{jk}U^*_{lk}e^{-2i\frac{m^2_kL}{4E}}~,
\end{equation}

\noindent where $L$ and $E$ are the distance traveled and energy respectively
of the $\nu$. The Majorana phases in the diagonal matrix of
Eq.~(\ref{eq:MixingMatrix}) cancel with their complex conjugates in the product $UU^*$.

Many authors have suggested that the Majorana phases might be observable if
precise values of \mee\ and \mb\ could be compared~\cite{aba03,pas03,sug03}.
\textcite{ell04} provided a graphical example of how the measurements from
oscillations, \BBz, beta decay, and cosmology might be combined to learn about
possible values for the phases of Eq.~\ref{eq:mass}. But any attempt would have to contend with
the problem that there are two such phases but only one known experimental
measurement that is sensitive to them. With only \BBz\ as a probe, no
unambiguous measurement of both $\alpha$'s is possible. Although Leptogenesis
depends on CP violation in the lepton sector~\cite{fuk86,buc02a}, it will not
be easy to use cosmological measurements to help quantify the $\alpha$'s,
because the relation between CP violation in the decay of heavy Majorana
neutrinos in the early Universe and the phases in the light-neutrino mixing matrix is
model dependent.

If \tot\ is zero, however, only one of the $\alpha$'s will contribute to \mee.
This might allow that $\alpha$ to be determined from measurements of \BBz.
There is still a problem, though:  extracting information about $\alpha$ will
require the calculation of \Mz\ with an accuracy that is currently out of
reach~\cite{barg02,ell06}.

\subsection{Sterile Neutrinos}
\label{Sec:Steriles}

The LSND~\cite{agu01} neutrino-oscillation result indicates a $\Delta
m^2$ scale of $\sim$1 eV$^2$. A value this large cannot be
incorporated into a 3-neutrino mixing scheme along with the
atmospheric and solar scale $\Delta m^2$ values. This inconsistency
is sometimes called the {\em Three} $\Delta m^2$ {\em Problem}:
with 3 neutrino masses there are only 2 mass differences. One
approach to solving this problem is to add some number of light
sterile neutrinos to the three active neutrinos known to exist.

Light sterile neutrinos might seem unlikely, but can be produced by the see-saw
mechanism~\cite{gel79,yan79,moh80} if a symmetry makes the Majorana mass
matrix singular \cite{chu98,chi00,gol00,ste05}.
\textcite{gos06} have reviewed
the consequences of light steriles of various kinds
for \BB. In particular they discuss a model,
originally suggested by \textcite{sor04} that
adds two sterile neutrinos. One solution reproduces
the LSND data while still matching the null results from other
short-baseline-accelerator neutrino oscillation experiments and does
not violate constraints
from cosmology. This solution ($\Delta
m^2_{41}$ = 0.46 eV$^2$,  $\Delta m^2_{51}$ = 0.89 eV$^2$, $U_{e4}$
= 0.090, $U_{e5}$ = 0.125) would provide a maximum additional
contribution to \mee\ of $\sim$ 20 meV if $m_1$ is
small.  The MiniBooNE experiment~\cite{baz00,agu07} has found
no evidence for the 2-flavor oscillation model as an explanation for
the LSND result, although the experiment sees a low-energy excess that is 
unexplained as of this writting. Whether models with 3 active neutrinos and 2 or more
sterile neutrinos can explain both sets of data is still being investigated (\emph{e.g.} \textcite{mal07}).
Sterile neutrinos, however have the virtue of saving
the heavy element nucleosynthesis process
in the neutrino-driven wind behind a supernova shock~\cite{fet02},
and are not ruled out in general by any experiments.

\subsection{Alternative Mechanisms}
\label{sec:AltMech}
If \BBz\ is observed, it will, as discussed in Sec.~\ref{sec:rate},
demonstrate that the neutrino mass
matrix has a nonzero Majorana component \cite{sch82}. But, as we also
discussed earlier, this does not imply that the decay rate is proportional to
the square of \mee, as it would be if the light-neutrino mass were the driving
mechanism.
After any observation of \BBz, thefore, it would become important to
find an experimental
signature that isolates the underlying physics. The decay rate for
\BBz\ depends on a
Lepton Number Violating Parameter (LNVP) --- \mee\ in the case of light
neutrino exchange produced by left-handed currents --- that will be different
for each possible mechanism. In
addition, the nuclear matrix element may also depend on the exchange
mechanism.

A number of suggestions for potential signatures of the
underlying physics have been published.
As noted above, if the weak interaction includes right handed currents, the
left-handed virtual neutrino could couple to the absorbing neutron
without an helicity flip.  \textcite{doi85}
suggested the use of the kinematic distributions to discern right-handed
currents from
light-neutrino-mass contributions and \textcite{ali06} discussed the
use of the angular distribution to distinguish the left-right symmetric
model from other possibilities. \textcite{tom00}  proposed  examining the
ratio of rates to the 2$^{+}$ excited state to that to the ground
state in the same isotope as a signature of right-handed currents.

Many alternative exchange mechanisms involve heavy exchange
particles ($\simgt 1$ TeV), such as heavy neutrinos~\cite{hir96},
leptoquarks~\cite{hir96b}, supersymmetric particles~\cite{hir96c,ver02}
and scalar bilinears~\cite{kla03c}. As noted already,
electron-kinematics will not in general depend on the heavy particle
exchanged~\cite{pre03}. \textcite{sim02} showed that the relative
rates of the decay to the first excited $0^+$ state and the ground
state might distinguish among light-neutrino exchange, the
heavy-neutrino exchange, and supersymmetric-particle exchange.
\textcite{pre06} estimated the relative contributions to
$\Gamma^{0\nu}$ from light neutrinos and heavy particle exchange.

Some models lead to complicated expressions for decay rates.
Singlet neutrinos in extra dimensions can mediate \BBz~\cite{bha03}, but
the Kaluza-Klein spectrum of
neutrino masses spans values from much less than to much greater than the
nuclear Fermi momentum. One cannot, therefore, factor the decay
rate into a product of nuclear and particle factors.
Futhermore, the decay rate depends on unknown
parameters such as the brane-shift parameter and the extra-dimension
radius. Finally,
mass-varying neutrinos (MaVaNs) might lead to decay rates that
depend on the density of the matter in which the process
occurrs~\cite{kap04}.

\textcite{cir04}  recognized the potential of using $\mu \rightarrow
e$ and $\mu \rightarrow e\gamma$ in combination with \BBz\ to decide whether
light-neutrino exchange is the dominant mechanism. Certain supersymmetric
particles and heavy Majorana neutrinos could be produced at
the Large Hadron Collider. Seeing these particles, or not seeing them,
could help us determine what is responsible
for \BBz.

\section{CALCULATING NUCLEAR MATRIX ELEMENTS}
\label{sec:NuclTheory}
\subsection{Nuclear-Structure Theory}

The better we can calculate the nuclear matrix elements that govern double-beta
decay, the more accurately we can extract an effective neutrino mass, or a
heavy-particle mass.  Some scientists have tried to minimize or play down the
uncertainty in the matrix element in an attempt to strengthen claims regarding the precision with which
\mee\ can be determined. In contrast, physicists in other fields have
occasionally exaggerated the uncertainty to downplay the impact \BBz\ would
have regarding conclusions about neutrino mass.  Here we try to assess this complex issue as
objectively as possible.  To begin, we must review the accomplishments,
difficulties, and main lines of inquiry in the theory of nuclear-structure,
which is currently in an exciting period.

There are a large number of nuclei and many phenomena to observe in each
one.  Nuclear-structure theory has tended to concentrate on properties that are
easiest to measure:  spectra, electromagnetic transition rates, and cross
sections for the scattering of electrons or hadrons.  Traditional
nuclear-structure theory divides the chart of nuclides into different
regions --- some
in which the nuclei are deformed, some in which they are spherical, others in
which the behavior is more complicated --- and adopts different models for each
kind of nucleus.  Increased computing power is allowing us to
gradually abandon these models, which usually consider the dynamics of only one
or a few nucleons explicitly, lumping the rest into a collective core with only
a few degrees of freedom.  Calculations now are increasingly  ``{\em ab
initio}'', attempting to solve the many-body Schr\"{o}dinger equation directly
from empirical
two- and three-nucleon interactions, which are obtained from fits to
$NN$ phase shifts and energy levels in the triton and $^3$He.  Monte Carlo
techniques now yield accurate wave functions with such
interactions in nuclei as heavy as $^{12}$C (see \textcite{pie01} and
references therein), and controlled
approximation schemes that treat all the nucleons with the same
interactions ({\em e.g.} the no-core shell model
\cite{nav00} and the coupled-clusters approach \cite{kow04,wlo05}) currently
work well for bound-state energies in
nuclei up to about $^{16}$O, with the closed-shell nucleus $^{40}$Ca in range.

Unfortunately, these calculations aren't likely to reach atomic numbers relevant to \BB\ soon, and even
if they did, would focus first on spectra and other observables generated by
one-body operators before considering the difficult two-body decay operator
discussed in Section \ref{sec:rate}.  The other main line of structure theory,
however, is geared towards systematic (though less accurate) predictions in
arbitrarily heavy nuclei.  Its framework is mean-field theory (treating all the
nucleons) and extensions such as the random phase approximation (RPA),
restoration of symmetries artificially broken by the mean field, and the
generator-coordinate method \cite{ben03}.  These techniques, which can involve
modifying the equations in ways that improve phenomenology but at first sight
appear dubious, are closely related to density-functional theory, which leads
naturally to similar mean-field-like equations and has been spectacularly
successful in electronic systems.  But as in lighter nuclei, the initial focus
is on nuclear spectra, with some attention to simple transitions.  Though other
observables can be calculated, it can be difficult to judge the accuracy of the
results, because systematic data with which to test calculations is not
available.

These two lines of research ---- ab initio and mean-field calculations --- are
both developing rapidly and naturally absorb most of the energy of the
nuclear-structure-theory community.  Relatively few researchers focus on
double-beta decay, which involves complicated operators in structurally complex
nuclei such as $^{76}$Ge, and for good reason:  there are so many more general
problems that have frustrated theory for years but are now becoming solvable.
As a result, the models and techniques that have been applied to double-beta
decay are more limited in scope and power than they might be.  They fall
primarily into two categories, the Quasiparticle Random Phase Approximation
(QRPA) and the Shell Model.  We discuss each below and try to assess the
accuracy of the corresponding results before looking at prospects for improved
calculations.

\subsection{QRPA}
\label{sec:QRPA}
The neutron-proton QRPA, first used in \textcite{hab67}, is an extension of the random-phase approximation, a
technique primarily for calculating the energies and excitation strengths of
collective states.  In calculations of double-beta decay, a number of
approximations are made even before the QRPA is applied.  Most
nuclear-structure calculations add to the Hamiltonian, a one-body
potential/mean-field, which is diagonalized first to provide a single-particle basis in which to
treat the two-body (and in principal three-body) Hamiltonian that comes from
$NN$ and $NNN$ data.   In a complete calculation, the effects of the one-body
potential are subtracted off later.  In most nuclear-structure calculations,
however, the potential is retained and the single-particle basis truncated.  In
double beta-decay QRPA calculations, the ``model space'' typically includes
states within 10 or 20 MeV of the Fermi surface.  The energies
of these states (or equivalently, the one-body potential) are adjusted to
reproduce properties of states with a single nucleon outside a closed shell.

The two-body interaction is also not treated exactly, again even before
application of the QRPA.  It is possible in principle to construct an effective
Hamiltonian and effective transition operators that, for low-lying states, have the same eigenvalues
and matrix elements in the truncated single-particle space as does the full Hamiltonian in the
full Hilbert space.  In applications of the QRPA, the effective interaction is
usually approximated by a G-matrix \cite{hjo95} , which takes into account the
exclusion of very high-lying levels but not those just above the model space or
those below, which are assumed to contain an inert core.  Then certain parts of
the interaction and operator are scaled to reproduce data:  all $0^+$ matrix
elements are sometimes multiplied by about 1.1 so that, when the interaction is
treated in the BCS approximation (the starting point of the QRPA) they
reproduce experimental odd-A --- even-A mass differences.  The parameter $g_A$
that is squared in  the double-beta operator is often taken to be 1.0 rather
than 1.26, and short-range correlations responsible for converting the bare
Hamiltonian into a $G$-matrix  must be inserted explicitly (although
approximately) into the operator's two-body matrix elements. Other more dramatic adjustments take place in the application of
the QRPA itself.

To understand the logic behind the neutron-proton QRPA, we first remove the
``Q'', addressing the simpler charge-changing RPA.  The two-body operator that
governs (in the closure approximation) the decay, in second-quantized form,
can be written $\hat{O} = \sum_{ijkl} O_{ijkl} p^\dag_i p^\dag_j n_k n_l$, where $p^\dag$
creates a proton, $n$ destroys a neutron, and the  coefficients $O_{ijkl}$
are two-body matrix elements.  One can rewrite this operator as a sum of
products of one-body operators $\hat{O} = \sum_{ijkl} O_{ijkl}(p^\dag_i
n_k)(p^\dag_j n_l)$.  In the charge-changing RPA the one-body $p^{\dag} n$
and $n^{\dag} p$ operators
are replaced by operators with the same
quantum numbers but boson commutation rules, both in the decay operator and
the Hamiltonian.  After the ``bosonized'' Hamiltonian
is diagonalized to obtain a representation of the intermediate-nucleus states
as a set of ``one-phonon'' states,  the bosonized one-body operators contained
in the decay operator
then connect those states with the mother and daughter ground states,
which are ``phonon vacua". (Actually there are two sets of intermediate
states, one corresponding to excitations of the mother vacuum and one to
excitations of the daughter.  The
sets are not identical, a fact that has so far been taken into account
only approximately.)  This ``quasi-boson'' approximation is expected to
work best for strongly excited intermediate states, where excitations with
different single-particle quantum numbers contribute coherently.  When the intermediate states
are summed over, the result is the ground-state to ground-state double-beta decay
matrix element.

The QRPA differs from the above only in that pairing between like particles is
taken into account through BCS theory, so that the mother and daughter nuclei,
are now ``quasi-particle phonon'' vacua.  They include correlations by breaking
number conservation and have the correct number of neutrons and protons only on
average.  The smearing of the Fermi surface induced by pairing has a large
effect on the matrix elements, which actually vanish for \BBt\ in the absence
of pairing because the transition to the daughter ground state is Pauli blocked.

Two-quasiparticle states, the creation operators for which are now the
bosonized objects, contain some components that are
particle(proton)-hole(neutron) excitations and others that are
particle-particle or hole-hole excitations (based on nuclei with two more or
less particles than the one we're actually interested in).  The interaction has
matrix elements in both channels, and these matrix elements are usually
treated as independent because a large change in, {\em e.g.}, a single
particle-particle matrix element translates into small changes in all the
particle-hole matrix elements. The particle-hole matrix elements are generally
left alone because their primary effect is on the giant Gamow-Teller resonance,
which is collective and reproduced pretty well by the QRPA.  The
particle-particle matrix elements, by contrast, affect low-lying states with
much less strength, and are multiplied by a parameter traditionally called
$g_{pp}$ that takes values between 0.8 and 1.2.  This parameter turns out to have
a large effect on double-beta decay, particularly the $2\nu$ variety and particularly in the QRPA~\cite{vog86,eng88}. It is needed
because the QRPA simplifies the particle-particle correlations in a way that
must be compensated.  The parameter is usually adjusted to reproduce
related observables such as $\beta^+$ or \BBt\ rates, which
also depend on it strongly.  But although adjusting the parameter increases
$\beta^+$ rates, for example, to better agree with experiment, it appears to
concentrate the strength of the $\beta^+$ operator at too low an energy.  And
if the parameter is increased a bit past its best value, an unphysical
deuteron condensate becomes lower in energy than the BCS-like quasiboson
vacuum, causing the QRPA to break down completely.  Some of this problem is due
to the artificial separation of particle-particle correlations into ``static''
correlations associated with isovector pairing that are in the BCS
quasiparticle vacuum, and ``dynamic'' correlations, associated with isoscalar
pairing (\emph{i.e.}\ neutron-proton particle-particle interactions in the $L=0$
channel) that appear in QRPA through the parameter $g_{pp}$ as a correction to the
quasiparticle vacuum.  A satisfactory treatment of both kinds of pairing on the same
footing, something we are still awaiting in QRPA-like methods, would help.

In the meantime, much of the QRPA work has revolved around two questions: 1)
How, besides the better treatment of pairing just mentioned, can the artificial
sensitivity of the QRPA to $g_{pp}$ be reduced, and 2) which observables should fix
$g_{pp}$ (beta decay, \BBt\  \ldots) and how
accurate will the resulting \BBz\ matrix elements be?  The first issue has been the subject of an impressively large number of
papers. One approach is ``Second" QRPA \cite{rad91,sto01}, in which the fact
that the quasiboson approximation is merely the first term in an exact
expansion is exploited by calculating the effects of the next term. Another
very popular approach is Renormalized QRPA (RQRPA)~\cite{har64,row68,toi95}.
The quasiboson approximation is equivalent to replacing commutators by their
ground-state expectation values, with the ground states treated in the BCS
approximation.
The RQRPA uses the QRPA ground states instead of BCS states to evaluate the
commutators.  Because the commutators in turn help fix the ground states, the
two are evaluated self-consistently.  A variant of this approach is the ``Full
RQRPA", in which the effects of isovector $np$ particle-particle interaction,
artificially strengthened to account implicitly for isoscalar pairing that
underlies the sensitivity to $g_{pp}$, are included in the BCS calculation that
defines the quasiparticles as well as in the subsequent QRPA calculation~\cite{sch96,sim97}.  (Isovector $np$ pairing was first introduced in this way
in the unrenormalized QRPA~\cite{che93,pan96}. Another extension is the
Self-Consistent RQRPA (SCQRPA) \cite{bob01}, in which the occupation numbers in
the RQRPA ground state are fed back into the BCS approximation and the BCS-RQRPA
sequence iterated until the BCS and RQRPA ground states are consistent.
All these methods reduce the dependence of the matrix elements on the strength
of the neutron-proton pairing interaction.

On the other hand, all these calculations also leave out correlations, and the
profusion of related methods, none clearly superior to the rest, is confusing.
Some people have even begun to treat the calculations as a statistical
sample from which to extract an error bar on the calculated matrix elements
\cite{bah04}, reaching the conclusion that the matrix elements are only known to within an
order of magnitude.
\textcite{ell04} tabulate results for $^{76}$Ge and discusses why a
statistical analysis is not a good idea: many of the calculations are
explicitly preliminary or make no attempt to reproduce related data. Here, we
merely note that recent papers  \textcite{rod03,rod06}, addressing issue 2) above,
argue that if the strength of the pairing interaction is adjusted to fit
experimental mass differences and (especially) if $g_{pp}$ is then adjusted to
reproduce measured \BBt\ rates, then almost all QRPA-like calculations,
independent of their choice of number of single-particle levels, interaction,
quenching of $g_A$, etc., give the same results to within about 30\%.  If
not, then they have either not included the induced-pseudoscalar terms in the
nucleon current or neglected to add short-range correlations to the wave
functions.   Though we have no way of knowing for sure that any of the
calculations give results that are right, it is comforting to think that they
agree with one another.  Not everyone is reassured, however,  for two reasons.
First, the claim that agreement with measured \Rt\ is more important
than agreement with other observables, such as single-beta decay rates from the
lowest lying states in the intermediate nucleus, can be disputed \cite{suh05}.  When single-beta
decay is used to adjust parameters, the resulting \Rz\ are different
because the QRPA is not able to reproduce both the beta decay and the \BBt\ at the same time.  Second, the size of matrix-element
quenching by short-range correlations, which affect \BBz\ but not
\BBt\ (for which the operator has no radial dependence), is under
debate.  Most authors have used the phenomenological correlation function of
\textcite{mil76}, but recent papers \cite{kor07,kor07a} argues that more realistic procedures, {\em e.g.}
the Unitary Operator Correlation Method (see, {\em e.g.}, \textcite{rot05}), produce substantially less
quenching.  Until these issues are resolved --- and the short-range-correlations
issue clearly can be with more theoretical work --- it is hard to say that the QRPA
gives unique predictions for the matrix elements (and therefore to the value of
\mee\ extracted from an experiment) to better than a factor of 2 or
so.

There are also systematic deficiencies in the way the QRPA has been applied, in
addition to the obvious simplification inherent in the method itself, that may
cause errors.   Almost all calculations so far have assumed that the ground
states are spherical, when we know in many cases, {\em e.g.}  $^{76}$Ge, that they
are not.  Some preliminary work on \BBt\ with a ``deformed QRPA'' exists
\cite{sim03,alv06}, but \BBz\ remains uninvestigated.  With today's
computers, however, there is no reason not to use a deformed QRPA.  The
inclusion of the continuum is another feature that has been implemented just
once, in a preliminary way \cite{rod06a}.  Here too, the state of the art in nuclear
structure is more advanced than what is being applied to double-beta decay.  In
mean-field-based nuclear structure studies, continuum versions of the QRPA that
treat all the nucleons (so that there is no inert core) have been applied many
times, and versions of these treatments that include deformation are beginning
or about to appear \cite{hag04}.  A Skyrme-HFB-continuum QRPA calculation of \BBz\,
which is clearly possible even with deformation included, would be
worthwhile.  Calculations of single-beta decay rates in that formulation, although
not yet with deformation, have been around for some time (see, {\em e.g.},
\textcite{eng99}).  Perhaps
one reason this more complete calculation has not been done for double-beta
decay is that it would still treat isovector and isoscalar pairing differently,
and as a result would still show sensitivity to $g_{pp}$.  What is really needed
to lessen reliance on parameter tuning is a more unified treatment of
particle-particle correlations.  The shell model, although it has its own set of
difficulties, does provide that.

\subsection{Shell Model}

The nuclear shell model was recently reviewed by expert practitioners in this
journal \cite{cau05}.  Compared with the QRPA it has only one disadvantage: the
(much) smaller number of single-particle states that can be included.  But because the
number is smaller, one can include correlations of arbitrary complexity within
the single-particle space.  In addition, it's possible to calculate essentially
any spectroscopic observable.  In the QRPA, by contrast, only energies of or
transitions from ground states can really be reliably calculated because the
correlations necessary to describe configurations more complicated than a
two-quasiparticle excitation of the ground state are not included.

Because of its sophisticated treatment of correlations, a good shell-model
calculation is harder to carry out than a QRPA calculation.  First, as in the
QRPA, the energies of single-particle states, which usually are assumed to have
harmonic-oscillator wave functions, must be fit to spectra in closed-shell + 1
nuclei  and an effective interaction appropriate to the model space must be
constructed.  Typically something like a G-matrix is a starting point but with
many extra terms subsequently added and fit to a wide range of energies and
transition rates (painstaking work) in nuclei near the beginning or end of the
shell, where Hamiltonian matrices are relatively small.  In the nuclei of
interest, the Hamiltonain matrices are much larger --- nowadays up to $10^{9}
\times 10^{9}$ or even larger --- and the Lanczos method is often used to
diagonalize them.  The transition operator is modified by inserting the effects
of short-range correlations, again as in the QRPA.  The axial vector coupling
constant is sometimes quenched, but not always.

The first modern shell-model calculations of \BB\ decay date from
\textcite{hax84} and references therein. Only a few truly
large-scale shell model calculations have been performed.  The heavy
deformed \BB\ nuclei, $^{238}U$, and $^{150}$Nd, for example,
require bases that are too large to expect real accuracy from the shell model.\footnote{Psuedo-SU(3)-based
truncations have been used for these nuclei ({\it
e.g.}, in Hirsch {\it et al} 1996). }.  Realistic
work has thus been restricted to $^{48}$Ca, $^{76}$Ge, $^{82}$Se, and
$^{136}$Xe \cite{cau96,hoi07} (and now, though still unpublished
\cite{pov06} in $^{116}$Cd, $^{128}$Te, and $^{130}$Te as well).
Less comprehensive calculations have been carried out in several
other nuclei \cite{suh97}.

The largest-scale shell-model calculations tend to produce matrix elements
that are a bit smaller than those from QRPA, by factors of up to 2 or 3.
In addition recent unpublished results \cite{pov06} appear to differ by 50\% or more from
earlier published results by the same authors \cite{cau96}, who are essentially the
only people working on the problem in a comprehensive way.  Some of the
uncertainties, {\em e.g.}, the quenching of $g_A$ and the effects of short-range correlations, are the same ones
that afflict the QRPA; others are connected with truncation and determination
of the effective Hamitonian.
%\begin{table}
%\caption{\bf Ignore this table!}
%\begin{tabular}{c|c|c|c}
% Nucleus &QRPA &SM &SM/1.3 \\
%\hline
%  76Ge & 2.3-2.4 &2.35 &1.80 \\
%   82Se &1.9-2.1& 2.26& 1.74 \\
%   96Zr& 0.3-0.4 &&\\
%   100Mo &1.1-1.2 &&\\
%   116Cd &1.2-1.4 &2.49&1.92\\
%   130Te &1.3& 2.13 &1.64\\
%    136Xe& 0.6-1.0 &1.77 &1.36
%\end{tabular}
%\end{table}

The shell model calculations can be improved in the short term, even without
going to larger spaces (which is in fact possible via, {\em e.g.} the factorization
method of \textcite{pap03}).   Perhaps the most important step that could be
taken now is a more systematic treatment of the effective decay operator
appropriate for the shell-model space.  Perturbative corrections to the
operator that account for the finite model-space size can be calculated, although
perturbation theory in the residual two-body interaction does not always
converge \cite{hjo95}.  Its use would nonetheless give us a better idea about
the size of corrections to the na\"{i}ve renormalization (quenching of $g_A$ and
artificial short-range correlations) that is the current state of the art.
Exploratory calculations in a solvable model indicate that the corrections may
be significant \cite{eng04}.  Though perturbative corrections to the Gamow-Teller
operator have been evaluated \cite{sii01}, nobody has looked at two-body
operators beyond the Hamiltonian.

\subsection{Prospects for the future}

Most good calculations give the same result for a given matrix element to
within a factor of 2 or 3, usually less.   Assuming that there is no
many-body physics that affects the result but is systematically absent from all
calculations, we can take that range to approximately represent the level of
uncertainty in our knowledge of the matrix element.  What are the prospects for
reducing that uncertainty?

In the short term we can make progress by being more careful --- by quantifying
uncertainty in the weak nucleon current, form factors, short-range correlations,
quenching, {\em etc.}. The size of short-range-correlation effects is currently in dispute, but
structure theorists now know enough about such effects to resolve the debate.
In the medium term, the best hope is a better shell-model calculation.  Here
what we really need is progress in constructing effective operators, both the
Hamiltonian and the decay operator.  Recent years have seen dramatic progress
in our ability to compensate for high-momentum physics that is cut out
(see, {\em e.g.}, \textcite{bog03}),
but reliably correcting for low energy excitations such as core polarization is
a longstanding problem.  Partial summation of diagrams, a tool of traditional
effective-interaction theory, is helpful but apparently
not foolproof.

In the long term these issues will be solved.  As already mentioned, the
coupled-cluster approximation, an expansion with controlled behavior, is being
applied in nuclei as heavy as $^{40}$Ca.  With enough work on three- and
higher-body forces, on center-of-mass motion, and on higher-order clusters, we
should be able to handle $^{76}$Ge.  The time it will take is certainly not
short, but may be less than the time it will take for expermimentalists to
see neutrinoless double-beta decay, even if neutrinos are indeed Majorana
particles and the inverted hierarchy is realized.  And the pace of theoretical
work will increase dramatically if the decay is seen.  Observations in more
than one isotope will only make things better.  Our opinion is that the
uncertainty in the nuclear matrix elements in no way reduces the attractiveness
of double-beta decay experiments.  Given enough motivation, theorists are
capable of more than current work seems to imply.

\section{EXPERIMENTAL ASPECTS}
\subsection{Background and Experimental Design}
\label{sec:background}

\begin{figure}
\vspace{9pt}
\begin{center}
\includegraphics[width=7.5cm]{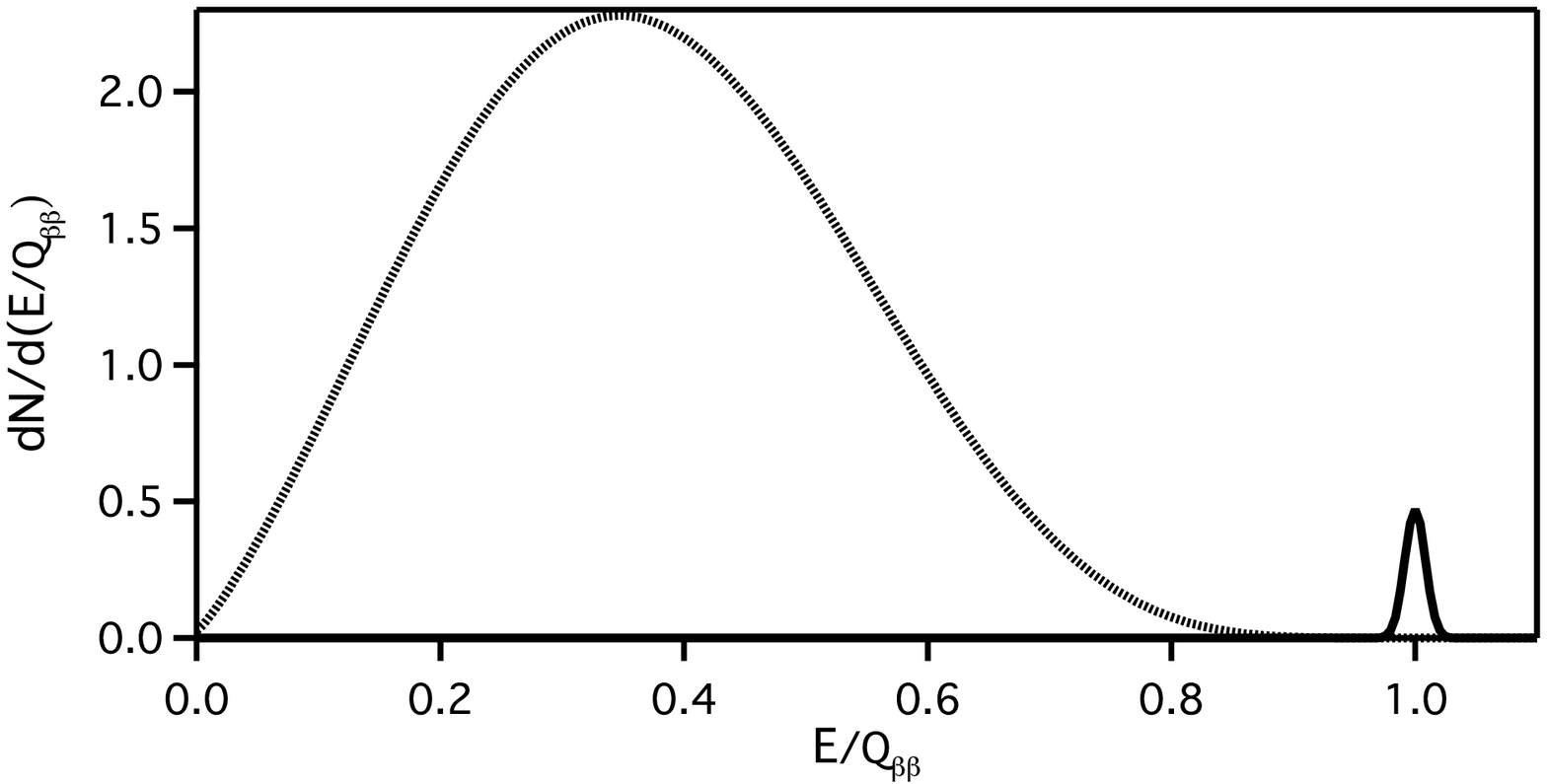}
\end{center}
\caption{The
distribution of the sum of electron energies for \BBt\ (dotted curve)
and  \BBz\ (solid curve). The curves were drawn assuming that \Rz\ is
1\% of \Rt\ and for a 1-$\sigma$ energy resolution of
2\%.}
\label{fig:SumSpectrum}
\end{figure}

Double beta decay experiments are
searching for a rare peak (see Fig.~\ref{fig:SumSpectrum}) upon a
continuum of background. Observing this small peak and demonstrating
that it is truly \BBz\ is a challenging experimental design task. The
characteristics that make an ideal \BBz\ experiment have been
discussed previously~\cite{ell02,zde02b,ell03}. Although no detector
design has been able to incorporate all desired
characteristics, each includes many of them.
(Section~\ref{sec:FutureExperiments} describes the various
experiments.) Here we list the desirable features:

\begin{itemize}
 \item The detector mass should initially be
large enough to cover the degenerate mass region (100-200 kg of
isotope) and be scalable to reach the inverted-hierarchy scale region
($\approx$ 1 ton of isotope).
 \item The \BBz\ source must be
extremely low in radioactive contamination.
 \item The proposal must
be based on a demonstrated technology for the detection of \BB.
\item A small detector volume minimizes internal backgrounds, which
scale with the detector volume. It also minimizes
external
backgrounds by minimizing the shield volume for a given
stopping power.  A small volume is easiest with an apparatus
whose source is also the detector. Alternatively, a very large
source may have some advantage due to self shielding of a fiducial
volume.
 \item Though expensive,
the enrichment process usually provides a good level of purification
and also results in a (usually) much smaller detector.
 \item
Good energy resolution is required to prevent the tail of the \BBt\
spectrum from extending
into the \BBz\ region of interest. It also
increases the signal-to-noise ratio, reducing the background in the
region of interest. Two-neutrino double-beta decay as background was analyzed by \textcite{ell02}
  \item Ease of operation is required because these
experiments usually operate in remote locations and for extended
periods.
 \item A large \qval\ usually leads to a fast \BBz\ rate and also
places the region of interest above many potential backgrounds.
\item A relatively slow \BBt\ rate also helps control this
background.
 \item Identifying the daughter in coincidence with the
\BB\ decay energy eliminates most potential backgrounds except
\BBt.
 \item Event reconstruction, providing kinematic data such as
opening angles and individual electron energies,
can
reduce background. These data might also help distinguish light- and
heavy-particle exchange
if a statistical sample of \BBz\ events
is obtained.
 \item Good spatial resolution and timing information can help reject
background processes.
 \item The nuclear theory is better understood
in some isotopes than others. The interpretation
 of limits or
signals might be easier
 for some isotopes.
\end{itemize}

Historically, most
\BB\ experiments have faced U and Th decay-chain isotopes as their
limiting background component. A continuum spectrum arising from
Compton-scattered $\gamma$ rays, $\beta$ rays (sometimes in
coincidence with internal conversion electrons), and  $\alpha$
particles from the naturally-occurring decay chains, can overwhelm any
hoped-for peak from the \BBz\ signal. This continuum is always
present because U and Th are present as contaminants in all
materials. The level of contamination, however, varies from material
to material and even within different production batches of the same
material. Over the years, great progress has been made identifying
the level of contamination in materials that compose detectors and
their shields. Once the location of the background-causing
contaminant was determined, either cleaner materials were
substituted or purification techniques were developed.  As a result
detectors are now fabricated from amazingly pure
components, some with activities as low as 1 $\mu Bq/kg$ or less.

As \BBz\
experiments reach for the \mee\ scale implied by the atmospheric
neutrino oscillations, the rate of the signal is anticipated to be a
few counts/(t y). To observe such a small peak superimposed on a
continuum of background, the background underlying the peak will need
to be 1 count/(t y) or less. This will require unprecedented low
levels of U and Th within the experimental apparatus. In fact,
the required levels of assay sensitivity are beyond what is
currently technically achievable. Research and development are
proceeding to improve assay capability in, for example, direct $\gamma$-ray
counting, surface $\alpha$- and $\beta$-ray counting, and mass
spectroscopy.

At such extreme low
levels of activity from U and Th, other components of the continuum
that were previously a secondary concern may contribute a significant
fraction to the background continuum. In fact it may be that no
single component will dominate and a number of contributors may form
a complicated mix. In the remainder of this subsection, we discuss
the possible contributors and strategies to mitigate their effects.

Double-beta-decay
experiments are conducted deep underground to avoid cosmic ray
interactions. At these depths, muons are the only surviving cosmic
ray particles, but their interactions can produce high-energy
secondaries of neutrons, bremsstralung $\gamma$ rays and
electromagnetic showers. The experimental apparatus itself is usually
contained within a detection system for muons. Hence any signal
produced by the muon itself or any of it's prompt emissions from
interactions in the apparatus will be eliminated by an
anti-coincidence requirement.

Neutrons, being neutral, can have
sizable penetrating power through a shield. They generate background
through $(n,n'\gamma)$ and $(n,\gamma)$ reactions or $(n,x)$
reactions that lead to radioactive nuclei.  In particular background
originating from  $(n,n'\gamma)$ reactions within the experimental
apparatus produce a rather non-descript spectrum. Any individual
level in a nucleus might be excited only weakly, but there are many
possible levels. Hence a significant $\gamma$-ray flux without
strong identifiable lines can appear, making,
neutron induced backgrounds difficult to diagnose. Neutrons
have two sources: fission and $(\alpha,n)$ reactions in
the cavity rock where U and Th levels can be relatively high, and
muon interactions in the rock. Fission and $(\alpha,n)$ neutrons have
energies less than about 10 MeV. These low-energy neutrons can be
moderated and shielded from the detector with layers of hydrogenous
material. The neutrons arising from muon interactions can have very
high energy ($>$1 GeV) and therefore may penetrate the shield and
induce a background-causing reaction near the detector. One
particularly dangerous reaction is $(n,n'\gamma)$ in Pb; it can
produce $\gamma$ rays with an energy very near to the \qval\ of
\nuc{76}{Ge}~\cite{mei07}. Even more onerous, is the
Pb$(n,n'\gamma)$-produced 3062-keV $\gamma$ ray, which has a
double-escape peak very near \qval. This single-site energy deposit
could be problematic for Ge detector experiments using Pb as a
shield. (See Section~\ref{sec:GeExperiments} for a discussion of the
Ge experiments.) Neutrons from muon-induced reactions can be reduced
by going deeper underground.

Fast neutrons, along with other hadrons, can
produce radioactive nuclei that may create background for
\BBz~\cite{bro90,avi92}. For example, \nuc{68}{Ge} and
\nuc{60}{Co}
are produced in Ge and \nuc{60}{Co} is also produced in Cu. The
production rates are significant on the surface of the Earth where
the hadronic cosmic-ray flux is large. Because these nuclei have long
half lives (0.8 and 5.3 y respectively), they will remain in the
apparatus and potentially create background after the materials have
been placed underground. These backgrounds can be mitigated by
storing material underground and letting the isotopes decay away (not
practical for Co) or by purifying the radioactive isotope from the
host material and then minimizing its exposure above ground. Both
strategies will be used in future efforts. Anthropogenic isotopes,
such as \nuc{207}{Bi}, have also been observed in \BB\
experiments~\cite{kla04b}.
Some exotic backgrounds have also been
considered. For example, solar neutrinos are unlikely to be a
significant background for most experimental
configurations~\cite{ell04}.

\subsection{Facility Requirements}
There is no doubt that a laboratory at significant
depth is required for \BB\ experiments. Just how deep is a difficult
question to answer, but clearly, deeper is better. However,
logistical concerns may induce a collaboration to choose a particular
laboratory. The depth requirement for \BB\ is set by the flux of
high-energy, muon-induced neutrons and subsequent reactions
that can lead to background. (See
Section~\ref{sec:background}.) Depth-requirement calculations are exceedingly
difficult to perform. However, two recent works~\cite{mei06,pan07}
have estimated the backgrounds in Ge-detector \BBz\ experiments. Both
found that the background rate at a depth equivalent to that of Gran
Sasso (3.1 km.w.e. flat overburden equivalent~\cite{mei06}) could be kept below
$\sim$1/(t y), but that significant veto efforts would be required.
The problematic fast neutron flux decreases with the muon flux,
which, in turn drops by a factor of 10 for every 1.5 km.w.e. in
depth. Mei and Hime~\cite{mei06} estimate that a depth of 5 km.w.e.
(flat overburden) would greatly reduce veto-system
requirements.

As the background requirements become more stringent,
increasingly heroic efforts will be required to minimize activity
internal to detector materials. In addition, as the experiments get
larger and more complicated, the required infrastructure at the
experimental site will become more demanding. For example, underground
manufacturing of Cu to avoid cosmogenic activation of \nuc{60}{Co},
clean room facilities for assembly of sensitive parts, strong
exclusion of Rn to prevent daughters from plating-out, and material
purification capability are a few of the high-technology requirements
for a remote underground location. Since many
experiments (in addition to those on \BB) will need an underground
site and similar capabilities, the importance of a major facility to
provide common infrastructure is clear. The
opportunities and requirements for a deep underground laboratory are
described in \textcite{bei06}.

\subsection{Measurements to Constrain Nuclear Matrix Elements}

Part of the \BB\ program is an effort to reduce uncertainty in the
nuclear-physics factors that, along with \mee, determine the rate of
\BBz.  What data will be most useful? A recent workshop~\cite{zub05}
addressed this question and summarized measurements that might help
clear up nuclear-physics issues. Such measurments, says the report,
should focus on high \qval\ isotopes (the decay rate is proportional
to \qval$^5$), even though low \qval\ isotopes such as \nuc{128}{Te}
and \nuc{238}{U} are of some interest. The 11 isotopes with a \qval\
greater than 2.0 MeV are: \nuc{48}{Ca}, \nuc{76}{Ge}, \nuc{82}{Se},
\nuc{96}{Zr}, \nuc{100}{Mo}, \nuc{110}{Pd}, \nuc{116}{Cd},
\nuc{124}{Sn},  \nuc{130}{Te}, \nuc{136}{Xe}, and \nuc{150}{Nd}.
Most of these isotopes are currently part of a \BB\ experimental
program.

The \qval\ for these nuclei are known to only
a few keV, with the exception of that in \nuc{76}{Ge}, which is known to
$\sim$ 50 meV. Precision measurements of the other Q-values would be
useful because energy resolution is improving. The energy
resolutions of the COBRA ($\sim$55 keV) and CUORE ($\sim$5 keV)
 experiments are larger than or comparable to the uncertainty
in the Q-values~\cite{aud03,red07} for their chosen isotopes (4 keV for
\nuc{116}{Cd} and 2 keV for \nuc{130}{Te}).
It would also be helpful to measure the masses of certain nuclei involved in
radiative EC-EC decay, because that  decay rate can be
resonantly enhanced if the energy of the 2P-1S atomic transition is
similar to the Q-value~\cite{suj04}. At present the masses of the
interesting nuclei are not known precisely enough to determine the degree
of enhancement.

Precise \BBt\ data in all nuclei used for
\BBz\ and single-$\beta$ data for the associated intermediate nuclei, where possible, are also needed.
Rates for both processes have been used to fix the parameter \gpp\ that
plays a crucial role in QRPA equations~\cite{rod03, rod06,suh05}. (See
Sec.~\ref{sec:QRPA}.)  More data would test internal consistency, and
help decide which \gpp\ calibrator is better.

In \BBt, and in \BBz\ as well in the closure approximation is not used, the expressions for the overall matrix element (see Eqs.(\ref{eq:2nu}) and (\ref{eq:preclosure})) contain matrix elements of the charge changing weak current from both the parent and daughter nuclei to states in the intermediate nucleus.  In the long-wavelength approximation, which the high-momentum virtual neutrino prevents for \BBz, these reduce to the Fermi and Gamow-Teller operators.  GT strength distributions from the parent nucleus in the relevant isotopes can be measured with charge-exchange reactions~\cite{eji00a,amo07}, and should be measured for all the 9 high-Q-value isotopes of experimental interest. Extracting GT strengths from the daughter \cite{amo07} and the strengths for higher-multipole analogs of the GT operator is critical for \BBz\ but  much harder because the relations between the cross sections and operator matrix elements are more complicated, and polarized beams are sometimes required.  Transitions from the parent can be studied, {\em e.g.}, with (p,n) or (\nuc{3}{He},t) reactions and transitions from the daughter with (n,p), (d,\nuc{2}{He}), or (t,\nuc{3}{He}). But even if the connection between cross sections and transition strengths can be properly made, the strength is still the square of a matrix element, so that the corresponding sign, important for \BB, is unavailable.

Muon capture is governed by matrix elements similar to those connecting
the intermediate and daughter nuclei in \BB. Because the muon is heavy, states
with any multipole can be populated during capture, and partial muon
capture rates to specific states can be linked to the corresponding
parts of the \BB\ matrix elements~\cite{kor02}. \textcite{vol05}
proposed
charged-current neutrino-nucleus scattering as a technique to constrain \Mz. By
using both $\nu$ and
$\overline{\nu}$, one can probe matrix elements of one-body operators to
intermediate states from both the parent and daughter nucleus.

The study of pair correlations of nucleons in
the ground state of the parent and daughter nuclei allows one to
probe aspects of
the wavefunctions that are important for \BB\ matrix elements even when the
closure approximation is used. Double-beta decay can
be represented as the removal of two neutrons followed by
the addition of two protons, and calculations show that when the
representation
is decomposed into multipoles, the $J^\pi=0^+$ channel
is the most important. Recent results of (p,t) reactions for the Ge-Se
system~\cite{fre07} have shown no evidence of pairing
vibrations that have been found in low-A systems. \nuc{76}{Ge} and
\nuc{76}{Se} have quantitatively similar neutron pairing correlations,
and calculations will be easier there than they would be otherwise.

Other processes, such as pion double-charge exchange, electromagnetic
transitions to isobaric analogue states~\cite{eji06a}, and \BBt\ to excited
state in the daughter~\cite{aun96,suh98} can also help constrain \BBz\ matrix
elements.  But despite the usefulness of all these auxiliary measurements, none
will provide a magic bullet.  Such data will help judge the quality of
calculations, but not reduce uncertainty to near zero.  For that, the best data
are probably \Tz\ themselves, the same quantities one is trying to calculate.
If the lifetimes are measured in several nuclei, one can check directly whether
the calculations work, most simply by fixing \mee\ from one nucleus and seeing
whether the other lifetimes are correctly predicted.  Of course, this assumes
that light-neutrino exchange rather than some other mechanism is driving the
decay.

\section{FUTURE PROGRAM}
\label{sec:FutProg}
\subsection{Previous Experiments}

Since the first direct observation of \BBt\ in 1987~\cite{ell87}, decay rates
in a number of other isotopes have been measured, either directly or with
geochemical/radiochemical techniques.
The recent
work of Rodin {\em et al.}~\cite{rod03,rod06} has rekindled interest in the
precise measurements of \Tt. \textcite{bar06}, in a review of the
\BBt\ experiments, has recommended {\em average} \Tt\ values, which we quote
in Table \ref{tab:TwoNeutrino}.

Here we just note that because \Rt\ is proportional to \qval$^{11}$, whereas
\Rz\ is proportional to \qval$^5$, \Rz\ might be larger than \Rt if \qval\ is
small and geochemical measurements of the decay rates that do not distinguish
between \BBz\ and \BBt\ might still produce a competitive limit on \mee. This
is the case in \nuc{128}{Te}, where the \Tz\ given in Table
\ref{tab:TwoNeutrino} produces a limit on \mee~\cite{ber93} that is only
slightly worse than the best limits given in Table \ref{tab:PastExperiments}.
The nucleus \nuc{238}{U} is another case~\cite{tur91} in which the
radiochemical experiment could not distinguish \BBz\ and \BBt. \nuc{238}{U}
decays to \nuc{238}{Pu}, which in turn $\alpha$-decays with a 87.7-y half life.
A sample of U salt had been stored for 33 years and was milked for its Pu
content. By counting the \nuc{238}{Pu} $\alpha$, the \BB\ half life was
determined. Because of its low \qval\, the high observed decay rate (compared
to the theoretically predicted \Tt) has been interpreted as evidence for \BBz.
Unfortunately, the experiment is difficult to repeat because of the special
nature of the U-salt sample available to the experimenters.

The entire history of \BB\ measurements up to about 2001, including \BBt, \BBz,
and Majoron modes, can be found in \textcite{tre02}. The best experiments to
date have been in Ge. The Heidelberg-Moscow~\cite{kla01a} and
IGEX~\cite{aal02a,aal04} experiments have provided the best limits on \Tz\ by
building detectors with the lowest backgrounds.

\begin{table}[htdp]
\caption{A list of the values of \Tt\ for various isotopes. These values
are recommended by \textcite{bar06} as the best interpretation of the
experimental data. One should heed discussion in \textcite{bar06} before
using the values in a quantitative way.}
\begin{center}
\begin{tabular}{cccc}
Isotope                          & \Tt\   (years)                                                  & Isotope                          & \Tt\   (years)    \\
\hline
\nuc{48}{Ca}                &   $(4.2^{+2.1}_{-1.0}) \times 10^{19}$      &\nuc{128}{Te}                &   $(2.5 \pm 0.3) \times 10^{24}$                    \\
\nuc{76}{Ge}                &   $(1.5 \pm 0.1) \times 10^{21}$                &\nuc{130}{Ba} EC-EC(2$\nu$)                &  $(2.2 \pm 0.5) \times 10^{21}$\\
\nuc{82}{Se}                &   $(0.92 \pm 0.07) \times 10^{20}$            &\nuc{130}{Te}                &  $(0.9 \pm 0.1) \times 10^{21}$     \\
\nuc{96}{Zr}                  &   $(2.0 \pm 0.3) \times 10^{19}$                &\nuc{150}{Nd}                &  $(7.8 \pm 0.7) \times 10^{18}$\\
\nuc{100}{Mo}                &  $(7.1 \pm 0.4) \times 10^{18}$               &\nuc{238}{U}                  &  $(2.0 \pm 0.6) \times 10^{21}$ \\
\nuc{116}{Cd}                &  $(3.0 \pm 0.2) \times 10^{19}$              \\
\end{tabular}
\end{center}
\label{tab:TwoNeutrino}
\end{table}%

\begin{table*}[htdp]
\caption{A list of recent  \BBz\ experiments and their 90\%-confidence-level (except as noted) limits on \Tz. The \mee\ limits, if provided, are those quoted by the authors,
who each made choices about which calculated \Mz\ to use.}
\begin{center}
\begin{tabular}{|c|c|c|c|c|}
\hline
Isotope                &  Technique                                             & \Tz\                                                  & \mee\ (eV)       & Reference  \\
\hline
\hline
\nuc{48}{Ca}     & CaF$_2$ scint. crystals             &$>1.4 \times 10^{22}$ y               &   $<$7.2-44.7   & \textcite{oga04}\\
  \nuc{76}{Ge}   & \nuc{enr}{Ge} det.                       &    $>1.9 \times 10^{25}$ y          & $<$ 0.35           & \textcite{kla01a} \\
  \nuc{76}{Ge}   & \nuc{enr}{Ge} det.          &    $(2.23^{+0.44}_{-0.31}) \times 10^{25}$ y (1$\sigma$)   & $0.32\pm0.03$ & \textcite{kla06}\\
  \nuc{76}{Ge}   & \nuc{enr}{Ge} det.                       &    $>1.57 \times 10^{25}$ y            & $<(0.33-1.35)$ & \textcite{aal02a} \\
  \nuc{82}{Se}    &Thin metal foils and tracking               & $>2.1 \times 10^{23}$ y              & $<$(1.2-3.2)        &  \textcite{bar06a}\\
\nuc{100}{Mo}   &Thin metal foils and tracking               & $>5.8 \times 10^{23}$ y             & $< $(0.6-2.7)         &  \textcite{bar06a}\\
\nuc{116}{Cd}   &\nuc{116}{Cd}WO$_4$ scint. crystals  & $>1.7 \times 10^{23}$ y              & $<$1.7       & \textcite{dane03}\\
\nuc{128}{Te}    & geochemical                                         & $>7.7 \times 10^{24}$ y               & $<$(1.1-1.5)         & \textcite{ber93} \\
\nuc{130}{Te}    & TeO$_2$ bolometers                     & $>3.0 \times 10^{24}$ y              & $<$(0.41-0.98)        &  \textcite{arn07}  \\
\nuc{136}{Xe}   &  Liq. Xe scint.  & $>4.5 \times 10^{23}$ y\footnote{See footnote 4 in \protect\textcite{ber02}} &  $<$(0.8-5.6) & \textcite{ber02}    \\
\nuc{150}{Ne}   & Thin metal foils and tracking &$>3.6 \times 10^{21}$ y               &                                                 & \textcite{bar05} \\
\hline
\end{tabular}
\end{center}
\label{tab:PastExperiments}
\end{table*}%

\subsection{Overview of the Future Program}
In the inverted hierarchy with a \ml\ near 0 meV, \mee\ will be near 20-50 meV.
The \Tz\ resulting from this neutrino mass will be near $10^{27}$ y with a
resulting count rate of few/(t y). On the other hand, if the recent claim of
\mee\ $\sim$ 400 meV is borne out, the half life will be nearer to $10^{25}$ y,
with a count rate of a few 100/(t y). A precision measurement (20\% or better)
requires a few hundred events.  To make such a measurement in a 2-3
year run, it seems prudent to build an experiment containing 100-200 kg of
isotope that can be expanded to 1 ton at a later time. But collecting
statistics in a reduced background experiment is not the only goal.  The recent
claim for \BBz\ is controversial. Future claims must provide supporting
evidence to strengthen the argument~\cite{ell06}. This evidence might include
any number of the following.

\begin{itemize}
\item To show that \BBz\ likely exists, one needs a combination of:
  \begin{itemize}
  \item a clear peak at the correct \BBz\ energy,
  \item a demonstration that the event is a single-site energy deposit,
  \item measured event distributions (spatial, temporal) representative of \BBz,
  \item a demonstration that the measured decay rate scales with isotope fraction.
  \end{itemize}
\item To present a convincing case, one needs:
  \begin{itemize}
  \item an observation of the 2-electron nature of the \BBz\ event,
  \item a demonstration that the  kinematic distributions (electron energy sharing, opening angle) match those of \BBz,
  \item to observe the daughter nucleus together in time with the \BBz\ decay,
  \item to observe excited-state decay with parameters indicating \BBz.
\end{itemize}
\item To remove all doubt, many of the above \BBz\ indicators should be
measured in several isotopes.
\end{itemize}

\noindent The proposals summarized below in Section \ref{sec:FutureExperiments} address these possibilities in different ways.

\subsubsection{The Number of Required Experiments and Their Precision}
If \BBz\
is observed, the qualitative physics conclusion that neutrinos have a Majorana
nature will have a profound effect on the development of models for mass
generation. Quantifying the physics associated with
lepton-number violation will be more challenging because of the
uncertainty in \Mz.  If the matrix elements were precisely known,
one could compare \Rz\ measurements in several nuclei to address the question
of the underlying mechanism. If, in addition, the mechanism was certain, one
could determine the LNVP from a lone \Rz\ measurement.

Recently a number of papers have appeared~\cite{bil04a,bah04a,dep06,geh07} that
try to quantify, in light of matrix-element uncertainties, the number and type
of experimental results needed to draw quantitative conclusions. Although the
papers make different assumptions, the concensus is that if \BBz\ exists,
measurements of \Rz\ in at least 3 different isotopes are warranted.  All the
analyses assume that existing calculations provide an estimate of the
difference in rates produced by different mechanisms, and require that the
experiments determine the rate to within 20\%.  This last
number comes from the level of variation in calculated matrix elements.

\subsubsection{Kinematic Distributions} \label{Sec:KinDist} If \BBz\ is
observed, future experimenters will want to study the kinematic distributions
associated with the outgoing electrons. The NEMO 3 experiment~\cite{arno05} has
shown the power of this type of analysis with \BBt. The collaboration used the
spectrum of electron spectra from \BBt in \nuc{100}{Mo}, taken one at a time
(the {\em lone electron spectrum}), to show that the decay proceeds
predominately through the $1^+$ ground state of the intermediate \nuc{100}{Tc}
nucleus. The effect on the spetrum of this ``single-state dominance" is
small~\cite{sim01}, so $10^5$ events were required for the analysis. But the
analysis demonstrates that important information contained within kinematic
distributions.

\begin{figure}
\vspace{9pt}
\begin{center}
\includegraphics[width=7.5cm]{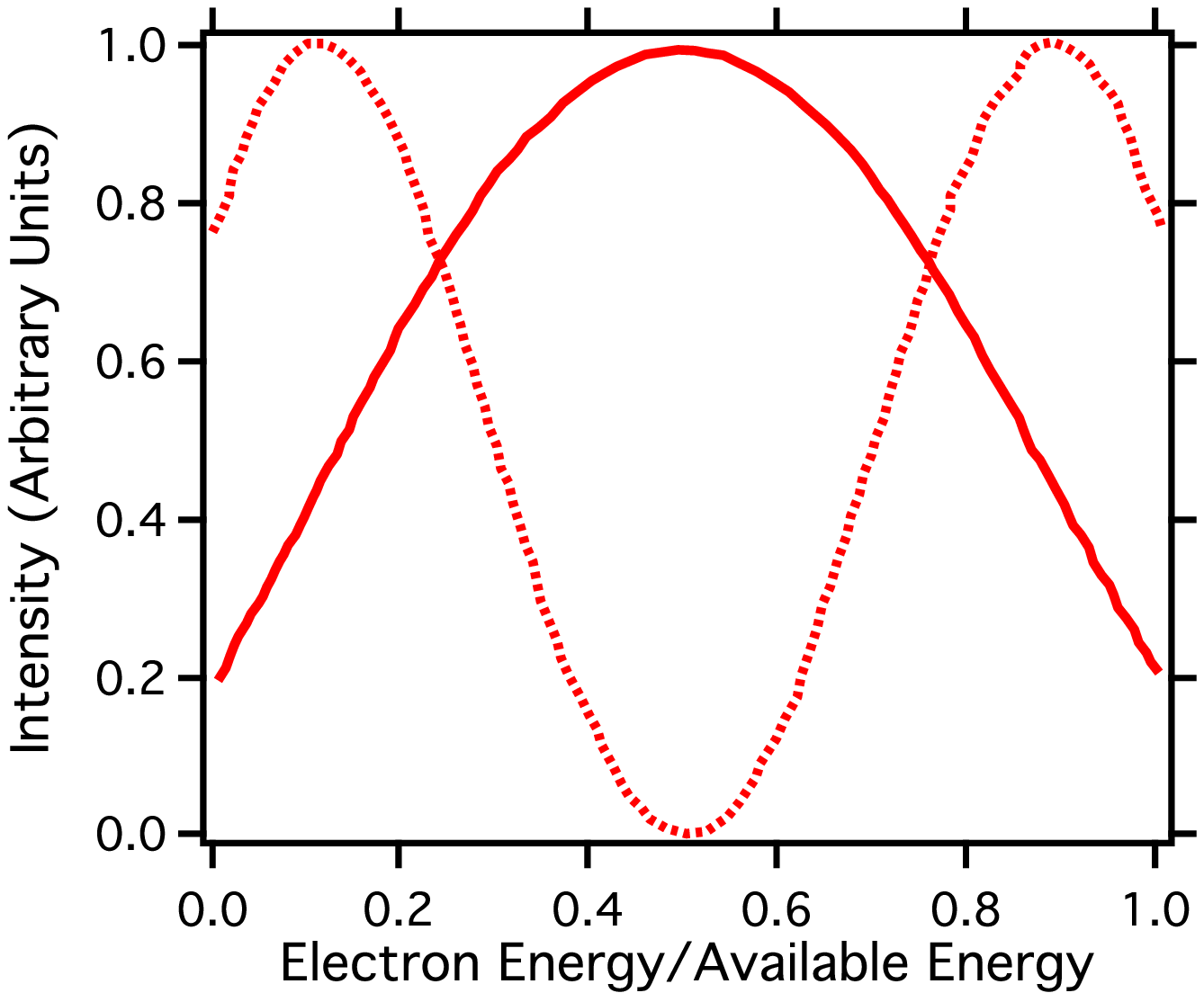}
\end{center}
\caption{The distribution of the electron energies taken one at a time
for \BBz. The solid curve is for light-neutrino exchange and the dotted
curve is for right-handed currents in both the leptonic and hadronic
currents.~\cite{doi85}.}
\label{fig:LoneEnergy}
\end{figure}

Figure \ref{fig:LoneEnergy} shows an example of the kinematic distributions for
\BBz. In this Figure, the lone electron spectra are shown for two possible
exchange mechanisms. The {\em mountain shaped} curve is for light-neutrino
exchange mechanism and the {\em valley shaped} curve is for right-handed
interactions in both the leptonic and hadronic currents~\cite{doi85}. It is
clear from this Figure, that only a few 10s of events would be needed to show
that one of these mechanisms dominates the decay. However, a large sample of
events would permit an analysis to constrain the fraction of a competing
mechanism.  The opening-angle distribution is another observable that is
sensitive to aspects of the underlying exchange mechanism.

The only detectors able to register these types of distributions involve
thin-foil, tracking-volume sandwiches.  The MOON~\cite{nak06} and
SuperNEMO~\cite{bar04a} proposals are examples. There are competing demands in
the design of such experiments.  one would like a thick foil for a large source
mass, but a thin foil to minimize energy loss and scattering so the
integrity of the distributions will be retained. As a result, such experiments
will have to be large with many electronic channels in order to collect the
required numbers of events.

\begin{table*}[htdp]
\caption{A summary list of the \BBz\ proposals and experiments.}
\begin{center}
\begin{tabular}{|c|c|c|c|c|c|}
\hline
Experiment         &   Isotope             & Mass  &  Technique                                      & Present Status                            & Reference  \\
\hline
\hline
CANDLES         &  \nuc{48}{Ca}    & few tons & CaF$_2$ scint. crystals               & Prototype               &  \textcite{ume06} \\
CARVEL            &\nuc{48}{Ca}      &1 ton        & CaWO$_4$ scint. crystals               & Development       & \textcite{zde05}        \\
COBRA              &  \nuc{116}{Cd}    & 418 kg & CZT semicond. det.                      & Prototype             &  \textcite{zub01}   \\
CUORICINO      &  \nuc{130}{Te}    & 40.7 kg & TeO$_2$ bolometers                  & Running              &  \textcite{arn05} \\
CUORE              &  \nuc{130}{Te}    & 741 kg & TeO$_2$ bolometers                   & Proposal              &  \textcite{ard05}             \\
DCBA                 & \nuc{150}{Ne}    & 20 kg    &\nuc{enr}{Nd} foils and tracking  &  Development     & \textcite{ish00}  \\
EXO-200            &  \nuc{136}{Xe}    & 200 kg &Liq. \nuc{enr}{Xe} TPC/scint.       & Construction        & \textcite{pie07}             \\
EXO                    &  \nuc{136}{Xe}    & 1-10 t   &Liq.  \nuc{enr}{Xe} TPC/scint.       & Proposal              &  \textcite{dan00a,dan00b}             \\
GEM                   &  \nuc{76}{Ge}      & 1 ton    & \nuc{enr}{Ge} det. in liq. nitrogen & Inactive                & \textcite{zde01} \\
GENIUS             & \nuc{76}{Ge}       & 1 ton    & \nuc{enr}{Ge} det. in liq. nitrogen & Inactive               & \textcite{kla01a} \\
GERDA              &  \nuc{76}{Ge}    & $\approx$35 kg    &\nuc{enr}{Ge} semicond. det.  & Construction    &  \textcite{sch05}             \\
GSO                   & \nuc{160}{Gd}    & 2 ton     &Gd$_2$SiO$_5$:Ce crys. scint. in liq. scint. & Development & \textcite{dane00,wan02} \\
\MJ\                    &  \nuc{76}{Ge}    & 120 kg   &\nuc{enr}{Ge} semicond. det.         & Proposal              &  \textcite{gai03}             \\
MOON               &  \nuc{100}{Mo}    & 1 t & \nuc{enr}{Mo} foils/scint.                          & Proposal              &  \textcite{nak06}             \\
SNO++              &  \nuc{150}{Nd}    & 10 t & Nd loaded liq. scint.                               & Proposal              &  \textcite{che05}             \\
SuperNEMO      &  \nuc{82}{Se}    & 100 kg & \nuc{enr}{Se} foils/tracking               & Proposal              &  \textcite{bar04a}             \\
Xe                      & \nuc{136}{Xe}    &1.56 t     & \nuc{enr}{Xe} in liq. scint.                   &Development      & \textcite{cac01} \\
XMASS             & \nuc{136}{Xe}    & 10 ton  & liquid Xe                                                & Prototype            & \textcite{tak04} \\
HPXe                 & \nuc{136}{Xe}    &  tons  & High Pressure Xe gas                         & Development       & \textcite{nyg07} \\
\hline
\end{tabular}
\end{center}
\label{tab:FutureExperiments}
\end{table*}%

\subsection{The Experiments}

\label{sec:FutureExperiments}
In this section, we describe a number of proposed experimental programs for
\BBz. A summary of the proposals is given in Table~\ref{tab:FutureExperiments}.
\subsubsection{CUORE (Cryogenic Underground Observatory for Rare Events)}
\label{sec:cuore}

Before addressing the CUORE experiment, it is appropriate to discuss the
general principles of cryogenic, or thermal, detectors, and the pilot
experiment, CUORICINO.  That detector is an array of bolometers made from a
material containing the parent  decay isotope. In CUORICINO and CUORE the
individual bolometers are $5\times5\times5$ cm$^3$ single crystals of TeO$_2$.
They are made from natural-abundance Te, which is 33.8\% \nuc{130}{Te}. They
are operated at very low temperatures where they have tiny specific
heats. They are weakly thermally coupled to copper support frames that act as
the heat bath. When an event leaves energy in the crystal, the temperature
rises and that rise is measured by a sensitive thermistor, which produces
a signal proportional to the energy deposited. The heat leaks back through the
weak thermal couplings and the signal decays as the temperature returns to its
baseline.

This technique was suggested for \BB\ decay searches by \textcite{fio84}, and
applied earlier by the Milano group in the MIBETA experiment~\cite{arn03}
before CUORICINO~\cite{arn05} . The CUORICINO bolometers are dielectric and
diamagnetic, and are operated at temperatures between 8 and 10 mK~\cite{arn03,
arn04b,arn05}. According to the Debye Law, the specific heat of TeO$_2$
crystals is given by $C(T)=\beta(T/\Theta_D)^3$, where  $\beta$=1994 J/(K mol)
and $\Theta_D$ is the Debye temperature. In these materials,  $C(T)$ is due
almost exclusively to lattice degrees of freedom.  The constant $\Theta_D$  was
specially measured by the Milan group for $5\times5\times5$ cm$^3$ TeO$_2$
crystals as 232 K~\cite{arn03}, which differs from the previously published
value of 272K~\cite{whi90}. The specific heat follows the Debye Law down to 60
mK. The heat capacity of the crystals, extrapolated to 10 mK, is 2.3x10$^{-9}$
J/K. With these values of the parameters, an energy deposition of a few keV
will result in a measurable temperature increase $\Delta T$.  The temperature
increase caused by the deposition of energy equal to the total \BBz\ decay
energy, \qval\ = 2530 keV, would be 0.177 mK. To obtain usable signals for such
small temperature changes, very sensitive thermistors are required. In
CUORICINO, $\Delta T$ is measured by high resistance germanium thermisters
glued to each crystal. More details can be found in \textcite{fio84}, and in
subsequent publications~\cite{all97,all98}.

The thermistors are heavily doped high-resistance germanium semiconductors,
with an impurity concentration slightly below the metal-insulator transition.
High-quality thermistors require a very homogeneous doping concentration. For
the CUORICINO thermistors (Neutron Transmutation Doped (NTD)), this was
achieved by uniform thermal neutron irradiation throughout the
entire volume in a nuclear reactor. The electrical conductivity
of the devices depends very sensitively on the temperature because they use
variable range hopping (VRH) mechanisms. The resistivity
varies with temperature according to $\rho=\rho_0\exp{(T_0/T)^{\gamma}}$, where
the constants $\rho_0$, $T_0$  and $\gamma$ all depend on the doping
concentration. In the case of thermistors operating with VRH mechanisms,
$\gamma=1/2$.

Thermistors can be characterized by their sensitivity, $A(T)$, defined as
follows:  $A(T)=|d(ln R)/d(ln T)| = \gamma(T_0/T)^{\gamma}$, where the
resistance is $R(T)=R_0\exp{(T_0/T)^{\gamma}}$. The parameter $R_0=\rho_0(d/a)$,
where  $d$ and  $a$ are the distance between the contacts and the cross section
of the thermistor, respectively. The values of  $R_0$, $T_0$ and $\gamma$  must
be measured for each thermistor. This is done by coupling the thermistor to a
low-temperature heat sink with a high heat-conductivity epoxy.  The base
temperature of the heat sink is between 15 and 50 mK. A current flows through
the device and a V-I load curve is plotted. The curve becomes non linear
because of the power dissipation and dynamic resistance, causing the slope of
the I(V) curve to change from positive to negative. The optimum operating
biasing current occurs where $dI/dV$  increases rapidly with increasing bias
voltage, $V_b$, maximizing the signal to noise ratio. The parameters of
each thermister are determined from a combined fit to a set of load curves
at different base temperatures. The
characterization process was described in deatil earlier for Si thermistors~\cite{all99},
and the same process is used for the CUORICINO Ge thermistors.

    The thermistors in CUORE will be the same as those used in CUORICINO; they
were produced by the UC Berkeley/LBNL group~\cite{hal82}. It is necessary to
optimize the neutron doping of the Ge. This is done by using
foils of metal with long-lived $(n,\gamma)$ radioactive daughter
nuclides irradiated along with the Ge to calibrate the neutron
flux. Accordingly, the neutron exposure can be
accurately evaluated without having to wait for the intense radiation of the
\nuc{71}{Ge}  in the sample to decay. Following the decay period, the Ge is
heat treated to repair the crystal structure, and cut into $3\times3\times1$ mm
strips. Electrical connections are made by two 50 $\mu$m  gold wires, ball
bonded to metalized surfaces on the thermistor. The thermistors are glued to
each bolometer with 9 spots of epoxy, deposited by an array of pins.

\begin{figure}
\vspace{9pt}
\begin{center}
\includegraphics[width=7.5cm]{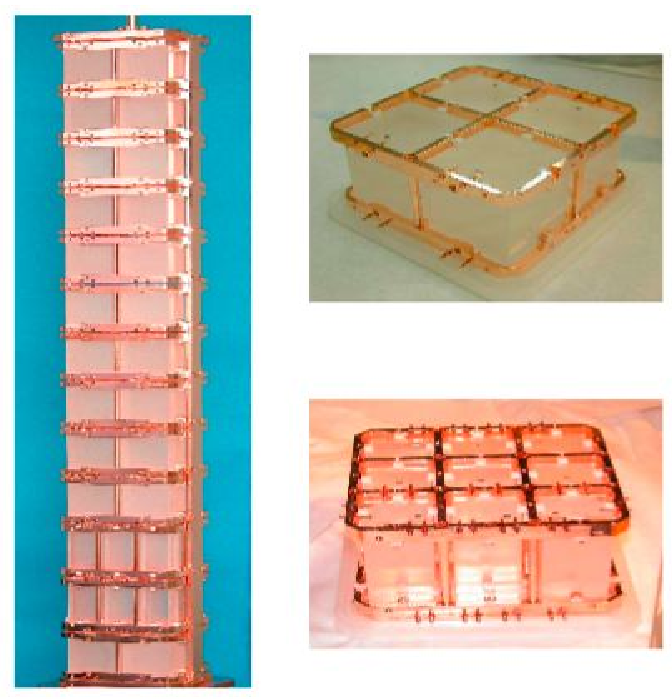}
\end{center}
\caption{The CUORICINO structure. Figure courtesy of the CUORE collaboration.}
\label{fig:cuoricino}
\end{figure}

Figure \ref{fig:cuoricino} shows the CUORICINO structure. Each of the upper 10
planes and the lowest one consists of four $5\times5\times5$-cm$^3$ crystals of
natural isotopic abundance, as shown in the upper right hand figure,
while the 11th and 12th planes have 9 $3\times3\times6$-cm$^3$ crystals, as
shown in the lower right hand figure. In the $3\times3\times6$ cm$^3$ planes
the central crystal is fully surrounded by the others.

The crystals for CUORE will be prepared the same way as for CUORICINO. The
crystals were grown with pre-tested low-radioactivity material by the Shanghai
Institute of Ceramics of the Academy of Science (SICAS) and shipped to Italy by
sea to minimize activation by cosmic rays. They were lapped
with a specially selected low contamination polishing compound. All these
operations, as well as the mounting of the tower, were carried out in a
nitrogen atmosphere glove box in a clean room. The mechanical structure is made
of OFHC Copper and Teflon, both of which were previously tested to be sure that
measurable radioactive contaminations were minimal. Thermal pulses are recorded
by the NTD Ge thermistors thermally coupled to each crystal. The gain
of each bolometer is calibrated and stabilized by a 50-100 k$\Omega$
resistor attached to each absorber and acting as a heater. Heat pulses are
periodically supplied by a calibrated pulser~\cite{arn04b}. The tower is
suspended from the 50 mK plate of the cryostat through a 25 mm copper bar
attached to a steel spring. The steel spring provides mechanical isolation from
vibration of the cryostat that can result in heating and spurious pulses in the
detector.

\begin{figure}
\vspace{9pt}
\begin{center}
\includegraphics{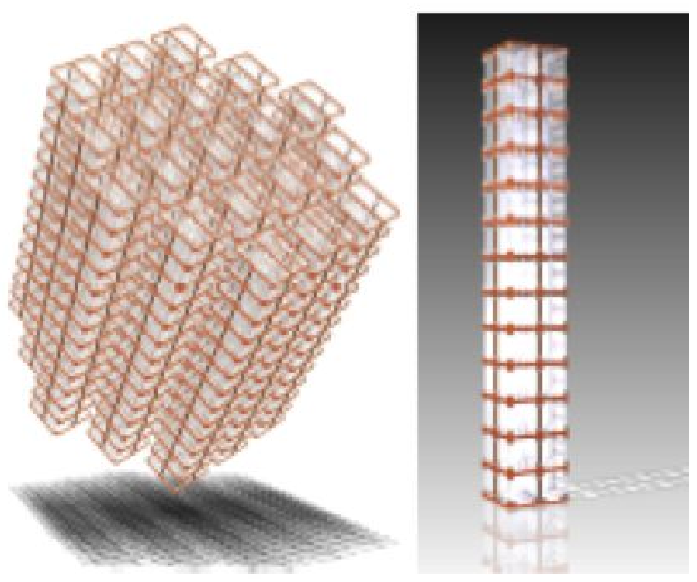}
\end{center}
\caption{The CUORE structure. Figure courtesy of the CUORE collaboration.}
\label{fig:cuore}
\end{figure}

The CUORE detector will be an array of 19 towers, each similar to the CUORICINO
tower except that all crystals will be $5\times5\times5$ cm$^3$ (See Fig.
\ref{fig:cuore}). It will have 988, 760-g bolometers, containing a total of
about 750 kg of TeO$_2$, or $\approx$200 kg of \nuc{130}{Te}. The expected
background is projected from the CUORICINO background, and the improvements
thus far achieved by the collaboration, as well as the effect of coincidence
cancellation due to the granularity of the detector. The predicted background
rate for the first phase of the experiment is 0.01 counts/(keV kg y).
Substituting into Eq. (\ref{eq:Sensitivity}) with $n_{\sigma}$ = 1, $a=0.348$,
$\epsilon=0.84$, $W=162$, $M=760$ kg, $t=10$ y, $\delta(E)=7$ keV  and $b=0.01$
counts/(keV kg y) the predicted sensitivity is \Tz(\nuc{130}{Te})$\approx 2.5
\times 10^{26}$ y. According to three recent nuclear structure calculations,
$F_N=(4.84^{+1.30}_{-0.64})\times10^{-13}$/y~\cite{rod06},
$F_N=2.57\times10^{-13}$/y\cite{cau06}, or
$F_N=5.13\times10^{-13}$/y~\cite{civ03a,civ03b}. These results and the
sensitivity above yield \mee-limits of 47 meV, 53 meV, and 45 meV, respectively.  A version of
CUORE that is isotopically enriched to 80\% in \nuc{130}{Te} would reduce these
values by a factor of 1.54 to
 31 meV, 34 meV, and 29 meV, respectively, which
covers the inverted hierarchy if \ml\ is 0 meV. The aim of the CUORE
collaboration is to reduce the background further.

    The CUORE experiment is approved and funded by the Instituto Nazionale de
Fisica Nucleare (INFN) in Italy, and was recently approved for construction and
put in the 2008 fiscal-year budget of the US Department of Energy. However
construction has already begun in Laboratori Nazionali del Gran Sasso (LNGS),
in Assergi, Italy with Italian funds. Data acquisition is scheduled to begin in
2011. A detailed description is given in \textcite{ard05}. The recent results
from CUORICINO with a total exposure of 11.8 kg-y  of  \nuc{130}{Te} provide a
lower limit of \Tz\ $\geq 3.0 \times 10^{24}$ y~\cite{arn07}. Using the nuclear
structure factors listed above, the bounds on  \mee\ are  0.42, 0.58, and 0.40 eV, respectively. The mission of the first phase of CUORE, then,
is to improve the half-life sensitivity by a factor of ~100, and that of  \mee\
by one order of magnitude.

\subsubsection{EXO (The Enriched Xenon Observatory)}
\label{sec:exo}

EXO is a proposed large next generation experiment that will search for the
\BBz\ in \nuc{136}{Xe}. The goal is to use between one and ten tons of Xe
enriched to 80\% in \nuc{136}{Xe} in a tracking Time-Projection Chamber (TPC).
The intent of the collaboration is to trap and
identify the daughter \nuc{136}{Ba} ion by laser spectroscopy. The
decay results in a \nuc{136}{Ba}$^{++}$  ion that will rapidly capture an
electron, leading to leading to \nuc{136}{Ba}$^{+}$ ion which is stable in Xe. These
ions can be positively identified via their atomic spectroscopy by optical
pumping with blue and red lasers. This technique was suggested by
\textcite{moe91} and is described in the context of this experiment by Danilov
{\em et al.}~\cite{dan00a,dan00b}. To exploit this method of positively
identifying the daughter nucleus, it will be necessary to capture the   ion at
the decay site and transfer it to an ion trap by
optical pumping. Alternatively, the ion could possibly be identified at the
decay site by directing the lasers. If either of these techniques works,
backgrounds that do not result in the correct ion will be eliminated. The
development of these techniques is a significant research and development
challenge, but one with a very high potential payoff.

\begin{figure}
\vspace{9pt}
\begin{center}
\includegraphics[width=8cm]{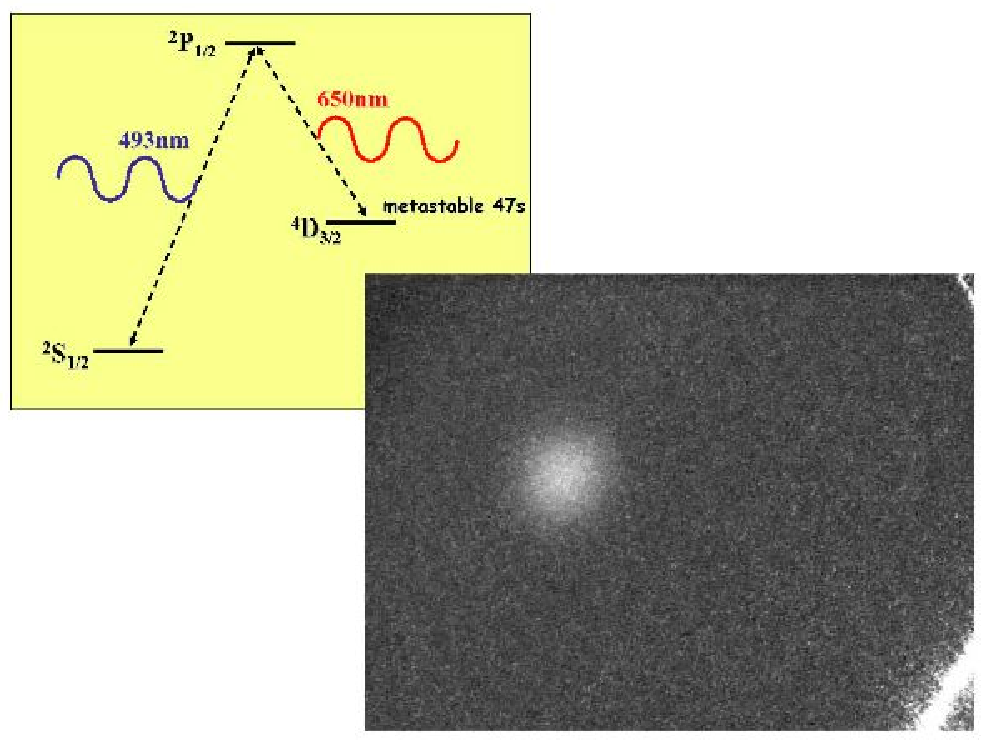}
\end{center}
\caption{The atomic level structure of a Ba ion indicating the levels of interest and a picture of the emission of a lone ion within an ion trap. Figure courtesy of the EXO collaboration.}
\label{fig:BaIon}
\end{figure}

The atomic level structure of the Ba ion is shown in Fig. \ref{fig:BaIon}. The
493 nm ground-state transition is very strong allowing the ions in the
$6^2S_{1/2}$ ground-state to be optically excited to the $6^2P_{1/2}$
excited state.  The decay of this state has a 30\% branching ratio to the
$5^4D_{3/2}$ metastable state. The Ba$^+$ ion is
identified by irradiating this $6^2P_{1/2}$ state with 650 nm red laser light,
exciting the ion back to the $6^2P_{1/2}$  state, which decays with a 70\%
branching ratio to the ground state by emitting 493 nm blue light. The
$6^2P_{1/2}$  excited state has a mean life of 8 ns and when saturated can emit
$\sim$10$^7$ 493-nm photons per second. This wavelength is compatible with the
maximum quantum efficiency of bialkali photomultiplier tubes. The bright
glowing spot shown in Fig. \ref{fig:BaIon} is a photograph of an actual
excitation in a test trap.

    The isotopic enrichment of tons of xenon is technically feasible because
xenon is a noble gas that can be directly introduced into a mass-separating
centrifuge with no chemistry. In addition, the residual xenon can be returned
for normal industrial use. This represents a significant cost saving in the
acquisition of the raw xenon. The xenon gas can also be continuously purified
during the operation of the experiment, in particular to eliminate the
ubiquitous radioactive radon and krypton isotopes.

\begin{figure}
\vspace{9pt}
\begin{center}
\includegraphics[width=8cm]{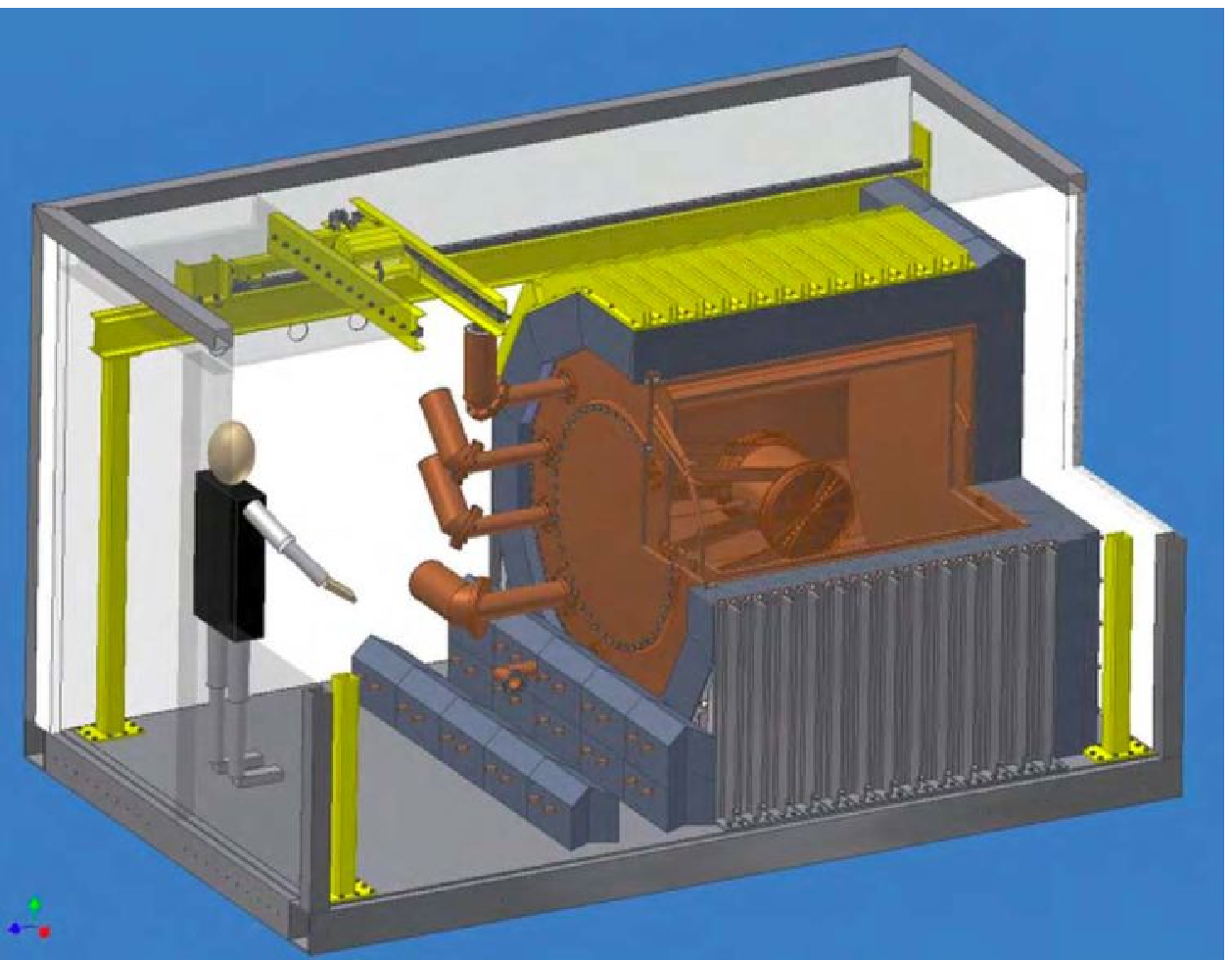}
\end{center}
\caption{An artists conception of the EXO-200 experiment. Figure courtesy of the EXO collaboration.}
\label{fig:EXO200}
\end{figure}

There are two activities that will presumably lead to the final construction of
the ton-level EXO with barium-ion tagging. One is the development of the
tagging procedure discussed above, and the other is the development of a
high-resolution liquid xenon TPC with good tracking ability.  The EXO
collaboration has approval and funding to construct a 200-kg liquid TPC with Xe
enriched to 80\% in  \nuc{136}{Xe}. This project is called EXO-200. It will not
have the barium-ion tagging feature; however, it is intended as an R\&D program
for the TPC itself. The enriched isotope has been purchased and is in house.
This construction of EXO-200 is presently underway, and the experiment will be
installed and operated in the Waste Isolation Pilot Project (WIPP) site in
Carlsbad, New Mexico. This site has an overburden of 1600 m.w.e.  to reduce the
backgrounds associated with cosmic rays. The detector will have an estimated
115 kg of fiducial volume contained in a thin-walled copper vessel 40 cm in
length. The copper was supplied by a German company, and has been assayed by
Inductively Coupled Plasma Mass Spectroscopy (ICPMS).  It contains the
following limiting levels of radioactivity:  \nuc{40}{K} $\leq$ 6 ppm,
\nuc{232}{Th} $\leq$ 0.5 ppt  and  \nuc{238}{U} $\leq$ 0.3 ppt. The copper for
the TPC was stored at the Deutches Electron Synchrotron Laboratory (DESY) in
Hamburg, Germany, in a concrete bunker to reduce activation from energetic
neutrons produced by cosmic ray muons. It has now been shipped to Stanford. It
is possible that in the future, the collaboration will use Teflon for the
vessel; however, the first vessel will be made from this copper. An artist's
conception of the EXO-200 is shown in Fig. \ref{fig:EXO200}.

The high voltage cathode will be located in the middle of the TPC volume.  The
design calls for up to 3.5 V/cm. At each end there are 114 x-wires and 114
y-wires (actually at 60$^0$ pitch with respect to the x-wires). These are for
charge collection, energy, and position information necessary for the
reconstruction of the trajectories of the electrons. A 1-MeV $\beta$ particle
creates about 50000 electrons in the liquid xenon. One interesting property of
xenon gas and liquid is that they are efficient scintillators. The
scintillation light will be detected by 258 large area photodiodes on each side
of the TPC. The use of both collected charge and scintillator light improves
the energy resolution. The combination was tested in a small chamber, and the
data imply that EXO-200 will have an energy resolution of about 1.5\%.

The installation in the WIPP site is scheduled to occur in mid 2007. The
cryostat has been constructed and tested at Stanford University at the time of
this writing. The EXO Collaboration anticipates reaching a sensitivity of about
0.3 eV for \mee. The actual design of the full-size EXO detector will occur
after the research and development of the barium-tagging scheme are completed.

\subsubsection{Next Generation \nuc{76}{Ge} Double-Beta Decay Experiments: \MJ\ and GERDA}
\label{sec:GeExperiments}

As stated earlier, the most sensitive \BBz\  experiments so far are the
Heidelberg-Moscow~\cite{kla01a} and the IGEX~\cite{aal02a,aal04} experiments. In
addition, the claimed observation discussed earlier was based on a
reanalysis of the Heidelberg-Moscow data~\cite{kla01a,kla03b,kla04a,kla04b,kla06}.
The goals of the \MJ\ and GERDA \nuc{76}{Ge} experiments are twofold, first to
confirm or refute the claimed evidence, and second, in the case of a positive
result, to make a measurement of the half life with a 20\% or better
uncertainty.  If the claim is not correct, the goal is to make the
most sensitive search for this decay. We will discuss the \MJ\ experiment first
because it will use improved ultra-low background cryostat technology similar
to that used in the IGEX experiment, which, except for the copper, was similar
to that used in the Heidelberg-Moscow experiment. By contrast, the GERDA
Collaboration is pursuing a novel technique of cooling and shielding the bare
Ge detectors by directly immersing them in liquid argon~\cite{heu95} and water
as low-Z shielding. This technique was the cornerstone of a proposal by the
GENIUS (GErmanium NItrogen Underground Setup) collaboration~\cite{kla97,kla01b} and
was tested in an initial experiment at Gran Sasso in the GENIUS test
facility~\cite{kla03a}. Similarly the GEM (Germanium Experiment for neutrino
Mass) proposal~\cite{zde01} also involved bare Ge detectors, but with
pure water forming the shield instead of liquid nitrogen. Neither the GEM
nor GENIUS collaboration is currently active. In the end, the \MJ\ and GERDA
collaborations envision a joint one-ton experiment, using whichever of their two
experimental techniques proves to be the most appropriate to reach the final
goals.

\nuc{76}{Ge}  decay has several significant advantages. First, all
three recent nuclear structure calculations are in fair agreement and predict
large nuclear structure factors, $F_N$. They are: $F_N= 1.22^{+0.10}_{-0.11}
\times 10^{-14}$ /y~\cite{rod03,rod06}; $F_N= 4.29 \times
10^{-14}$/y~\cite{cau06}; and $F_N= 7.01 \times 10^{-14}$
y~\cite{civ03a,civ03b}. $F_N$ is proportional to the square of the nuclear
matrix element, so the matrix element differences themselves are not large, whereas
in previous estimates they varied by a factor of 3. It is generally believed
that more reliable shell model calculations will be possible in the future for
this nucleus. In addition, \nuc{76}{Ge} experiments have the lowest background
demonstrated thus far. The background rates at IGEX and Heidelberg-Moscow were
$\approx$0.06 counts/(keV kg y), after the application of pulse-shape analysis.
Many of the backgrounds have been identified, and both future experiments are
projecting significant reductions. Finally, Ge detectors have the best energy
resolution of any of the proposed experimental techniques for double-beta decay
experiments. Resolutions of 3 keV at the decay energy of 2039 keV, or $\sim$0.15\%
full width at half maximum (FWHM), have been achieved. The main difference
between the \MJ\ and GERDA proposals are the method of cooling the detectors,
and shielding and background control. Both propose to use detector segmentation
and pulse shape discrimination to reduce backgrounds. The goals of both
proposed experiments is to probe the quasi-degenerate neutrino mass region
above 100 milli-electron (meV) region in the initial phases. In a subsequent
phase, and possibly with collaboration between \MJ\ and Gerda, the goal is a
one-ton experiment to explore the inverted hierarchy down to about 20 meV.

    To achieve these sensitivities it is necessary to fabricate the detectors from germanium enriched to 86\% in \nuc{76}{Ge}. This material can be produced in Zelenogorsk, Russia,  in the required quantities via centrifuge. The Ge metal is converted into a GeF$_4$ gas that is very stable at high temperatures, but must be converted into an oxide or metal after centrifugation. This is the same type of starting material that was used in both the Heidelberg-Moscow and IGEX experiments. No observable internal contamination of the Ge was found in either of those experiments.

{\bf The Proposed \MJ\ Experiment:} The original \MJ\ concept was for eight
modules with fifty-seven, 1.1 kg segmented Ge detectors each, for a total of
$\sim$500 kg of Ge, enriched to 86\% in \nuc{76}{Ge}~\cite{gai03}. The
detectors within each module would be arranged in a hexagonal configuration of
19 towers, each three detectors high, inside of a copper cryostat. The
cryostats will be electroformed as in the case of the IGEX cryostats. An
intense R\&D project is underway to reach the following levels of radio-purity
for the Cu: \nuc{214}{Bi}: 0.3 $\mu$Bq/kg, and \nuc{232}{Th}: 0.1 $\mu$Bq/kg.
These two isotopes produce the most serious of backgrounds, for example
the 2615-keV $\gamma$ ray from \nuc{208}{Tl} at the end of the \nuc{232}{Th}
chain. The background associated with this $\gamma$ ray is difficult to
eliminate sufficiently by analysis as discussed below. For this reason, the
reduction of thorium, as well as the \nuc{214}{Bi}, in the copper is a
critical step.

    The electroforming of the copper cryostat parts begins with the selection
of pure materials. The copper is electroplated on stainless steel mandrels from
a copper sulfate solution. The solution is prepared with semiconductor-grade
acids. The copper sulfate is purified by re-crystallization. This procedure was
tested by artificially contaminating the copper sulfate, and then
re-crystallizing it several times, measuring the radioactivity at each step.
The solution is constantly re-circulated through micro filters to remove oxides
and precipitates. It is also circulated through a barium scavenge filter to
remove radium. The plating tanks have a cover gas to reduce oxides. The copper
parts are machined periodically to prevent the formation of porous copper,
which can lead to vacuum leaks and poor structural strength. While this is a
time consuming process, the payoff is copper cryostat parts with less than 1
$\mu$Bq/kg of \nuc{232}{Th}.

\begin{figure}
\vspace{9pt}
\begin{center}
\includegraphics[width=8cm]{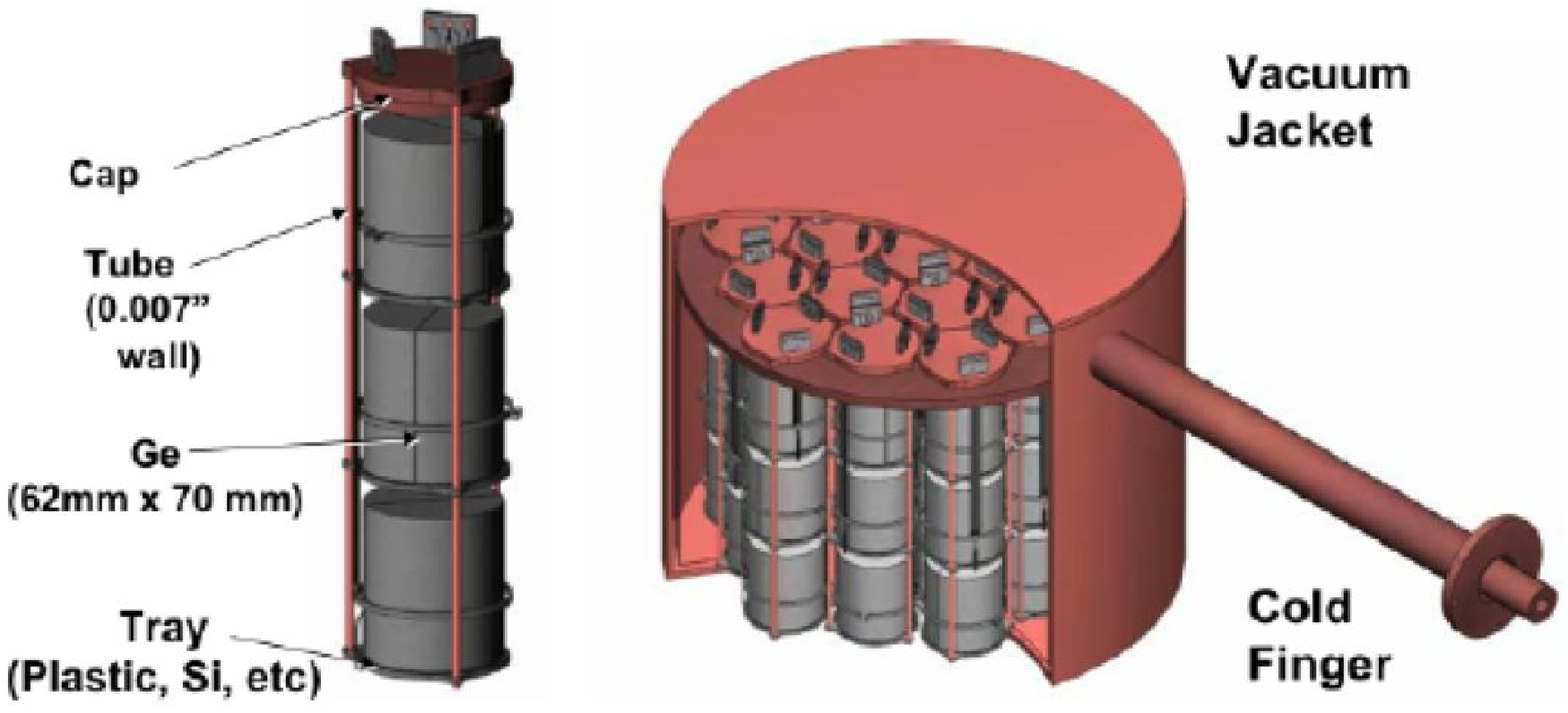}
\end{center}
\caption{The left panel shows a concept drawing of 1 of 19 {\em strings} of Ge detectors that would be contained within a \MJ\ module shown in the right panel. Figure courtesy of the \MJ\ Collaboration.}
\label{fig:MajTower}
\end{figure}

Each module of 57 detectors comprises 19 three-detector towers as shown in
Fig.~\ref{fig:MajTower}. The final design of the individual Ge detectors has
not yet been completed; however, the base plan has 1.1-kg segmented detectors
with a geometry similar to that of the Canberra CLOVER detectors. Some initial
studies concerning the joint use of segmentation and pulse shape analysis with
these detectors has been encouraging~\cite{ell06a}. Background events from
\nuc{56,57,60}{Co}, \nuc{65}{Zn}, and \nuc{68}{Ge}, formed in the Ge crystals
by spallation reactions from high-energy neutrons produced by cosmic ray muons
can be identified event-by-event by the deposition of energy in more than one
detector segment, and by pulse-shape discrimination.

\begin{figure}
\vspace{9pt}
\begin{center}
\includegraphics[width=7.5cm]{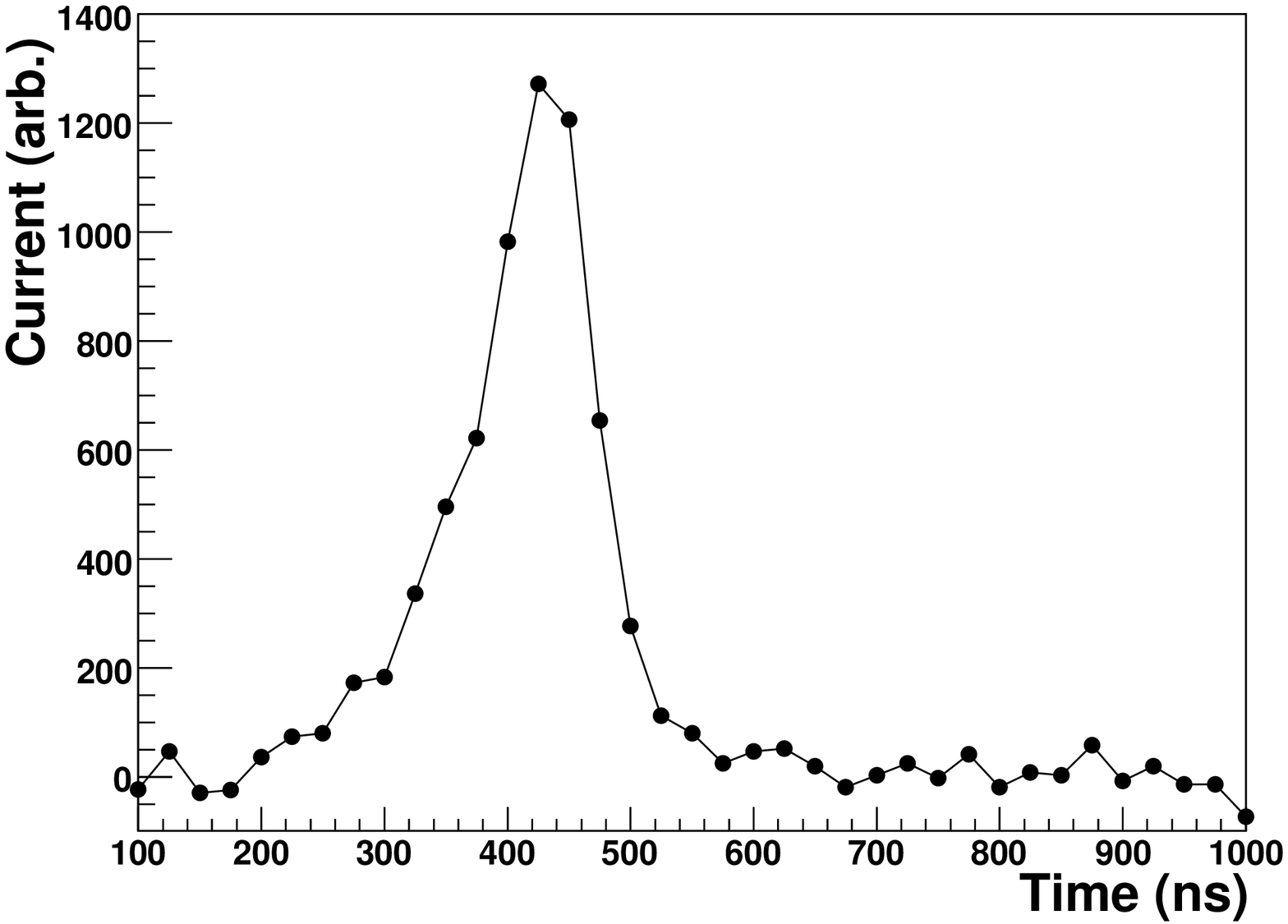}
\includegraphics[width=7.5cm]{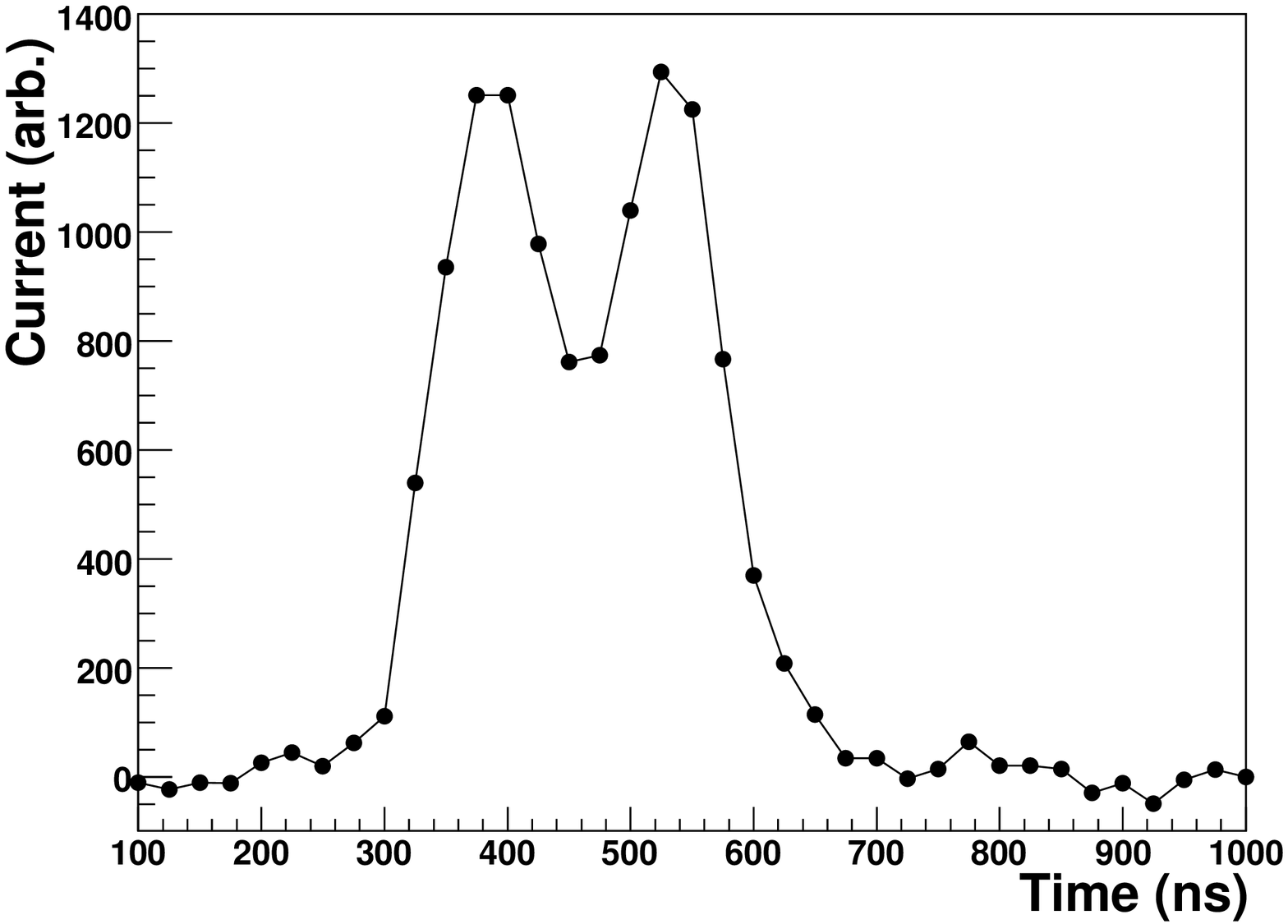}
\includegraphics[width=7.5cm]{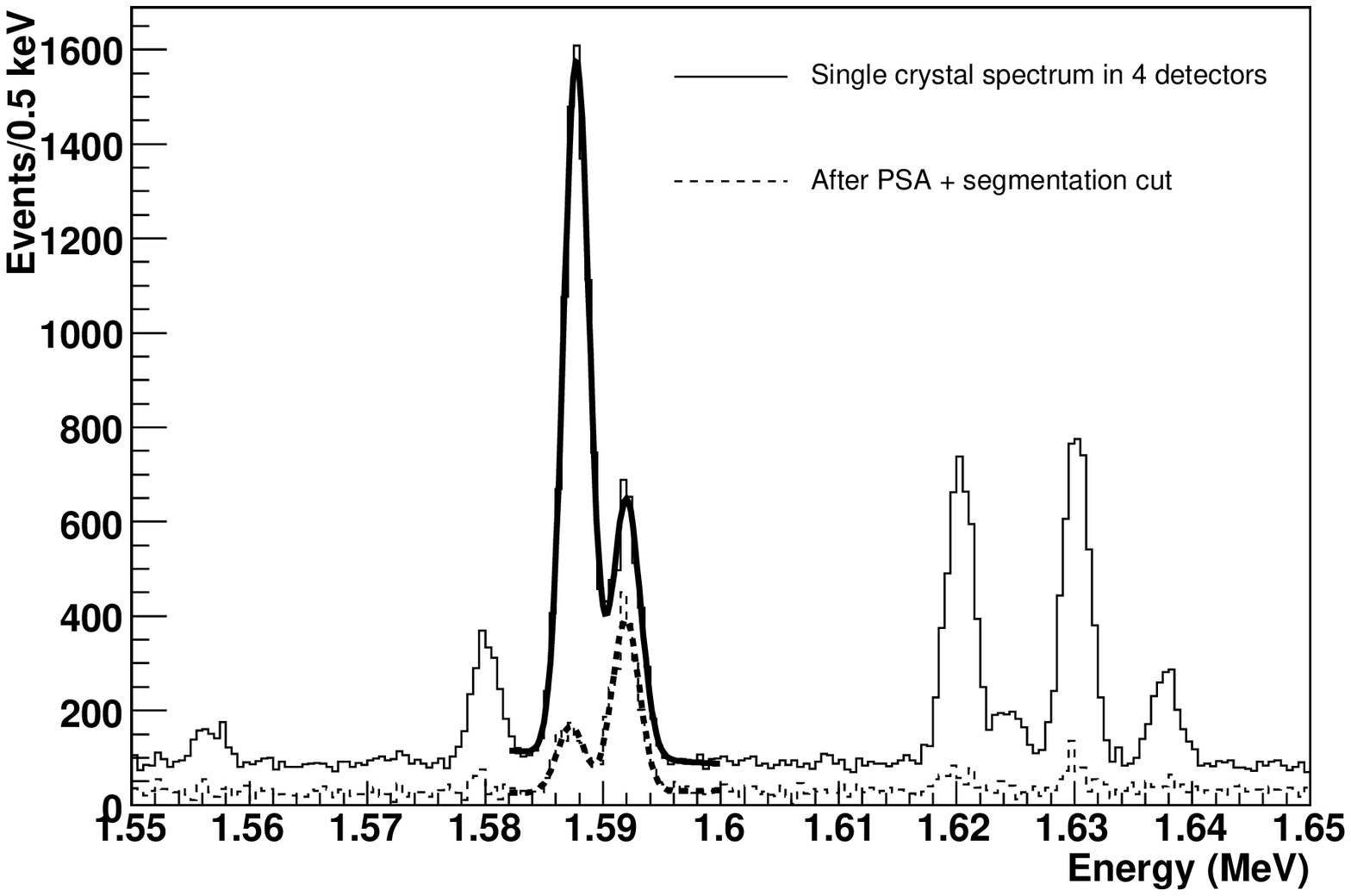}
\end{center}
\caption{The top panel shows the waveform of a candidate single-site
energy deposit in a Ge detector. The middle panel shows a candidate
multiple-site energy deposit of similar energy. The bottom panel shows how
the spectrum in the region near 1.6 MeV changes as pulse shape analysis
and segmentation cuts are employed. Figure courtesy of the \MJ\ Collaboration.}
\label{fig:MajPSA}
\end{figure}

Another significant source of background is externally produced $\gamma$ rays.
Gamma rays that can deposit 2 to 3 MeV in a Ge detector most likely scatter
several times before either being absorbed or leaving the crystal. This creates
multiple sites where clouds of electrons and holes are formed. The result is a
complex displacement current pulse that can be distinguished from one that
results from a single-site event like that expected from \BBz. Sample
experimental pulses are shown in Fig.~\ref{fig:MajPSA}.  The figure displays an
experimental spectrum of background after segment cuts and pulse-shape
discrimination have been applied.  All of the $\gamma$ ray lines have been
significantly reduced except for the 1593-keV double-escape-peak line of the
2615-keV $\gamma$ ray in \nuc{208}{Tl}. This well-known line is from the single
site creation of electron-positron pairs, in which both annihilation $\gamma$
rays completely escape the detector. One also sees that about 1/3 of the
continuum remains because it results from single Compton scatters.

Accordingly, by lowering the background in the copper cryostat parts, by
carefully screening and selecting the front-end electronic parts for low
background, and by applying segmentation and pulse-shape discrimination, the
\MJ\ Collaboration hopes to reduce the background in the region of
interest at 2039 keV by a factor of ~150 below that of IGEX or
Heidelberg-Moscow. The target half-life sensitivity, after an exposure of 0.46
ton-years, is $5.5 \times 10^{26}$ y. If we use the central value of the  nuclear 
structure factor, $F_N= 1.22 \times 10^{-14}$/
y~\cite{rod03,rod06}, the lifetime corresponds to a
\mee\  $\sim$61 meV to a 90\%
C.L. If the claimed observation is correct~\cite{kla04a}, then this exposure
would result in a measurement of the half life with an accuracy of ~10\%.

\MJ\ is a collaboration of about 70 physicists from 16 institutions in
Canada, Japan, Russia and the USA. At the time of this writing, it has been
approved for major research and development.

\begin{figure}
\vspace{9pt}
\begin{center}
\includegraphics[width=8cm]{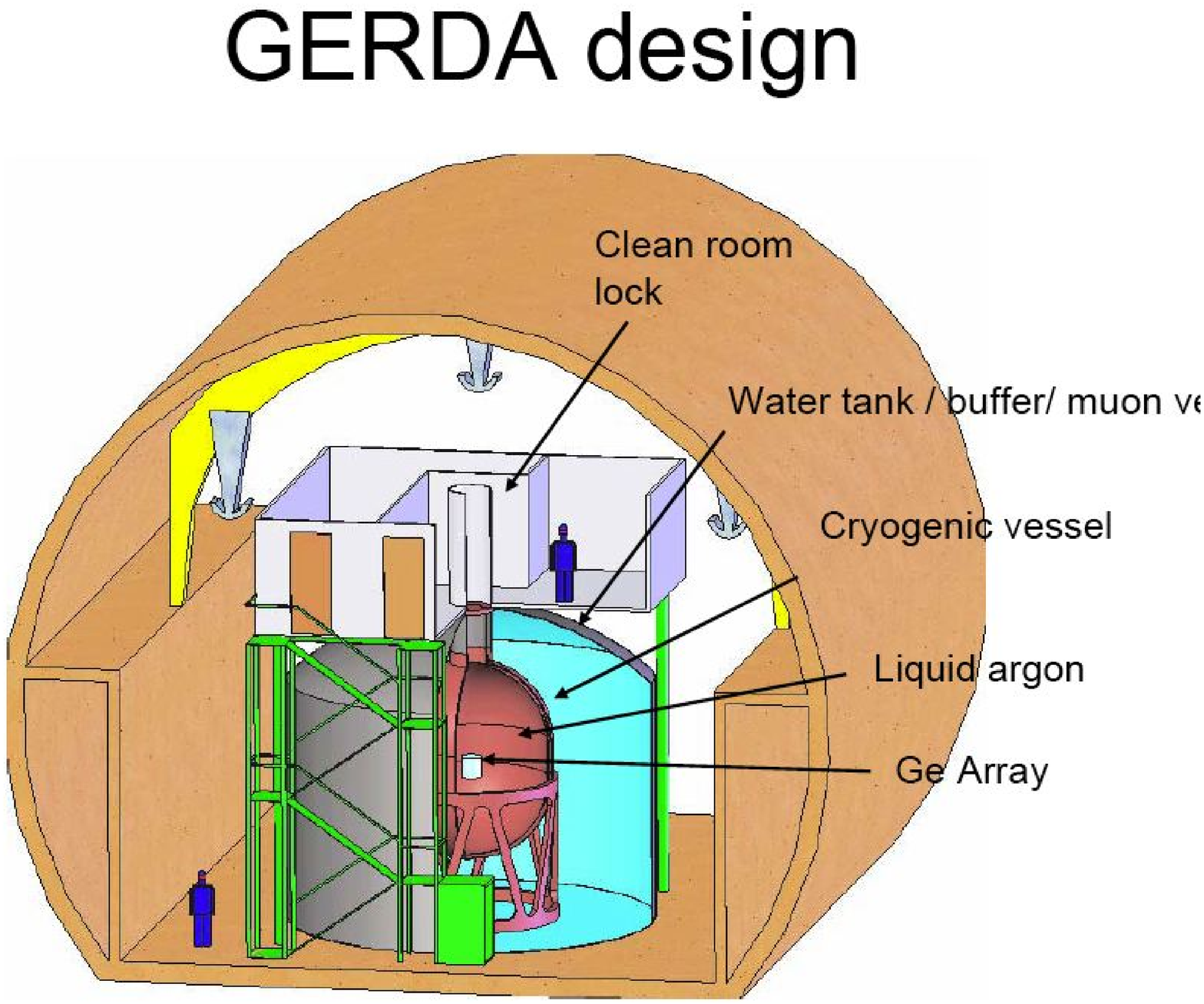}
\end{center}
\caption{An artist's conception of the GERDA experiment.}
\label{fig:GERDADesign}
\end{figure}

{\bf The GERDA Experiment (GERmanium Detector Array):} The Germanium Detector
Array (GERDA) is a planned array of Ge detectors fabricated from germanium
enriched to 86\% in \nuc{76}{Ge}, cooled and shielded by direct immersion in
liquid argon or nitrogen~\cite{abt04,sch05}. The facility is presently under
construction in the Laboratori Nazionali del Gran Sasso (LNGS) in Assergi,
Italy. It is planned in three phases. The first phase will proceed with the 5
detectors from the Heidelberg-Moscow (HM) Experiment, with 11.9 kg of enriched
Ge, and the 3 detectors from the IGEX Experiment with 6 kg of enriched Ge. All
of these detectors are closed-end p-type semi-coaxial detectors, and are
unsegmented. In their copper cryostats, these detectors achieved background
rates of $\approx$0.06 counts/(keV kg y) after pulse shape discrimination that
reduced the number of $\gamma$-ray events producing multi-site pulses. The
detectors have been underground since the early 1990's, the HM detectors in
LNGS, and the IGEX detectors in the Laboratorio Subteranio de Canfranc in
Spain. The goal of GERDA is to reduce the background to levels between 0.01 and
0.001 counts/(keV kg y) in the initial phases via ultra-clean materials,
pulse-shape analysis, segmentation, and live and bulk shielding with liquid
argon. An artist's conception of GERDA is shown in Fig. \ref{fig:GERDADesign}.

\begin{figure}
\vspace{9pt}
\begin{center}
\includegraphics[width=7.5cm]{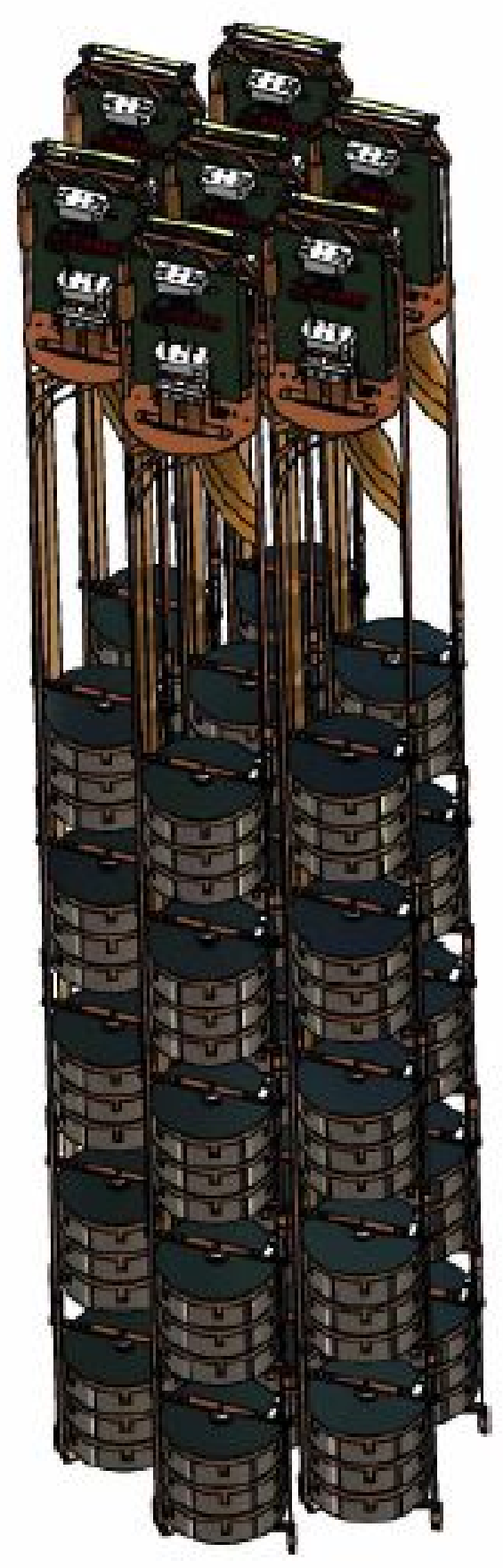}
\end{center}
\caption{Artist's conception of the GERDA Phase II tower.}
\label{fig:GERDATower}
\end{figure}

Phase I, comprising an active mass of 17.9 kg, is supposed to achieve a
half-life sensitivity of $\approx 3 \times 10^{25} $ y at (90\% CL). For Phase
II, an additional 37.5 kg of Ge, enriched to 87\% in \nuc{76}{Ge}, has
been delivered to the Max-Planck-Institute in Munich in the form of GeO$_2$. It
will be stored underground until fabrication begins. The collaboration is
testing one 6-fold segmented p-type detector, and one 18-fold segmented n-type
detector for possible use in Phase II~\cite{abt07}.
The goal of Phase II is to reach a half-life sensitivity of $\approx 1.4 \times
10^{26}$ y (90\% CL) and an exposure of 100 kg y. If we use the central value
of the nuclear structure factor $F_N= 1.22 \times
10^{-14}$/y~\cite{rod03,rod06},  this corresponds to a \mee\ of
$\sim$124 meV. In the case that the claimed observation by \textcite{kla04a}, is
correct, Phase I would observe it and Phase II would make an accurate
measurement of the lifetime. Fig.~\ref{fig:GERDATower} shows an artist's
conception of a
Phase II detector tower that would hang directly in the liquid argon.

The GERDA cryostat is a vacuum-insulated stainless cryostat with an inner
copper liner. The outer cylinder is 4.2 meters in diameter and 8.9 meters
long. The inner volume is 70 m$^3$, the mass of the empty vessel is 25 metric
tons and it will contain 98 tons of liquid argon. There will be a clean room
over the vessel and a rail system to lower and position the individual detector
strings.

Monte Carlo calculations were performed in collaboration with members of the
\MJ\ computation group to predict the background levels. GERDA will operate in
Hall-A of the LNGS with an overburden of $\approx$ 3600 m.w.e.  Accordingly, it
will have a muon veto shield to reduce backgrounds associated with cosmic-ray
muons. Given the low-Z material used as shielding in the vicinity of the
germanium array, the muon induced background will be 0.0001 counts/(keV kg y)
or less~\cite{pan07}. In addition to segmentation cuts, and pulse-shape
analysis, anti-coincidence signals between individual Ge detectors can be used
to tag and eliminate external $\gamma$ rays.  Coincidences from $\gamma$ rays in
decay chains can be used to further reduce backgrounds.  Finally, R\&D toward
the use the scintillation light of liquid argon as an active veto system is
underway.  The technique may be implemented in a later phase of
GERDA~\cite{dim06}.

The ultimate goal of a world wide collaboration would be a half-life
sensitivity of 10$^{28}$ y, corresponding to  $\approx$25 meV. This would
presumably be an experiment with the order of one ton of enriched germanium.
This endeavor may very well be a collaboration between the GERDA and \MJ\
groups, depending on which technology proves to be the best with respect to
background, detector stability, {\em etc.}. There exists a memorandum of
understanding to that effect, and it is clear that there will very probably be
only one Ge experiment in the world of that magnitude.

GERDA is a collaboration of 80 physicists from 13 institutions in
Belgium, Germany, Italy, Poland, and Russia. It was approved in November, 2004,
and is under construction at the time of this writing.

\subsubsection{COBRA (Cadmium telluride 0-neutrino Beta decay Research
Apparatus):}
COBRA~\cite{zub01} is an R\&D program developing CdZnTe
semiconductor crystal detectors (usually referred to as CZT detectors) for
\BBz. CZT detectors contain 9 isotopes that undergo \BB. The isotopes
\nuc{116}{Cd}, \nuc{130}{Te}, \nuc{114}{Cd}, \nuc{70}{Zn}, and \nuc{128}{Te}
all undergo $\beta^-\beta^-$, whereas the isotopes \nuc{64}{Zn}, \nuc{106}{Cd},
\nuc{108}{Cd}, and \nuc{120}{Te} undergo $\beta^+\beta^+$, $\beta^+$EC, or
EC-EC. The critical nucleus is \nuc{116}{Cd} as it has the highest \qval\ and
therefore the greatest sensitivity to \BBz. Although the \qval\ of
\nuc{130}{Te} is also high, it is below that of \nuc{116}{Cd} and therefore the Cd \BBt\ would
create a significant background for any \BBz\ study of Te. The isotopic
abundance of \nuc{116}{Cd} is 7.5\%, however, necessitating enrichment
for a competitive experiment. The existence of a number of $\beta^+$EC
isotopes within the detector and the high granularity of the experiment permit
a variety of studies in addition to the primary \BBz.

\begin{figure}
\vspace{9pt}
\begin{center}
\includegraphics[width=8cm]{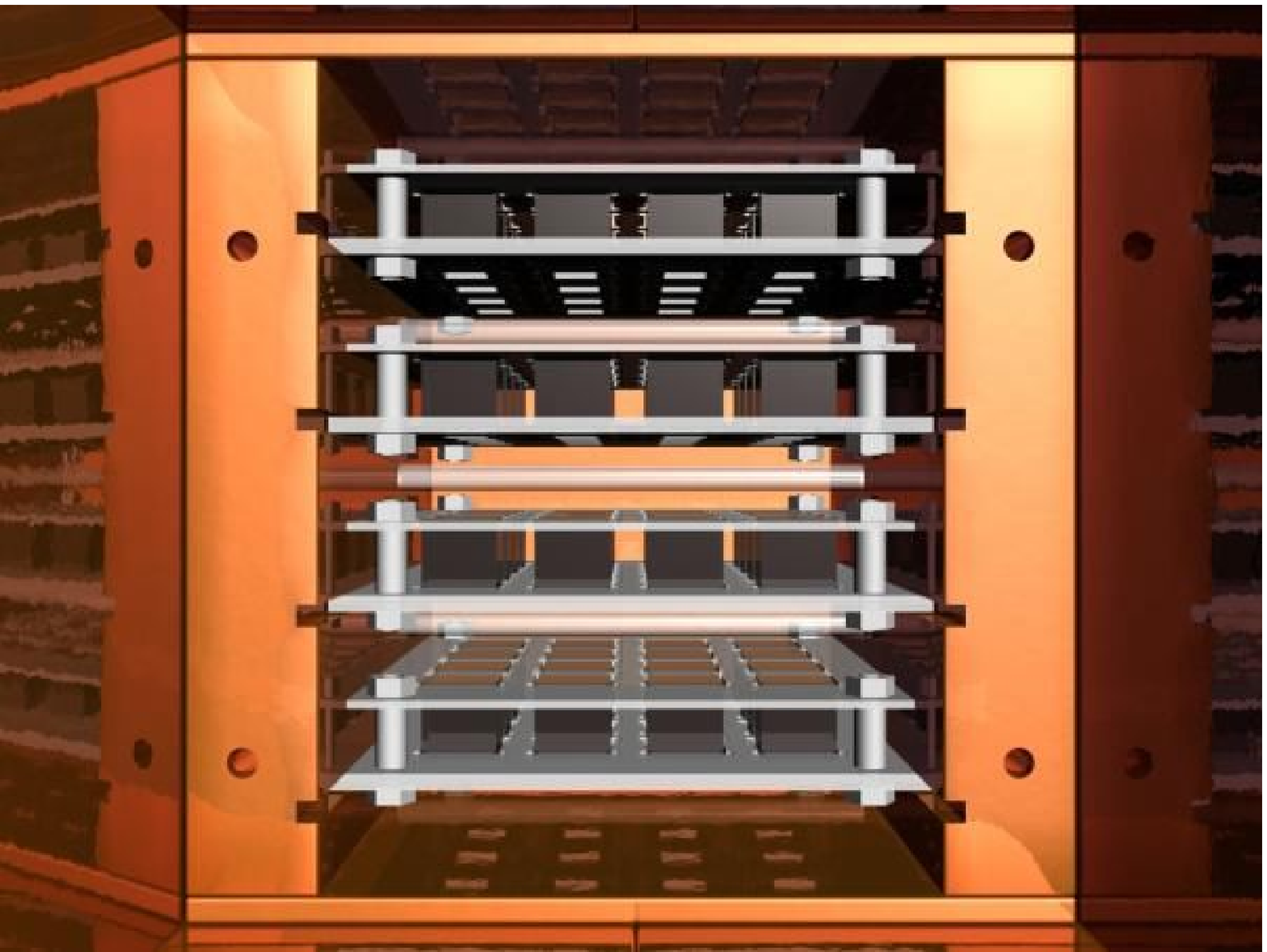}
\end{center}
\caption{An artist's conception of COBRA 64-detector setup. Figure courtesy of the COBRA collaboration.}
\label{fig:COBRALayout}
\end{figure}

\begin{figure}
\vspace{9pt}
\begin{center}
\includegraphics[width=8cm]{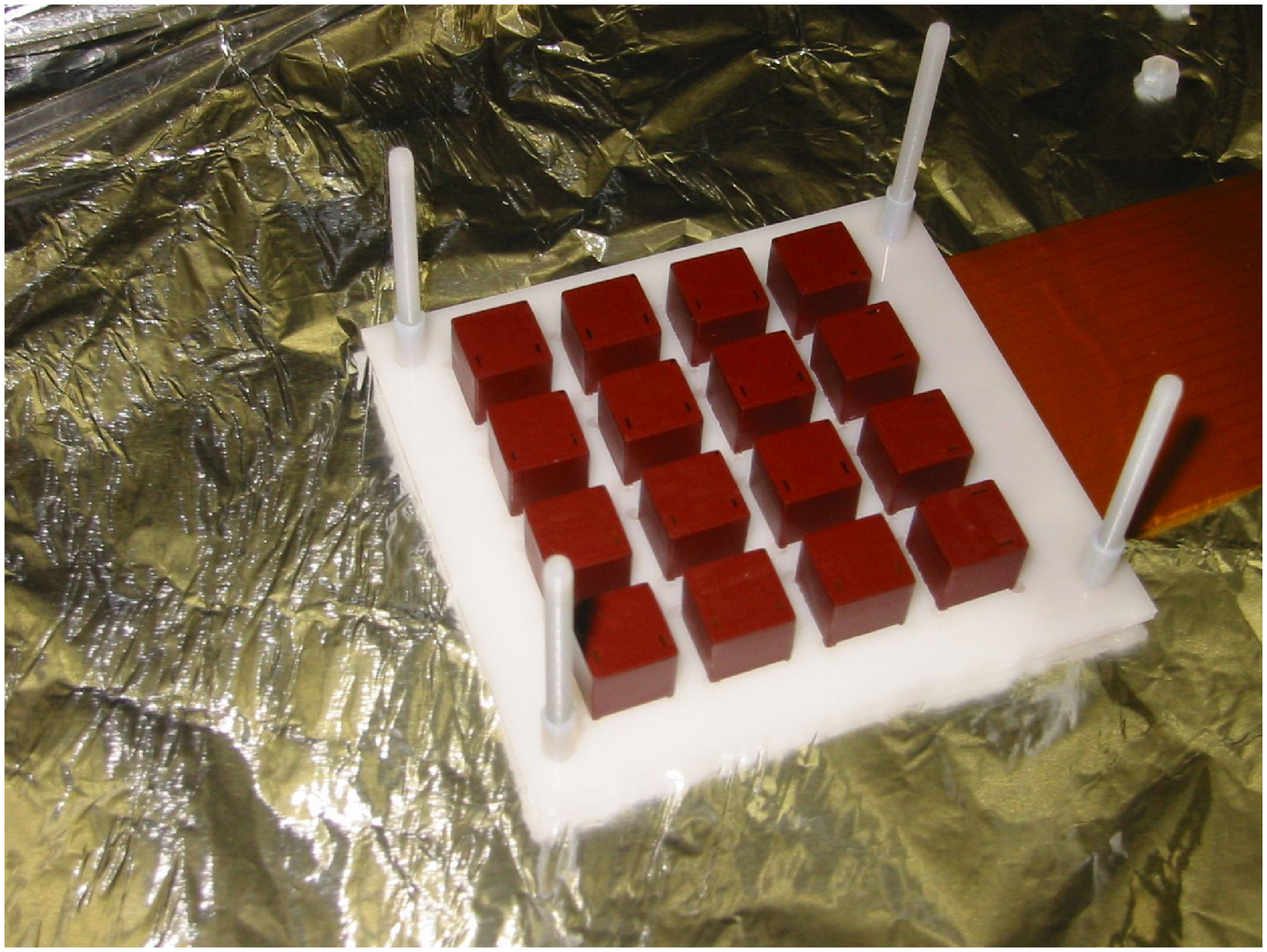}
\end{center}
\caption{A photograph of a detector holder containing 16 detectors, courtesy of the COBRA collaboration.}
\label{fig:COBRADet}
\end{figure}

The proposal is to operate 64,000 1-cm$^3$ CZT detectors
(\nuc{116}{Cd}$_{0.9}$Zn$_{0.1}$Te) with a total mass of 418 kg. The detectors
would be fabricated with Cd enriched to 90\% in isotope 116, so about 44\%, or
183 kg,  of the detector mass is the isotope of interest. The energy resolution
(FWHM) of the co-planar grid (CPG) detectors is better than 2\% at the
endpoint. If the background can be kept below 1 count/(keV t y), a half life
sensitivity of better than $10^{26}$ y is anticipated. To improve the
resolution of the detectors, cooling and new grid designs are being
investigated. The collaboration has operated a $2 \times 2$ detector prototype
and is assembling a $4 \times 4 \times 4$  prototype array at Gran Sasso
(Fig.~\ref{fig:COBRALayout}), with the first 16 detectors running since summer
2006 (Fig.~\ref{fig:COBRADet}).

Initial data~\cite{gos05} for the $2 \times 2$ detector array, including a
measurement of the 4-fold forbidden $\beta$ decay of \nuc{113}{Cd} and new
limits on positron decay modes of \BB, have been published. The collaboration
is also studying pixellisation of the detectors to improve spatial resolution
within a single detector for background rejection by particle identification.

\subsubsection{MOON (The Molybdenum Observatory of Neutrinos)}
\label{sec:moon}
The MOON experiment is unique because it has two applications. First, it is
a tracking detector to search for the double-beta decay of \nuc{100}{Mo} and
other isotopes. The energy and angular correlations of individual $\beta$ rays
are used to identify the $\nu$-mass term of the \BBz\ decay.  In addition it
will detect solar neutrinos, from the low-energy pp-neutrinos all the way
up to \nuc{8}{B} neutrinos. This dual function results from the unique
nuclear structure of the triplet \nuc{100}{Mo}, \nuc{100}{Tc}, and
\nuc{100}{Ru}~\cite{eji00,eji05}. In this discussion we will concentrate on
MOON as a double-beta decay experiment~\cite{eji06,nak06,nak06a}.

The MOON collaboration involves members from 12 institutions in Japan, the
Czech Republic, Russia and the US. The goal is to achieve an effective
sensitivity to \mee\ of $\sim$50 meV with one ton of \nuc{100}{Mo}. This
isotope has a high \qval\ (3034 keV), above the highest energy and troublesome
2615-keV $\gamma$ ray in \nuc{208}{Tl}. It has a correspondingly large
phase-space factor, about 7 times larger than that of \nuc{76}{Ge}, and a
natural isotopic abundance of 9.6\%. While its QRPA nuclear matrix
element is slightly smaller than that of \nuc{76}{Ge}, its predicted rate is
3.6 times faster~\cite{rod06}. On the other hand, for the value of  \gpp\ used
in the calculations of Civitarese and Suhonen, this ratio is 5.84. Neither the
old nor recent Shell Model calculations\cite{cau96,cau06} treat the case of
\nuc{100}{Mo}. The above two recent QRPA calculations differ by a 
factor of 1.6, and with other values of \gpp, as much as a factor of 4. This
makes a great difference in the predicted effectiveness of \nuc{100}{Mo} \BB\
experiments. Thus depending on the chosen value for \gpp, the predicted \BBz\
rate might vary by a factor of 4. It should also be pointed out that the MOON
technique, where the source does not comprise the detector, could also be used
with other isotopes, \nuc{82}{Se} and \nuc{150}{Nd}, for example. The choice is
made by considering \Mz, \qval, the \BBt\ rate, and other characteristics of a
given isotope.

The experimental technique is based on the ELEGANT V
design~\cite{eji91,eji91a}. The apparatus consists of multi-layer modules, each
with a thin film of enriched Mo, sandwiched between two position-sensitive
detector planes (thin MWPC chambers or thin PL fibers) and two solid
scintillator (plastic) plates, with all other modules acting as an active
shield. Precise localization of the two $\beta$-ray tracks enables one to select
true signals and reject background. In fact, background from radioactive
impurities and from neutrons associated with cosmic-ray muons is evaluated
by Monte-Carlo simulations and found to be acceptable ($<$ 0.3 /(t y)). One of
the key challenges in this type of experiment is achieving adequate energy
resolution to be able to recognize  \BBz\ events from background, and also from
the irreducible background from \BBt\ events. The half life for  \BBt\ of
\nuc{100}{Mo} was measured very accurately by the NEMO experiment as \Tt\ $=
7.11 \pm 0.02(stat.) \pm 0.54(syst.) \times 10^{18}$ y~\cite{arno05}. The theoretically
predicted \Tz\ of \nuc{100}{Mo}, for a \mee\ = 50 meV, is as long as $3.8
\times 10^{26}$ y~\cite{rod06}, although other \Mz\ calculations
indicate \Tz\ nearer to $10^{26}$ y. The \BBt\ rate is then between 7 and 8 orders of
magnitude faster than the predicted \BBz\ rate. Obviously, energy
resolution is a key issue in all such experiments. In the case of MOON, it is
being addressed by an intensive R\&D program~\cite{nak06}.

\begin{figure}
\vspace{9pt}
\begin{center}
\includegraphics{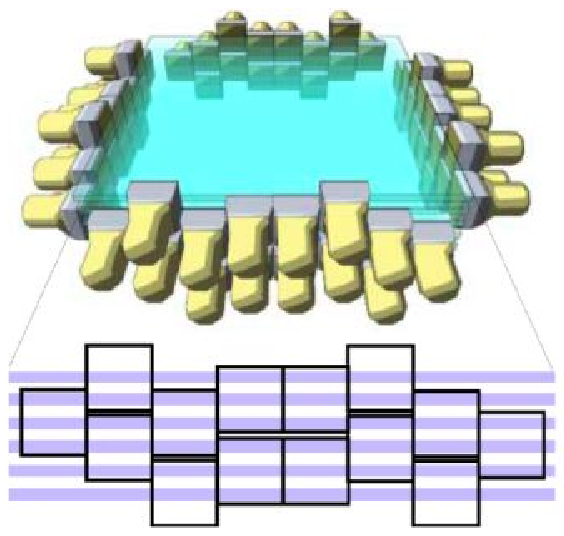}
\end{center}
\caption{A schematic of the MOON-I scintillator layout, courtesy of the MOON collaboration.}
\label{fig:MOONScint}
\end{figure}

\begin{figure}
\vspace{9pt}
\begin{center}
\includegraphics{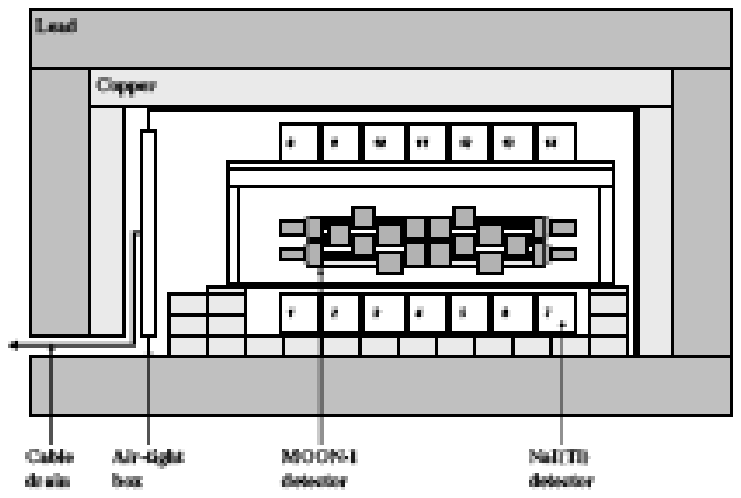}
\end{center}
\caption{A schematic of the MOON-I configuration, courtesy of the MOON collaboration.}
\label{fig:MOONOverview}
\end{figure}

The MOON-I prototype has the general features shown in
Fig.~\ref{fig:MOONScint}. The three 48 cm $\times$ 48 cm molybdenum films are
about 47 g each and are enriched to 94.5\% in \nuc{100}{Mo}. They are supported
and covered by aluminized Mylar films, 6 $\mu$g/cm$^2$. This prevents optical
cross talk between the scintillators. The six plastic scintillators are 53 cm
$\times$  53 cm $\times$  1 cm,  and are viewed by 56 Hamamatsu, R6236-01
KMOD photomultiplier tubes that are relatively low in \nuc{40}{K}. They are
optically coupled to the plastic scintillators with silicon ÒcookiesÓ, which
are made from silicon rubber 3 mm thick. The general geometry of the MOON-I
configuration is shown in Fig.~\ref{fig:MOONOverview}. The pattern of PMT
coupling to the scintillators shown in Fig.~\ref{fig:MOONScint} allows the {\em
hit-scintillator} to be identified by the PMT hit pattern. The measured energy
resolution is  $\sigma=4.8\pm0.2\% (\Delta E/E)$ at 1 MeV. This would yield a
$\sigma \approx 2.9$\% and FWHM of approximately 200 keV at the \qval\ of 3034
keV. The collaboration is attempting to improve it to $\sigma \approx 2.2$\% (as 
they have done in a small prototype) by adjusting the position dependent response.

The MOON-I experiment was performed with the detector inside the ELEGANTS-V
shield in the Oto underground Labotatory with an overburden of 1300
m.w.e~\cite{eji91,eji01}. At this level the muon flux was measured to be $4
\times 10^{-7} \mu$/(cm$^2$ s). There are NaI(Tl) detectors above and below the
MOON-I array to detect background $\gamma$ rays. The shield consists of 15 cm
of lead outside of 10 cm of oxygen-free, high conductivity copper. The shield
was flushed with dry N$_2$ that reduced the activity from Radon to a level of
125 mBq/m$^3$.

The MOON research and development program continues with the main goals of
reducing background as indicated by simulations (which have not yet been
reported), improving the energy resolution, and improving the position
resolution for tracking.  The sensitivity of any \nuc{100}{Mo} experiment to
\mee\ cannot really be
determined until more reliable nuclear matrix elements are available.
Experimental studies with charge exchange reactions are in progress for
\nuc{82}{Se}, \nuc{100}{Mo}, and \nuc{150}{Nd}, which are candidates for use
in MOON~\cite{eji00a}.

\subsubsection{NEMO-3 and SuperNEMO} 
\begin{figure}
\vspace{9pt}
\begin{center}
\includegraphics[width=8cm]{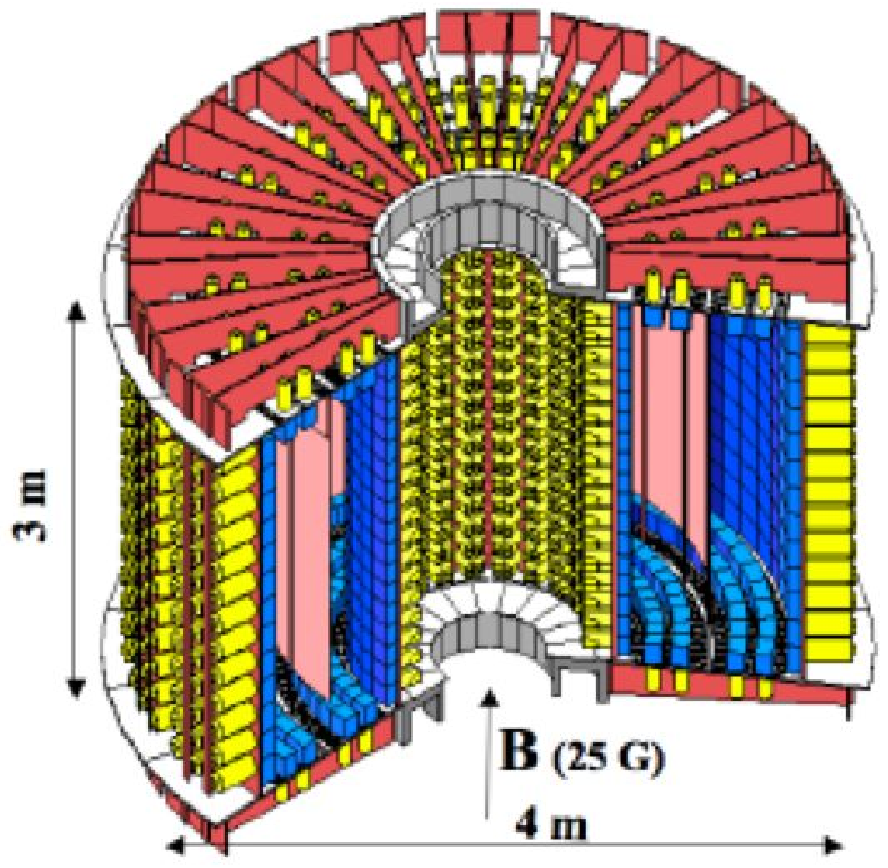}
\end{center}
\caption{A schematic of the NEMO detector, courtesy of the NEMO collaboration.}
\label{fig:NEMOOverview}
\end{figure}

{\bf NEMO-3:} The SuperNEMO
detector~\cite{bar04a} will be a tracking detector, as is its currently
operating predecessor, the NEMO-3 detector~\cite{arno05a}.  We discuss the
NEMO-3 experiment~\cite{arno05,bar06a} and its results to provide background.
NEMO-3 is the third generation of NEMO \BB\ detectors. An artist's conception is
shown in Fig. \ref{fig:NEMOOverview}. The detector has 20 segments of thin
source planes, with a total area of 20 m$^2$, that can support about 10 kg of
source material. It has a 3-dimensional readout wire drift chamber, with 6180
cells that operate in the Geiger mode for tracking.  The detector gas is He,
with 4\% ethyl alcohol, 1\% argon, and 0.1\% H$_2$O.  The tracking volumes are
surrounded by 1940 plastic scintillator-block calorimeters. The scntillator
detectors operate with thresholds of $\approx$30 keV, and have efficiencies of
50\% for 1-MeV $\gamma$ rays. The energy resolutions range from 11\% to 14.5\%
FWHM at about 1 MeV. The resolution is one parameter that will be given a lot
of attention in the design and construction of SuperNEMO. Good energy
resolution is critical for the discovery of processes
with long half lives. The source planes hang vertically in a cylindrical
geometry inside of a magnetic solenoid that generates a $\sim$25 G magnetic
induction field. Tracking in the magnetic field allows the differentiation
between electron and positron tracks, with only a 3\% chance of confusing the
two.

The detector is surrounded by 18 cm of low-background iron to reduce the
external $\gamma$-ray flux. Fast neutrons from the laboratory environment
are suppressed by an external shield of water, and by wood and polyethylene
plates. NEMO-3 is located in the Modane Underground Laboratory in Frejus,
France with an overburden of 4800 m.w.e.. The air in the experimental area is
constantly flushed and has a radon-free purification system serving the
detector volume. It reduces the radon to $\approx$200 mBq/m$^3$, and has a
capacity of 150 m$^3$/h. The measured radon activity in the detector was 4-5
mBq/m$^3$.

One of the great advantages of the NEMO-3 detector is that it can make
measurements on many different nuclei, and in fact has made
measurements on seven isotopes, and two blanks, simultaneously. The isotopes
were
\nuc{100}{Mo} (6.914 kg), \nuc{82}{Se} (932 g), \nuc{nat}{Cu} (621 g),
\nuc{nat}{Te} (491 g), \nuc{130}{Te} (454 g), \nuc{116}{Cd} (405 g),
\nuc{150}{Nd} (37 g), \nuc{96}{Zr} (9.4 g), and \nuc{48}{Ca} (7 g). The
\nuc{nat}{Te} and  \nuc{nat}{Cu} were used to measure the external background.

The measurement of the  decay of \nuc{100}{Mo} made with NEMO-3 truly sets a
standard for such measurements~\cite{arno05}. The result was: \Tt\ = $7.11 \pm
0.02(stat) \pm 0.54(syst) \times 10^{18}$ y. New bounds were set on
\BBz~\cite{bar06a}: \Tz\ $> 5.8 \times 10^{23}$ y for \nuc{100}{Mo}, and $2.1
\times 10^{23}$ y for \nuc{82}{Se} (90\% confidence level). The bounds on the
effective Majorana mass of the electron neutrino are \mee\ $<$0.6-2.7 eV for
\nuc{100}{Mo} and 1.2-3.2 eV for \nuc{82}{Se}. The NEMO-3 detector has been
operating since February 2003. The collaboration projects that the sensitivity
to \Tz\ of \nuc{100}{Mo} will be $2 \times 10^{24}$ y by the end of 2009. This
would reduce the bound on  \mee\ to $<$0.34-1.34 eV. It is clear that by that
time, the detector will have been operating for seven years and the resulting
sensitivity will not be competitive with next-generation experiments that are
designed to probe the inverted-hierarchy mass scale. To remedy that situation,
the SuperNEMO Collaboration was formed.

{\bf SuperNEMO:} This proposed experiment is a vastly expanded tracking chamber
of a modular design. The parent isotope will either be \nuc{82}{Se} or
\nuc{150}{Nd}. At this time there does not exist a source of kilogram
quantities of Nd isotopically enriched in \nuc{150}{Nd}, whereas  \nuc{82}{Se}
has been enriched by the gas centrifuge technique in Russia. In
Table~\ref{tab:nemo} below we briefly compare the experimental
parameters of NEMO-3 to those of SuperNEMO.

\begin{table*}[htdp]
\caption{A comparison of the SuperNEMO design parameters with those of the NEMO detector.}
\label{tab:nemo}
\begin{center}
\begin{tabular}{c|c|c}
 NEMO                                                 &   Detector/Experiment                 &       SuperNEMO  \\
\hline
\nuc{100}{Mo}                                    &                Isotope                            &  \nuc{82}{Se} or \nuc{150}{Nd}  \\
                7 kg                                      &     Source Mass                            & 100-200 kg \\
                8\%                                       &  \BBz\ Detection Efficiency         &  $\approx$30\%   \\
\nuc{208}{Tl} $<$ 20  $\mu$Bq/kg & External Background                  &    \nuc{208}{Tl} $<$ 2  $\mu$Bq/kg  \\
\nuc{214}{Bi} $<$ 300  $\mu$Bq/kg & In the source foil                       & For \nuc{82}{Se}:\nuc{214}{Bi} $<$ 10 $\mu$Bq/kg \\
8\% at $\approx$ 3 MeV                     & Energy Resolution, FWHM       &       4\% at 3 MeV \\
        $2 \times 10^{24}$ y                   &     \Tz\   Sensitivity                         &    $> 10^{26}$ y  \\
0.3-1.3 eV                                            &  \mee\ Sensitivity                            &  $<$0.05-0.1 eV \\
\end{tabular}
\end{center}
\label{default}
\end{table*}%

\begin{figure}
\vspace{9pt}
\begin{center}
\includegraphics[width=8cm]{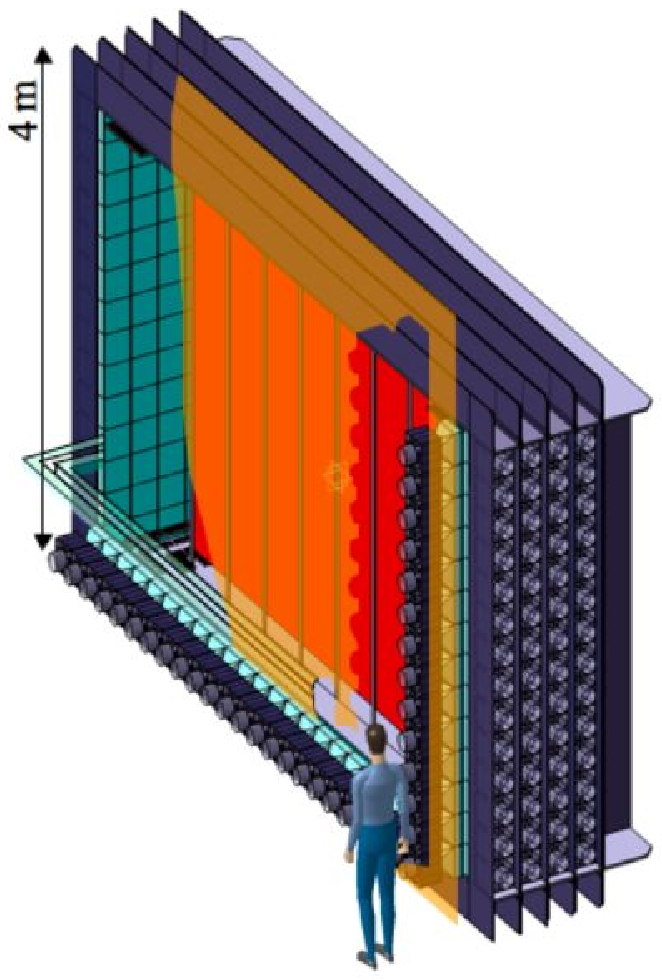}
\end{center}
\caption{A schematic of a SuperNEMO module. Figure courtesy of the NEMO collaboration.}
\label{fig:SuperNEMOMod}
\end{figure}

At the time of this writing, SuperNEMO is a collaboration of 26 institutes in
11 countries. An artist's conception of the detector modules is shown in
Fig.~\ref{fig:SuperNEMOMod}. The detector will
comprise 20 modules. Each module is designed to hold 5 kg of enriched isotope;
each has 12 m$^2$ of tracking volume, with 3000 channels of readout and 1000
photomultiplier tubes. The total detector will have 60,000 channels of drift
chamber readout, and 20,000 photomultiplier tubes.

One of the key issues is the background from the decay of \nuc{214}{Bi} to
\nuc{214}{Po} in the \nuc{238}{U} chain and \nuc{212}{Bi} to \nuc{212}{Po} in
the \nuc{232}{Th} chain. These are referred to as the BiPo backgrounds and a
prototype-test detector named BiPo is installed in the Laboratorio Subteranio
de  Canfranc (LSC) in Canfranc Spain to test the source foils. The BiPo
detector comprises two modules.  The intent is to test each source foil for the
BiPo backgrounds before installation in SuperNEMO.  The plan is to measure a 5
kg foil every month with a sensitivity of \nuc{208}{Tl} $< 2 \mu$Bq/kg  and
\nuc{214}{Bi} $< 10 \mu$Bq/kg.

    The proposed schedule is as follows: BiPo1 will run until 2008; BiPo will be under construction from mid 2007 until mid 2009. The final BiPo will be installed in LSC at the beginning of 2009. The construction of the first SuperNEMO module will start in early 2009 and be completed in mid 2010. The BiPo tests will continue throughout the construction of SuperNEMO. The full detector is scheduled to start operating at the beginning of 2012.

\subsubsection{CANDLES (CAlcium fluoride for studies of Neutrinos and Dark
matters by Low Energy Spectrometer)}

The proposed CANDLES experiment~\cite{ume06} is based on un-doped CaF$_2$
scintillators surrounded by a 4$\pi$  liquid scintillator live shield. The
scintillation light from pure CaF$_2$ crystals is in the ultra-violet region of
the spectrum and has a quite long decay time (1 $\mu$s) although the liquid
scintillator has a very short decay time ($\sim$10 ns).  One can employ this
difference to reject background signals from the liquid scintillator. A
wavelength shifter will be added to the liquid scintillator to match the
response of the photomultiplier tubes. The CaF$_2$ crystals are made with
natural-abundance calcium which is 0.187\% \nuc{48}{Ca}. The endpoint energy of
\nuc{48}{Ca}, 4.27 MeV, is the largest of all the \BB\ candidates, well above
the highest $\gamma$ ray from natural radioactivity, namely the 2615-keV
$\gamma$ ray in the decay of \nuc{208}{Tl}. It is interesting to note that
\nuc{48}{Ca} is a good candidate for shell-model analysis of the \BB\ matrix
elements.

\begin{figure}
\vspace{9pt}
\begin{center}
\includegraphics[width=8cm]{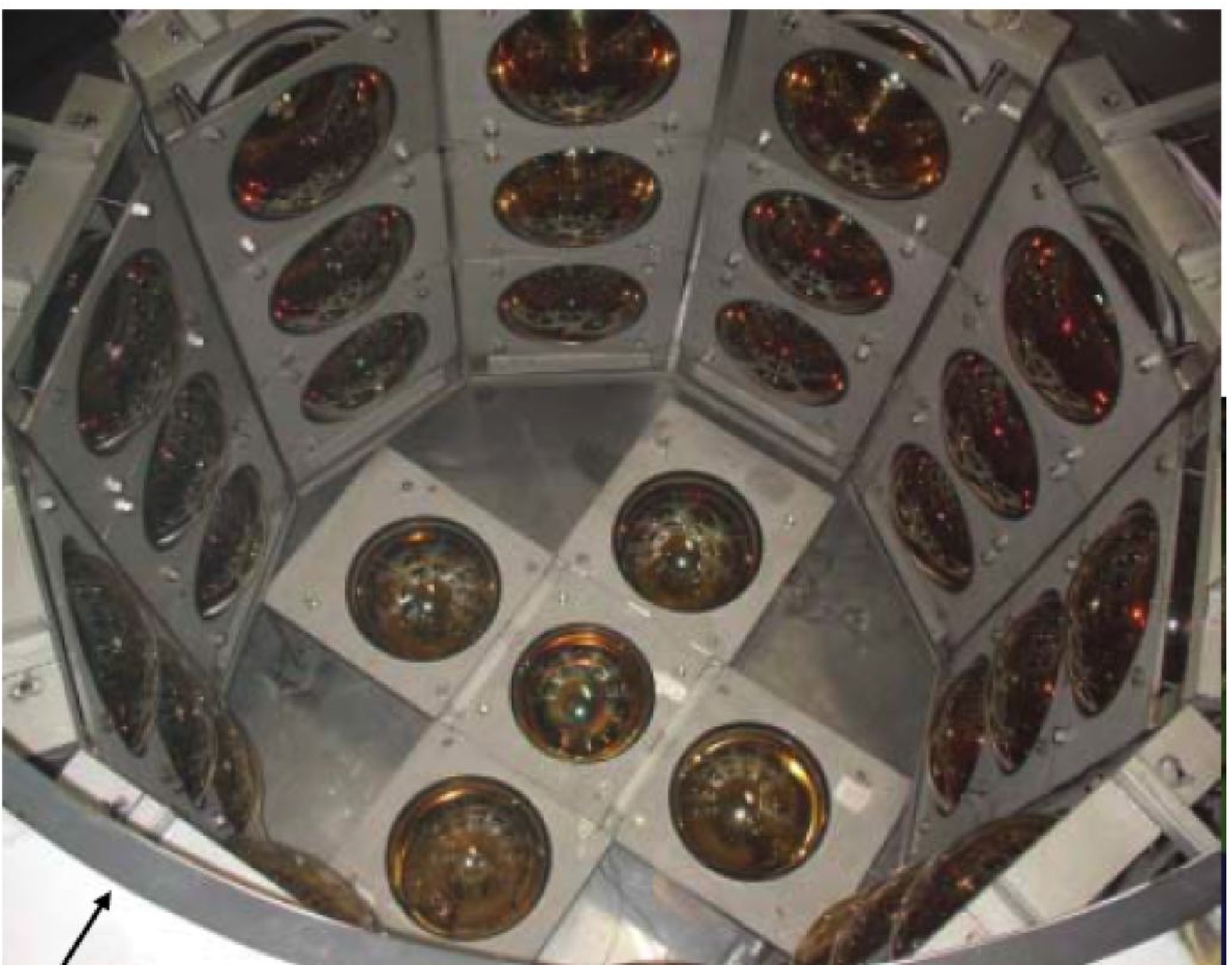}
\end{center}
\caption{A photograph of the CANDLES-III installation. Figure courtesy of the CANDLES collaboration.}
\label{fig:CANDLESOverview}
\end{figure}

The CANDLES project follows the ELEGANT VI experiments performed with CaF$_2$
scintillation detectors made with europium-doped CaF$_2$(Eu). In the ELEGANT VI
detector there were 23 CaF$_2$(Eu) crystals, each 4.5 $\times$ 4.5 $\times$ 4.5
cm$^3$ (290 g), surrounded by 46 CaF$_2$ (pure) crystals that acted as a shield
and as light guides. There was 6.7 kg of  CaF$_2$ containing 6.4 g of
\nuc{48}{Ca}. This in turn was surrounded by 38 CsI(Tl) crystals, each 6.5
$\times$ 6.5 $\times$ 25 cm$^3$, to act as a live shield. The entire array was
placed inside a bulk shield of copper, an air-tight box, lead, Li-loaded
paraffin and Cd, and B-loaded water. The experiment was located in the Oto
Cosmo Observatory with an overburden of 1400 m.w.e.. The energy resolution was
3.1\% at 4271 keV, and the detection efficiency was $\approx$49\%. The
experiment resulted in a new bound for \Tz($^{48}$Ca) $> 1.8 \times 10^{22}$
y~\cite{oga04}. According to recent shell model calculations~\cite{cau06},
\mee\  $<$ 23 eV. It will require several ton years for such an experiment to
be competitive. Currently CANDLES-III is under construction with 60 crystals
for a total mass of 191 kg at a sea level lab in Osaka (see
Fig.~\ref{fig:CANDLESOverview}).

The proposed CANDLES detector will comprise several tons of CaF$_2$(pure)
detectors, with almost no background, immersed in liquid scintillator that will
act as a veto detector. The target parameters for a 6.4-ton CANDLES are $<$ 3
$\mu$Bq/kg  of both \nuc{212,214}{Bi}, 3.5\% FWHM resolution, and 6 years of
counting with a total background of 0.3 per year under the expected \BBz\ peak.
The target sensitivity for this phase is \mee\ $\approx$0.1 eV. According to
shell model calculations~\cite{cau06}, this would require a sensitivity of \Tz\
$\sim$ 7.2 $\times$ 10$^{26}$ y.

The shell model calculations predict a value for the nuclear structure factor
$F_N$ (\nuc{48}{Ca}) = 3.61 $\times$ 10$^{-14}$/ y. The same model predicts
$F_N$(\nuc{76}{Ge}) = 3.22 $\times$ 10$^{-14}$/y and $F_N$(\nuc{130}{Te})= 2.13
$\times$ 10$^{-13}$/y. One can immediately appreciate the challenge of
performing an experiment with natural-abundance calcium. One would need about
400 times as many Ca atoms as Ge atoms, enriched to 86\% in \nuc{76}{Ge}, to
have the same decay rate.  In
comparison with \nuc{130}{Te}, one would need about 1000 times the natural
abundance Ca atoms as natural abundance of Te atoms to have the same theoretical
decay rates. The collaboration is investigating techniques for enriching Ca.

\subsubsection{Other Proposals} CARVEL~\cite{zde05} (CAlcium Research for VEry
Low neutrino mass) is a proposal to use isotopically enriched
\nuc{48}{Ca}WO$_4$ crystal scintillators  to search for \BBz\ of \nuc{48}{Ca}.
These scintillators, with natural calcium, have been tested for energy
resolution, internal background and pulse-shape discrimination that enables the
discrimination between alpha and gamma ray events. The energy resolution at
2615 keV is 3.8\% ($\approx$100 keV) which compares well with tracking
detectors. The total background between 3.8 and 5.4 MeV, after the elimination
of $\alpha$ background by pulse-shape discrimination, was 0.07 counts/(keV kg
y). With these parameters it was determined that with $\approx$100 kg of
\nuc{48}{Ca}WO$_4$, enriched to 80\% in \nuc{48}{Ca}, the half-life sensitivity
would be \Tz(\nuc{48}{Ca}) $\sim$ 10$^{27}$ y. This corresponds to a sensitivity in
effective neutrino mass of  \mee\ $\sim$ 55 meV~\cite{rod06}. With one ton of this
material, it was estimated that a half-life sensitivity of 10$^{28}$ y could
be reached, corresponding to \mee\ $\sim$ 0.02 eV. The difficulty with this
proposal is that there is no source of kg quantities of Ca enriched in
\nuc{48}{Ca}. While in principle Ca can be enriched by Atomic Vapor Laser
Isotope Separation (AVLIS), the effort and cost scale as the ratio of the final
to initial abundance. This cost would be more than 35 times that for
enriching an element with a 7\% natural abundance ({\em e.g.} \nuc{76}{Ge}).
This is a challenge for a ton-scale experiment.

DCBA~\cite{ish00} (Drift Chamber Beta-ray Analyzer)  is designed to search for
the
\BBz\ of \nuc{150}{Nd}. It comprises tracking chambers, a solenoid magnet and a
cosmic-ray veto detector. The source foils are vertical with a series of
high-voltage anode wires parallel to the source and the magnetic field. As the
$\beta$-particles are emitted they are given an extra velocity perpendicular to
the magnetic field, and are curved as they pass through the 1-atm He gas
mixture. The anode wires detect the drift electrons, while the cathode wires
across the chamber gather the drifted ions. The detector can measure the
momentum of each $\beta$ particle and can determine the vertex. Drift electrons
generate avalanche electrons on the anode wire plane. The pattern of avalanche
ions and the time of each signal are recorded with flash ADCs. From these data,
the x-position of the origin of the track is determined. The y-position is
fixed by the responding anode wire. The position on the anode wire of the
avalanche determines the z-position with pickup-wires, which are located
transversely near anode wires and pick up signals induced by the avalanche.
Each electron track is reconstructed in 3-dimensions, and the momentum and
kinetic energy are obtained from the curvature and pitch angle. DCBA-I will
contain natural Nd and will have $\approx 3.8 \times 10^{22}$ atoms of
\nuc{150}{Nd} (5.6\% natural abundance). It is expected to have a background of
$<$ 0.1 counts/(keV y). It will have an energy resolution of $\approx$100 keV.
 DCBA-II is planned to have $4.5 \times
10^{25}$ atoms of \nuc{150}{Nd}, in the form of Nd$_2$O$_3$ with Nd enriched to
90\% in \nuc{150}{Nd}. The expected half-life sensitivity is $4 \times 10^{24}$
y. According to recent QRPA calculations \cite{rod06} this would correspond to
\mee $\sim$ 0.12 eV. The R\&D for this detector is being carried out at KEK
with a test prototype DCBA-T.  On the basis of the that R\&D a new
project has been proposed, temporally named MTD (Magnetic Tracking Detector)
\cite{ish07}. The detection principle is the same as in DCBA, but the amount of
source installed is much larger. One module of MTD will contain natural Nd and
will have $\approx 6.7 \times 10^{24}$ atoms of \nuc{150}{Nd} (5.6\% natural
abundance) in the first experimental phase. The natural source plates will be
replaced with enriched material in the second phase. The enrichment of
\nuc{150}{Nd} is planned on a large scale with an international collaboration.

The CAMEO proposal~\cite{bell01} would install 1000 kg of \nuc{116}{Cd}WO$_4$
crystals in the liquid scintillator of BOREXINO~\cite{bell96} after its solar
neutrino program is complete. The described program would begin (CAMEO-I) with
thin ($\approx$15 mg/cm$^2$) metal isotopically-enriched Mo sheets sandwiched
between plastic scintillator and installed within the scintillator of the
BOREXINO Counting Test Facility (CTF). Better sensitivity could be achieved,
however, with an {\em active} source configuration of CdWO$_4$ crystals. For
CAMEO-II, 100 kg of crystals would be installed in the CTF. Measurements of the
intrinsic background of the crystals lead to an estimated \Tz\ sensitivity of
$\approx 10^{26}$ y. Installing 1 ton of crystals into the BOREXINO detector
could allow a sensitivity of $10^{27}$ y.  Because of the focus on solar
neutrinos at BOREXINO, this program is not currently active.

Another proposal that would exploit the BOREXINO detector or the CTF is to
dissolve \nuc{136}{Xe} in the scintillator~\cite{cac01}. In the most ambitious
design, a vessel of radius 2.7-m containing Xe and scintillator is itself
contained with the BOREXINO vessel (radius 4.25 m) filled only with
scintillator.  Outside this scintillating region is a pseudocumene buffer.
About 2\%-by-weight-Xe is soluble in scintillator with an appreciable
temperature dependence, so an estimated 1565 kg of Xe could be
used at 15$^o$ C.  Simulations indicate that phototube activity and \BBt\ will
dominate the background, resulting in a \Tz\ sensitivity of $2 \times 10^{26}$
y. This program is currently inactive.

The XMASS experiment~\cite{tak04} (Xenon detector for weakly interacting
MASSive particles) is a dark-matter search using liquid Xe as a target for WIMP
detection via nuclear scattering. The present detector uses 100 kg of natural
Xe with a fiducial volume of 3 kg viewed by phototubes. It is planned to expand
this to 1 ton and even possibly to 20 tons. This final configuration would
have a fiducial volume of 10 tons and would be used to study dark matter and
solar neutrinos. With such a large mass of Xe, the sensitivity to \BBz\ is
clearly of interest. The self shielding of the Xe is highly effective at
reducing the backgrounds at low energies ($<$500 keV) of interest to solar
neutrinos and dark matter. However, it is less successful at the higher energy
appropriate for \BBz. Present efforts are aimed at reducing the
backgrounds for the dark matter search. Results from this 100-kg prototype have
shown that the dark matter configuration will not work for \BB, which will
require a dedicated design. Work toward such a design is not a current priority
of the collaboration.

There are two proposals to use Ce-doped GSO scintillating crystals
(Gd$_2$SiO$_5$:Ce) to search for \BBz\ in \nuc{160}{Gd} (\qval\ = 1.73 MeV). In
one research effort, a 1.744-kg GSO crystal was used~\cite{wan02} to study
backgrounds. This crystal, coupled to a 2-inch phototube, was wrapped in Teflon
and black vinyl tape. The whole crystal was inserted into a well-type NaI
crystal. The system was contained within a Pb shield and a plastic-scintillator
cosmic-ray veto. The whole setup was situated in the bottom floor of a 7-story
building (12 m.w.e. overburden). The crystal was found to have a substantial
contamination from U- and Th-chain isotopes, which were not in equilibrium. In
addition, $\alpha$ decays of \nuc{152}{Gd} were present.  This program is no
longer being pursued. GSO crystals were also used in an experiment at the
Solotvina Underground Laboratory~\cite{dane01} at a depth of $\approx$1000
m.w.e.. In this test, a 635-g crystal was joined to a phototube by a 18.2-cm
plastic light guide. A passive shield of Cu, Hg and Pb surrounded the detector.
Internal radioactive contaminants within the crystal were the limitation of
this experiment, although the detector was cleaner than that discussed above.
This collaboration proposed putting GSO crystals within a large volume of
liquid scintillator (an idea similar to that of the CAMEO project) to help
control the background. Background levels will be a challenge for this
technology. Although GSO crystals are costly, the high isotopic abundance of
\nuc{160}{Gd} (22\%) would permit the use of natural Gd, which would be a great
relative cost reduction compared to other proposals, assuming that pulse shape
techniques could be used to suppress backgrounds. A modest R\&D program
continues on this project.

Finally, there is also a proposal~\cite{che05} to use the Sudbury Neutrino
Observatory (SNO)~\cite{bog00} for \BB, now that it has completed its solar
neutrino studies. The plan for the SNO detector is to replace the heavy water
with scintillator, beginning a program of low-energy solar-neutrino studies and
long-baseline-oscillation studies of reactor neutrinos, geoneutrinos, and \BB.
The parent isotope, situated in the liquid scintillator either as nanoparticles,
dissolved chemicals, or as absorbed gas, can be used as a source, with a
Nd-loaded liquid scintillator as the leading candidate.
The scintillator configuration is referred to as SNO+. A 1\% loading of natural
Nd in the scintillator would provide 10 tons of Nd, corresponding to more than
500 kg of $^{150}$Nd isotope. Simulations of this configuration indicate that
\BBt\ will be the dominant background because of the limited energy resolution.
However, if \mee\ is in the degenerate mass-scale region, the extremely high
count rate from this large isotopic mass would permit a statistical separation
of \BBt\ and \BBz. A development program for SNO+ is proceeding.

There are a number of research and development projects involving other
crystals that could be candidates for \BB\ detectors. The crystals under
investigation are CdWO$_4$~\cite{bard06}, PbWO$_4$~\cite{dane06},
YAG:Nd~\cite{dane05}, and ZnWO$_4$~\cite{dane05a}. A detector based on high
pressure Xe gas is also being considered~\cite{nyg07}.

\section{CONCLUSIONS}
The future for \BB\ is exciting. Technologies in hand will allow us to
measure at intriguing neutrino-mass levels.  Technological progress is
rapid, and developments in theory are improving our ability to
interpret the measurements. Strong \BB\ programs in both particle and nuclear
physics are the result.

The answer to the question: ``{\em Is the neutrino its own anti-particle?}",
is critical for theories of particle mass. It is also needed to help uncover
the reasons matter dominates over anti-matter in our Universe.
The neutrino mass will not only tell us about the
high scale at which the Standard Model breaks down ({\em e.g.}, through the
see-saw), but also will have
implications for the large-scale structure of the Universe.
Finally, lepton-number violaton is significant in its own right.

\BBz\ is the only practical method for investigating the
particle/anti-particle question.  And if the neutrino is its
own anti-particle, \BBz\ will have the best sensitivity to neutrino mass
of any laboratory technique. The \BB\ program outlined here,
consisting of several experimental results, will therefore greatly influence
the wider-particle physics endeavor.

Questions about how best to calculate the nuclear matrix elements
have led to developments in nuclear theory, despite the preoccupation
of most theorists with other problems.  Large-scale shell-model codes
are increasing in power, in part because of the need for better
matrix elements.  And a variety of experiments on nuclei are planned to
constrain those same matrix elements.  We should see some reduction in their
uncertainties.

Experimental progress in \BBz\ has also led to improved material-purification
techniques such as Cu electroforming, and to ongoing improvement in assay
techniques such as direct $\gamma$-ray counting and inductively-coupled
plasma-mass spectroscopy. The need for effective detectors has
led to many improvements in semiconductor ({\em e.g.} Ge, CZT) and bolometer
technologies that are now finding application in  basic science, medicine,
and homeland security. As researchers strive for 1-ton \BB\
experiments, advances in areas such as isotope enrichment and
detector production should follow.

Overall, the program in \BB\ has been and will continue to be very fruitful.

\section*{Acknowledgments}
The authors gratefully acknowledge useful comments from Alexander Barabash, Mark Chen, Hiro Ejiri, Gerry Garvey, Ryuta Hazama, Nobuhiro Ishihara, Andreas Piepke, Yoichiro Suzuki, Henry Wong, and Kai Zuber. A careful reading of the manuscript by Michael Smy, Hank Sobel, Vladimir Tretyak and Petr Vogel is greatly appreciated.

The work was supported by the US Department of Energy Grant DE-FG02-97ER41019, National Science Foundation Grants PHY-0139294 and PHY-0500337, and the Los Alamos National Laboratory Directed Research and Development program.

\bibliographystyle{apsrmp}
%\begin{thebibliography}{999}
\bibliography{RMPReferences}
%\end{thebibliography}

%\newpage

\end{document}